\documentclass[preprint,11pt]{aastex}
\usepackage[caption=false]{subfig}
\usepackage{float}
\shorttitle{VLA observations of Quasars}
\shortauthors{Gobeille et al.} 
\begin{document}
\title{VLA Observations of a Complete Sample of Radio Loud Quasars between redshifts 2.5 and 5.28: I. high-redshift sample summary and the radio images}

\author{Doug B.~Gobeille\altaffilmark{1,2}, John F.~C.~Wardle\altaffilmark{2}, C.~C. Cheung\altaffilmark{3}}

\altaffiltext{1}{Department of Physics, University of South Florida, Tampa, FL 33620-5700 USA}

\altaffiltext{2}{Department of Physics MS-057, Brandeis University, Waltham, MA 02454-0911 USA}

\altaffiltext{3}{Space Science Division, Naval Research Laboratory, Washington, DC 20375, USA}

\email{dgobeille@usf.edu, wardle@brandeis.edu, Teddy.Cheung@nrl.navy.mil}

\begin{abstract}

We present high resolution (arcsecond or better) observations made with the Karl G. Jansky Very Large Array of 123 radio-loud quasars with redshifts in the range $2.5 \leq z \leq 5.28$ that form a complete flux limited sample ($\geq 70$ mJy at 1.4 GHz or 5 GHz). Where possible, we used previous high resolution VLA observations (mainly A array at 1.4, 5 and 8 GHz) from the NRAO archive and re-imaged them (43 sources). For the remainder, new observations were made in the A array at 1.4 and 5 GHz. We show images of the 61 resolved sources, and list structural properties of all of them. Optical  data from the SDSS are available for nearly every source. This work represents a significant increase in the number of high redshift quasars with published radio structures, and will be used to study the properties and evolution of luminous radio sources in the high redshift universe.

\begin{center} {\em Version of \today} \end{center}
\end{abstract}


\section{Introduction}

This is the first of a series of papers that study the structure and evolution of radio-loud quasars over a very large range of redshifts. Here we present radio images made with the Karl G. Jansky Very Large Array (VLA) of 123 quasars with redshifts in the range $2.5 \leq z \leq 5.28$ that form a complete flux limited sample. High resolution (1 arcsec or better) radio images have been published for very few of these sources, so this work represents a significant increase in our knowledge of the radio structural properties of quasars in the high redshift universe. 

In 1988, Barthel and Miley presented evidence that the radio structures of quasars at redshifts greater than $z = 1.5$ exhibit clear differences compared to the structures of lower redshift quasars. Specifically, ``Distant quasars have a more bent, distorted appearance and are smaller than those nearby'' (Barthel \& Miley 1988). They suggested that both effects could be attributed to the effects of a denser and probably clumpier intergalactic medium at earlier epochs. Their $z \ge 1.5$ sample contained 80 steep-spectrum\footnote{The spectral index $\alpha$ is defined as $S_{\nu} \propto \nu^{-\alpha}$.} quasars, mostly observed by Barthel et al. (1988) with the VLA at 5 GHz. Further observations were made by Lonsdale, Barthel \& Miley (1993) 
with the VLA at 5 and 15 GHz. While the sample included  every then known high redshift, steep spectrum quasar north of $\delta = -30^{\circ}$, there was no clear flux limit and the completeness was unknown. The data for the 54 lower redshift sources were taken from the literature, and the sources were chosen to cover a similar range of luminosity to the high redshift sources, but they also do not form a complete sample. For brevity we will refer to this group of investigators as BML.

Beginning also in 1988, Hutchings and coworkers published three papers discussing the properties of radio loud quasars (Hutchings, Price \& Gower 1988; Neff, Hutchings \& Gower 1989; Neff \& Hutchings 1990). This group will be referred to as HNG. Their whole sample contains about 250 quasars, up to a redshift of 3.7. They chose sources to obtain a reasonably uniform coverage of the redshift-luminosity plane, but they also did not have a complete or rigorously defined sample. The VLA images for 40 quasars with $z \le 0.35$ were published by Gower \& Hutchings (1984). Those for 91 quasars with $0.35 \le z \le 1.00$ were published by Price et al. (1993). The images of the higher redshift quasars ($1.0 \le z \le 3.7$) discussed in those papers appear not to have been published. However, the data are in the NRAO archives, and we have re-imaged all those that lie within our sample area (see below).

BML and HNG had 33 sources in common, and while they agreed qualitatively on some results, they disagreed on their interpretation. HNG also found that ``Earlier epoch sources are smaller and slightly more bent,'' but found that these trends did not extend to the highest redshifts. BML found no dependence of  projected linear size on total radio {\em source} luminosity, while HNG found that projected linear size decreased with increasing radio {\em core} luminosity. This latter result can be understood in terms of beaming of the core emission, and the small angle to the line of sight of highly beamed sources. HNG appeared to discount beaming of the core emission as an important effect, and hence did not discuss this explanation. Instead, they favored an explanation in terms of core radio emission declining with time as the core-lobe distance increases.

Later, Blundell and coworkers (Blundell, Rawlings \& Willott 1999) studied the evolution of double radio sources using complete samples from the 3CR, 6C and 7C surveys. Complete samples, whose selection effects can be modeled and properly taken into account, are essential to distinguish whether a source property depends on redshift or luminosity. These were missing from the earlier studies. Blundell et al. studied classical FR II \cite{FR74} double radio sources without distinguishing between quasars and radio galaxies. They also selected their sources at relatively low frequencies (178 MHz for the 3CR sample and 151 MHz for the 6C and 7C sample), so the flat-spectrum radio cores made a negligible contribution to the observed radio emission. In section 7 of that paper, there is an extremely useful summary of previous work, in particular that pertaining to the variation of linear size with redshift.

The number of known radio sources at very high redshifts has mushroomed in recent years due to the contributions of deep surveys such as the FIRST survey (Becker, White \& Helfand 1995) and the Sloan Digital Sky Survey \citep[SDSS;][et seq.]{A03}. Unfortunately, high resolution radio imaging has not kept pace with the rate of high redshift identifications, leaving our picture of the high redshift radio universe very incomplete. The FIRST survey, for instance, used the B array at 1.4 GHz, and has a resolution of 5 arcseconds. The median angular size of double-lobed quasars is about 10 arcseconds or less for all redshifts greater than $z\approx 0.5$ \cite{WM74}, necessitating arcsecond or better angular resolution to determine their structure in sufficient detail, and to separate cores, jets and lobes. The CLASS survey \cite{M03}, searching for gravitational lenses, restricted itself to flat-spectrum sources and took only the briefest (30 sec) snapshots with the VLA in the A array at 8.4 GHz (0.2 arcsecond resolution). The data on steep-spectrum sources are even more sparse; very few have been imaged with adequate sensitivity and resolution (but see Pentericci et al. (2000) for VLA images of 27 $z>1.68$ radio galaxies).

The high redshift universe is important for understanding the formation and evolution of quasars. At redshifts of 4 to 5 we can observe clues to the formation of the earliest quasars and their central engines, as well as study how their properties change over cosmic time. We also know that the high redshift environment is very different than that of a low redshift source. The density of the intergalactic medium (IGM) scales with redshift as $(1+z)^3$ and the energy density of the cosmic microwave background radiation scales with redshift as $(1+z)^4$. While a higher IGM density at higher redshift might be expected to lead to smaller overall linear sizes, it will not by itself cause bends and distortions. Bent structures could suggest any of the following: a clumpy IGM, significant host galaxy velocities with respect to the IGM, or changing axes of the central engines. This is the natural scenario of galaxy and cluster formation through mergers at early epochs. Bent and distorted radio structure may therefore be a signature of young systems that are still in the process of formation (e.g. Overzier, Miley \& Ford 2007).

\section{A New High Redshift Sample} 

At redshifts greater than $z = 2.5$ the number of sources in the literature with well-determined (better than arcsecond resolution) radio structures drops rapidly, and the selection criteria become increasingly heterogeneous. We therefore wanted to construct a  sample of high redshift radio-loud quasars with a uniform flux limit and uniform optical data, including spectroscopic redshifts, for imaging with the VLA in the A configuration. We therefore chose an area of sky in common with the FIRST and SDSS fields, ensuring excellent optical data on every source, as well as wide field radio images. The overlapping areas of these surveys results in a sample field from 7 to 17.5 hours in right ascension and 0 to 65 degrees in declination.

Because the sample was to be used for more than one purpose, we chose a flux density limit of 70 mJy at {\em either} 1.4 GHz {\em or} 5 GHz. A query of the NASA/IPAC Extragalactic Database (NED) for sources in our field at high redshift ($z \geq 2.5$) produced 137 results. This list was cross-correlated with the NRAO archive where it was found that although there were adequate observations for 43 sources, the other 94 had no observations at all or observations that were too brief to be useful. In 2008 we used the VLA (proposal AW0728) to observe the 94 sources in the A array at 1.4 and 5 GHz. Fourteen sources were rejected from the sample for various reasons (they are listed in Table 3): three sources for being gravitational lenses, three for poor flux measurements, one that was misidentified, one that had an incorrect redshift listed in NED, and six that were high redshift radio galaxies. We ended up with a sample of 123 quasars with redshifts ranging between 2.5 and 5.28, with 48 of them at $z > 3.0$. The effect of selecting at both 1.4 and 5 GHz is that there are nine sources in Table 1 whose flux density at 1.4 GHz is below 70 mJy. One of them (J0744+2119) has a flux density listed in NVSS of 72 mJy but it is clearly variable and dropped below 70 mJy at the time of the VLA observations. The other eight have strongly self absorbed spectra and are brighter at 5 GHz than at 1.4 GHz. They are all unresolved by the VLA except for J1058+0443, which is an extremely core dominated triple source. Finally, the edges of the area covered by SDSS are somewhat ragged. There are therefore five sources 
in the field defined above that do not have SDSS spectra, but all of them do have spectroscopic redshifts and fit our selection criteria in every other respect.

The redshift distribution of the 123 sources is shown in Figure 1.

---Figure 1 here. ------

In comparison, BML had 11 sources at $z > 2.5$ and two at $z > 3.0$; HNG had 27 sources at $z > 2.5$ and 11 at $z > 3.0$, of which 16 remain in our survey. The Blundell et al. (1999) sample constructed from complete sub-samples from the 3CR, 6C and 7C surveys contains 8 sources at $z > 2.5$ and only 3 at $z > 3.0$. When combined with lower redshift sources in the literature our sample should make epoch-dependent properties much more apparent, and is large enough that we can make cuts by luminosity.

The sample discussed in the present work is selected for the most part at 1.4 GHz, which is a high enough frequency that emission from the flat-spectrum cores cannot be ignored. Since core radio emission (which is widely believed to be beamed) becomes increasingly dominant at high frequencies, and even more so at large redshifts due to the K-correction, we will want to construct complete sub-samples  that are free of orientation bias. This can be done by constructing sub-samples of sources whose lobe radio emission (widely believed to be un-beamed) is sufficient by itself to exceed the flux limit of the sample. This is analogous to what was done for 3CR quasars by Hough and Readhead (1989). 

The present paper is the first of a series of papers that investigate the evolutionary properties of radio-loud quasars over a wide range of redshift. Here we present VLA observations of the high redshift ($z \ge 2.5$) flux-limited complete sample.

\subsection{Sample Completeness}

Our sample has both radio and optical flux limits. The radio flux limit of 70 mJy at 1.4 GHz represents a strong source in the FIRST survey and is far above both the noise limit and the confusion limit (their limiting flux density is about 1 mJy). The completeness of our sample therefore hinges mainly on the algorithms used by SDSS to select quasar candidates for spectroscopic observation. 

The quasar selection algorithm (which determines whether or not to take a spectrum) is described in detail by Richards et al (2002). They do not try to see redshifts over 5.8 (which is where the minimum transmission between the {\em i-} and {\em z-} band filter curves at 8280 \AA \,\, is straddled by Ly $\alpha$ emission) and they require flux in at least two optical bands. After editing bad data, they preferentially target point-source matches to FIRST radio sources (within 2 arcsec) without reference to the optical colors. The optical flux limit is set by the number of optical fibers per plate (typically 80) that are assigned to quasar candidates. That limit is {\em i} = 19.1 for redshifts up to $z = 3.0$, and {\em i} = 20.2 for redshifts $z \geq 3.0$. Our sample therefore has different optical flux limits for the 75 quasars with $2.5 \leq z < 3.0$ and the 48 quasars with 
$z \geq 3.0$.

This algorithm should be extremely efficient at finding radio-loud quasars, because of the positional match with a FIRST radio source, and the absence of any color bias. At low redshift, the extended emission of a quasar might be significantly resolved by FIRST (but the radio core, coincident with the optical position will still be visible). We note that Best et al. (2005) achieved 95\% completeness by comparing a SDSS spectroscopic galaxy sample with both FIRST {\em and} the lower resolution NVSS survey \cite{CCG98}.

If a radio source is significantly resolved by the FIRST survey with its resolution of 5 arcsec, then the total flux density at 1.4 GHz could be underestimated, leading to its exclusion from our sample if the true flux density is close to our 70 mJy limit. However, at redshifts above $z = 2.5$ resolution effects should not be a problem. The observations by Barthel et al. (1988) include 11 quasars with $z \geq 2.5$, and the largest of these has a total angular extent of 12.3 arcsec. In the observations reported here of 123 quasars with $z \geq 2.5$, the largest total angular extent we have found is 31 arcsec (J0736+6513). What is important here is that although the total extent is well resolved in FIRST, the two lobes and the core are all compact and each would be unresolved or nearly unresolved (e.g. see Fig. 2(d)), and the total flux density would not be significantly underestimated. There are four additional sources whose total extents are in excess of 15 arcsec, but again their individual features are all relatively compact and their total flux will not be significantly underestimated. It therefore seems reasonable to suggest that the completeness of our sample is not adversely affected by not making use of NVSS in addition to FIRST, and therefore it should be similar to the 95\% completeness of the Best et al. (2005) sample.

Because of the relatively high finding frequency of our sample, there is an additional effect which is not easy to evaluate. Every source in our sample exhibits emission from a compact radio core coincident with the optical quasar, and many are strongly or completely core-dominated. The core emission is generally thought to be beamed and is expected to exhibit variability on a timescale of months to years. Thus a source may at different times be above or below the 70 mJy flux density limit. We simply use the 1.4 GHz total flux density listed in NED from FIRST or NVSS. The net effect of source variability is a slight over-representation of sources just above the flux limit. But since our primary interest is the radio morphologies rather than, for instance, the source counts, $N(\>S)$, source variability is not a significant concern to us.

Finally, Best et al. claim a {\em reliability} of 98.7\%. Completeness refers to sources that are not in the sample but should be, according to the stated selection criteria. Reliability refers to sources that appear to satisfy the selection criteria but should not be included in the sample. This can be for a variety of reasons: the majority of sources that we rejected on subsequent scrutiny were either gravitationally lensed, confused, or identified with high redshift galaxies. All fourteen rejects are listed in Table 3. We are not able to make a numerical estimate of the sample reliability in the manner of Best et al., but we have made every effort to find all the sources that should not be in the sample.

\section{Observations and Data Reduction}
\label{s:obs}

The data used to image the radio structure of 43 of the sample sources were taken from the NRAO archive of VLA observations. For sources with inadequate or no archival observations (80 sources), we used our own observations (AW0728) with the A array at 1.4 and 5 GHz. All primary source flagging and calibration was done in the NRAO software package AIPS. All the images were made with the Caltech software package DIFMAP (Shepherd 1997). 

The archival observations contain a wide variety of observing frequencies, array configurations and times on source. In most cases they consisted of brief snapshots lasting from 2-10 minutes, though some data sets with a minute or less of data were used if no other data were available. For the AW0728 observations, the total time on source was 15 minutes, broken into three five minute scans at different hour angles.

The typical dynamic range of the images is several hundred to one, with about one third of the images exceeding 1000:1. All images had thermal noise close to the theoretical limits (Ulvestad et al. 2009). We estimate that thermal noise contributes $\approx 1\%$ errors on all but the weakest components and briefest snapshots. The flux density scale was set in our own observations and in most of the archival observations by observing 3C286. We estimate that the total error in flux density  is $\leq 5\%$ in nearly all cases, and we apply this uniformly throughout the sample.

Sometimes, data sets from different arrays were combined within AIPS and imaged to yield a hybrid multi-array image in order to image better diffuse and large-scale structure. Although it is possible that the shorter observations may have missed some weak, larger scale extended structure, comparing our images with the lower resolution maps available from FIRST suggests that we are missing very little extended structure.

\section{Results and Images}
\label{s:results}

In this section we present the whole sample of 123 high redshift ($z \geq 2.5$) sources. Images are shown for the 61 sources that are appreciably resolved in  Figures (2) to (8). The image chosen to represent each source represents the best available combination of array and frequency to display the source morphology.  All images are centered on the quasar core component as best determined by optical identification and spectral index. In every image a cross marks the position determined by SDSS. The half-power restoring beam is shown as a filled ellipse in the lower left hand corner of each image. The source name and frequency are given in the figure captions. Contour levels and peak flux densities are listed in Table 2, and notes on individual sources are given in \S 4.1. Contours are logarithmic and increase by factors of $\sqrt{2}$.

The whole sample is listed in Table 1 along with some of the source properties measured from the images. 

\textit{Column (1)} - J2000 Source name in IAU format.

\textit{Column (2)} - Redshift from NED. Nearly all of them come from the SDSS, and where possible we have used the improved values listed in Hewett \& Wild (2010).

\textit{Columns (3) and (4)} - Right Ascension and Declination of the optical quasar. 

\textit{Column (5)} - {\em i}-band magnitude taken from the SDSS. For a few sources noted with asterisks*, they are $R2$-band magnitudes from the USNO-B1.0 (Monet al al.\ 2003).

\textit{Column (6)} - Morphological type as described in \S 4. Cores (C), Cores with a small extension (E), Doubles (D), and 
Triples (T). The classification will be discussed further, below. Note that every source exhibited a compact radio core coincident with the optical quasar.

\textit{Column (7)} - Total observed source flux density at 1.4 GHz: the total source fluxes were obtained from the final images. 

\textit{Column (8)} - Core flux density at 1.4 GHz. These were also obtained from the final images. In columns (7) and (8), a conservative estimate of the error is $\pm 5\%$, which should be applied to all the listed flux densities.

\textit{Columns (9) and (10)} - $\theta_B$ and $\theta_F$: the angular distances from the core to the brighter hotspot and from the core to the fainter hotspot, respectively.

\textit{Column (11)} - Bending Angle. This is defined as the complement of the angle between the lines from the peaks in the radio source extremities to the radio core.

\textit{Column (12)} - References to previously published images. We include references to Neff and Hutchings (1990) even though they do not show their images. We also include references to VLBI images, though these are not exhaustive.

\subsection{Comments on Individual Sources}

\indent \textit{J0730+4049:} The weak emission to the north-west of the source in the 5 GHz image is visible as a slight  extension of the core in the 1.4 GHz image. The 5 GHz VLBA Imaging and Polarization Survey \cite{HT07}, hereafter referred to as VIPS, sees a 20 mas jet pointing towards this weak emission.
\newline 
\indent \textit{J0733+2536:} The faint feature 6 arcsec directly south of the core in the 5 GHz image is likely an imaging artifact which could not be removed. It is not present in images at other frequencies and should be ignored.
\newline
\indent \textit{J0733+2721:} This source is a core dominated double separated by roughly 10 arcseconds from an unrelated triple source identified only as SDSS J073321.12+272054.4. NED lists a {\em g} = 22.7 magnitude for it but gives no other information. It has the appearance of a foreground radio galaxy. Both sources are visible in the FIRST image cutout.
\newline 
\indent \textit{J0736+6513:} The faint hotspot to the SW can barely be seen in the 5 GHz images, and not at all at 8 GHz.
\newline 
\indent \textit{J0801+1153:} This 1.4 GHz image shows more extended structure than in the originally published 5 GHz image, and hints at a jet pointing to the east.
\newline 
\indent \textit{J0801+4725:} A compact double source; the extended emission is only just visible in the 5 GHz image. 
\newline 
\indent \textit{J0805+6144:} Another compact double source, the extended emission disappears in the 15 GHz image. The VIPS image shows a 15 mas jet in position angle $120^{\circ}$.
\newline 
\indent \textit{J0807+0432:} This source has a beautiful, curving, 4 arcsecond-long jet. The NW hotspot cannot be seen in the 15 GHz image.
\newline 
\indent \textit{J0824+2341:} A compact double source, the extended emission is only visible in the 1.4 GHz data.
\newline 
\indent \textit{J0833+0959:} A double source, the extended emission is only visible in the 1.4 GHz data. Global VLBI observations at 5 GHz \cite {PFG99} see two components of a weak, curving jet pointing to the east.
\newline 
\indent \textit{J0833+1123:} A rather compact source (roughly 0.8 arcsec) whose extended emission is also visible in the 
1.5, 5 and 15 GHz images.
\newline 
\indent \textit{J0833+3531:} This source is challenging to describe due to the single low resolution image available. It is probably a triple source with complex extended structure.  Recent JVLA observations, currently being reduced, will clarify its structure.
\newline 
\indent \textit{J0905+4850:} The minor extension of the hotspot visible in the 5 GHz image is more pronounced in the 1.5 GHz data. The VIPS image shows a core that is slightly elongated in the north-south direction.
\newline 
\indent \textit{J0909+0354:} A compact double source; the component to the NW is also detectable at 1.5 and 8 GHz, but not at 15 GHz. Global VLBI observations at 5 GHz \cite {PFG99} see a faint component to the north east of the core.
\newline 
\indent \textit{J0915+0007:} The full triple nature of this source is only evident in the 1.4 GHz image. At 5 and 8 GHz only a short core extension to the west is visible.
\newline 
\indent \textit{J0918+5332:} This triple source  exhibits a prominent bent jet. The 1.4 GHz map shows diffuse emission around the jet and SW hotspot.
\newline 
\indent \textit{J0934+3050:} This triple source is small, bent and very asymmetric.
\newline 
\indent \textit{J0934+4908:} The SE hotspot of this source is also visible at 5 and 8 GHz, but not at 15 GHz. The VIPS image shows a core that is slightly extended in position angle $230^{\circ}$.
\newline 
\indent \textit{J0941+1145:} A triple source, the extended structure is visible at both 1.4 and 5 GHz, but only the core is visible at 8 and 15 GHz. Global VLBI observations at 5 GHz \cite {PFG99} see a small extension to the core to the north-east.
\newline 
\indent \textit{J0944+2554:} This compact and highly bent jet is quite unusual, and can be seen clearly also at 5 and 15 GHz. 
\newline 
\indent \textit{J0947+6328:} This image of a triple source is made from 1.4 GHz B array observations. At this time no other data exist.
\newline 
\indent \textit{J0958+3922:} Earlier optical positions were ambiguous as to which radio component is the core. An improved position and also the component spectral indices make it clear that this is a triple source.
\newline 
\indent \textit{J1007+1356:} This source has a prominent jet which does not point directly towards either lobe, giving an unusual Z-shaped morphology. The jet and southern lobe are also visible at 8 and 15 GHz, but the northern lobe vanishes at 15 GHz.
\newline 
\indent \textit{J1016+2037:} This is a very compact double source. The southern extended flux is also visible at 8 GHz. The VIPS image shows a broad, 15 mas jet in position angle $30^{\circ}$.
\newline 
\indent \textit{J1036+1326:} This nearly collinear triple shows extended structure in all data sets. There is a hint of a jet to the NW in the 8 GHz image.  EVN observations at 1.6 GHz show an unresolved core \cite {CF10}.
\newline 
\indent \textit{J1049+1332:} This curved triple source is visible at 1.4 and 5 GHz. Only the core is visible at 8 and 15 GHz.
\newline 
\indent \textit{J1057+0324:} A colinear triple source. The elongated point spread function makes it difficult to discern detail.
\newline 
\indent \textit{J1058+0443:} A bent triple source. Only the core is visible at 8 GHz.
\newline 
\indent \textit{J1127+5650:} This is a core-dominated compact double source. The elongated point spread function makes it difficult to discern detail. The VIPS image shows diffuse structure to the north-east.
\newline 
\indent \textit{J1204+5228:} This is an interesting source. Some authors have classified it as a high redshift radio galaxy, but the SDSS spectrum shows broad lines of C III, C IV and Ly $\alpha$. Pentericci et al (2000) suggest that the brightest component, seen at the center of our image, is a hotspot despite its flat spectrum. However, the SDSS position is coincident with this component, as is also the VLBI core observed by Helmboldt et al. (2007) in the VIPS survey. They see a 15 mas jet pointing north in position angle $-10^{\circ}$. We classify this source as a very asymmetric triple source with the northern (and fainter) hotspot close to the core. The diffuse flux to the SW in the 5 GHz image has no easy explanation. It is plainly visible at 1.4 GHz but not at 8 or 15 GHz. The SE hotspot is not detected at 15 GHz.
\newline 
\indent \textit{J1213+3247:} This is a small, very core-dominated triple.
\newline 
\indent \textit{J1217+3305:} This is another compact double; the extended emission shown in the 1.5 GHz image can also be seen at 8 GHz but not at 15 GHz. We do not have data at 5 GHz.
\newline 
\indent \textit{J1217+3435:} Presently we have no other images of this source. The structure on the NW side of the source probably consists of a bright jet and a hotspot.
\newline 
\indent \textit{J1217+5835:} We have classified this unusual source as a highly bent triple, even though all the extended structure is seen on only one side of the core. The source appears to consist of a core with a jet that makes an almost 90 degree bend away from what is probably a counter hotspot with a spectral index $\alpha \approx 0.4$. The VIPS image shows a 25 mas jet in position angle $135^{\circ}$.This is one of the most core-dominated sources in the sample, indicating  a small angle to the line of sight, and consistent with exaggerated projection effects.
\newline 
\indent \textit{J1242+3720:} This is a slightly bent triple source. The northern hot spot is not visible at 5 GHz, and no extended emission is detected at 8 GHz. The VIPS image shows an almost unresolved core.
\newline 
\indent \textit{J1246+0104:} This is a double source with a single very bright hotspot. The weak extended structure in the vicinity of the hotspot is also visible at 1.4 GHz.
\newline 
\indent \textit{J1346+2900:} The 5 GHz image shows a very core-dominated compact double source. At 1.4 GHz there is faint emission close to the core on the NE side. At 8 GHz only the core is visible.
\newline 
\indent \textit{J1353+5725:} This source is a very collinear triple at both 5 and 1.4 GHz. EVN observations at 1.6 GHz show an unresolved core \cite {CF10}.
\newline 
\indent \textit{J1356+2918:} This unusual triple source shows a short jet pointing to the NE at 8.44 GHz. The 1.4 GHz image shows trailing diffuse extended structure to the W that is completely resolved out in the 8 GHz image. Unfortunately there are no 5 GHz data available. The morphology of this source is reminiscent of that of J1204+5228.
\newline 
\indent \textit{J1400+0425:} This is a  moderately compact triple source with a prominent jet in both the 5 and 8 GHz images.
\newline 
\indent \textit{J1405+0415:} This is a core dominated double source. The bright emission close to the core is seen to be elongated in the 8 and 15 GHz images, and is most likely a jet. Observations with VSOP and the VLBA show a strongly curving jet with an initial position angle of $-45^{\circ}$ \cite {YG08}. See also O'Sullivan et al. (2011) for global VLBI polarization sensitive observations.
\newline 
\indent \textit{J1429+2607:} This is an asymmetric triple source. There are no 5 GHz data; at 8 GHz the extended structure is completely resolved, except for a small extension to the core pointing towards the SE component. That this is a jet is confirmed by VLBI observations \cite{HT07} that show a fine 50 mas long jet pointing in the same direction (position angle $120^{\circ}$). 
\newline 
\indent \textit{J1429+5406:} This is a very small double source. Although it is core-dominated, the core spectral index is steeper than usual ($\alpha = -0.4$). The VIPS image shows a 30 mas jet in position angle $140^{\circ}$.
\newline 
\indent \textit{J1430+4204:} This is currently the highest redshift ($z = 4.715$) quasar to exhibit kiloparsec-scale X-ray emission \cite{C12}. The weaker component of this double is coincident with the extended X-ray emission and is most likely a jet knot. The VIPS image at 5 GHz shows a very compact core with a slight extension to the south-west \cite {HT07}. The jet is much more prominent in the USNO observations at 2.3 GHz made by Fey et al. (2004) and re-imaged by Cheung ({\em  op. cit.})
\newline 
\indent \textit{J1435+5435:} This compact double is only resolved at 5 GHz, and the extended structure is not visible in the 8 GHz data.
\newline 
\indent \textit{J1450+0910:} This source shows weak extended structure at 1.4 GHz, but none at 5 or 8 GHz. The core is very self absorbed at 1.4 GHz ($\alpha_{1.4}^{4.9} < -0.9$).
\newline 
\indent \textit{J1457+3439:} The 5 GHz image shows a fine, 2 arcsec long jet, the brighter inner part of which is visible at 8 GHz. This source is actually a triple source, but the extended emission to the north is only visible in 1.4 GHz image. The VIPS image shows a 10 mas jet in position angle $270^{\circ}$.
\newline 
\indent \textit{J1459+3253:} This source has a complex S-shaped morphology. The over-resolved diffuse emission to the west of the northern component, and to the south and east of the southern component are quite prominent in the 1.4 GHz image.
\newline 
\indent \textit{J1502+5521:} This compact triple is unresolved at L band.
\newline 
\indent \textit{J1510+5702:} The extended structure for this compact double is not visible at higher frequencies. This is currently the second highest redshift ($z = 4.309 $) quasar to exhibit kiloparsec-scale X-ray emission (Siemiginowska et al. 2003, Cheung 2004). The core component is extremely compact. Helmboldt et al. (2007), Pushkarev \& Kovalev (2012) and O'Sullivan et al. (2011) all see a short southern extension at frequencies between 2 and 8 GHz.
\newline 
\indent \textit{J1528+5310:} The extended structure for this compact double vanishes at higher frequencies. It is interesting that there are two other radio sources nearby: a large FR II radio source at a distance of 40 arcsec in PA $55^{\circ}$  (also visible on the FIRST cutout), and a fainter double source at a distance of 80 arcsec in PA $240^{\circ}$. SDSS shows faint optical objects at the location of both radio sources, which are probably galaxies. We show both the radio image of the quasar (figure 7 (f)) and a zoomed out view of the whole field (figure 7 (g))
\newline 
\indent \textit{J1540+4738:} The 6 arcsecond long jet in this source is also seen in a 5 GHz image made from a short fragment of data. 
\newline 
\indent \textit{J1559+0304:} This is an extremely asymmetric triple source with a ninety degree bend. The northern extension of the core in the displayed 1.4 GHz image is a discrete component at a distance of 2.6 arcsec. Global VLBI at 5 and 8.4 GHz shows a short extension of the core to the south-east \cite {OGG11}.
\newline 
\indent \textit{J1602+2410:} This triple has a prominent jet with four bright knots. An unusual feature is that the lobe/hotspot on the side with the jet is much fainter than the lobe/hotspot on the opposite side.
\newline 
\indent \textit{J1610+1811:} The NW component of this core dominated double is visible in the 1.4, 5 and 8 GHz images. Using a global VLBI array at 8 GHz, Bourda et al (2011) detected a single weak component to the north-west of the core.
\newline 
\indent \textit{J1612+2758:} The NW component of this small triple source is not  visible in the 8 GHz image.
\newline 
\indent \textit{J1616+0459:} The NW component of this compact double is not visible in the 15 GHz image. Global VLBI observations by O'Sullivan et al. (2011) and by Pushkarev \& Kovalev (2012) find the core to be very compact.
\newline 
\indent \textit{J1625+4134:} 4C +41.32 is a compact, core dominated, nearly collinear triple. The emission south of the core is not visible in the 8 GHz and 15 GHz images. The VIPS image shows a 35 mas jet in position angle $-5^{\circ}$. Pushkarev \& Kovalev (2012) follow the jet out to 80 mas with a global VLBI array at 2.3 GHz.
\newline 
\indent \textit{J1655+3242:} This lobe dominated double source features a prominent 3 arcsecond jet with a strong lobe/hotspot. It is striking that there is no detectable emission on the other side of the core above a level of 130 $\mu$Jy/beam.
Unfortunately no data at other frequencies are available.
\newline 
\indent \textit{J1704+0134:} We have tentatively called this an unresolved core in Table 1. If the component 39 arcsec to the NW is related, then this source would be more than twice as big as any other source in the sample. There is not a hint of any extended structure around it or the quasar at the center of the field of view. Its spectral index is $\alpha = 0.5$, and we believe it is most likely an unrelated background source. Since it is just outside the SDSS footprint, it is difficult to search for an optical counterpart, but NED lists no optical object at its position.
\newline 
\indent \textit{J1715+2145:} This compact triple source is unresolved at 1.4 GHz and barely resolved at 5 GHz.

\subsection{Rejected Sources}

The NED inevitably contains a number of misidentifications, positional inaccuracies, erroneous redshifts and poorly measured flux densities. These invariably come from older observations and are superseded by more recent data from, for instance, FIRST and SDSS. It is important to check each source to verify its inclusion in the sample.

This led to 14 sources in the original sample being rejected for various reasons. Checking can be quite time consuming, so as a convenience to other workers we list them in Table 3, together with a brief reason for why they should not be included. Where a source has been identified as both a galaxy and a quasar in NED we defer to the classification by SDSS.

-- Table 3 here  --

\section{Discussion}
\label{s:Discussion}

The observations reported here represent a significant increase in the number of high redshift radio-loud quasars for which high resolution images and data are available. In order to investigate the evolutionary properties of the sources, we have also constructed a low redshift ($z < 2.5$) quasar sample with the same selection criteria as the high redshift sample. The radio data for the low redshift sample were obtained entirely by re-imaging archival VLA data (Gobeille 2011). These data and images will be the subject of paper II, and we defer most discussion of the present observations till then. Here we will briefly discuss the source morphologies that we find in the high redshift sample, and how the morphology statistics are different from those in the low redshift sample.

\subsection{Morphological classes}

Describing a morphology by a single letter does not begin to do justice to the variety of structures seen in the images. Since each letter can represent several different physical possibilities, we will discuss each in turn.

{\em Cores}: Sources classified as cores (C) present the least informative morphology. These sources could represent several possibilities: (a) a simple compact source with no extended radio emission at all, because it is very young; (b) a more complex source which appears as a core due to insufficient resolution; (c) a more complex source whose extended structure is completely resolved; (d) perhaps most likely, a highly beamed source whose jets make a small angle to the line of sight and the extended emission is too faint to be seen because of limited dynamic range.

{\em Extended Cores}: Sources classified as extended cores (E) exhibit poorly defined structure around the core, usually on the order of a beamwidth away. HNG assumed all extended cores were doubles within their survey. We will keep them separate here since extended cores could be: (a) an intrinsically small double or triple source; (b) a double or triple source highly foreshortened by projection; (c) a core where we see the beginning of a jet, or perhaps the first knot of a jet. 

{\em Doubles}: We classify a source with an identifiable core and one detected lobe and/or hotspot as a double source (D). Doubles might be: (a) a triple source with insufficient sensitivity to see the structure on the other side of the core; (b) a triple source where one lobe is completely resolved; (c) a highly bent triple source where one lobe appears superposed on the core or even on the same side of the core as the other lobe, due to the effects of projection; (d) a new class of truly one-sided objects.

{\em Triples}: Triples (T) are sources with a core and a lobe and/or hotspot on both sides of the core. For these sources we list in Table 1 the core and total flux densities, and the angular distance from the core to the hotspot or end of each lobe. We also measure a bending angle, which may be useful as a measure of projection effects. Note that we have not listed flux densities for individual parts of the extended emission. This is because the extended emission consists of lobes, hotspots and jets, probably with differing spectral indices, and many of the archival observations are just brief snapshots that do not allow us to distinguish the different features with any certainty.

\subsection{The effect of the K correction on morphology}

Table 4 shows the number of sources (and percentage) in each morphological class for the high redshift ($z \geq 2.5$) and low redshift ($z<2.5$) samples. The low redshift images will be published as paper II, and are also available in Gobeille (2011). The low redshift sample has the same flux density and sky area selection criteria as the high redshift sample and contains 168 sources. It is not complete but it should be representative. The images we use are all archival data from the VLA archives, re-imaged by us.

-- Table 4 here --

Table 4 shows striking difference between the high and low redshift samples. For instance, in the high redshift sample only 28\% of the sources show a triple morphology whereas 80\% of the low redshift sample are triples. This is not a resolution effect {\em per se} because the map scale in kiloparsecs per arcsecond does not vary much with redshift. In the concordance cosmology it changes from 8.2 to 6.2 kpc per arcsec for $2.5 \leq z \leq 5.28$. Also, in the high redshift sample 47\% of the sources are unresolved cores, but in the low redshift sample only 7\% are unresolved.

The differences between the high and low redshift samples is easily understood in terms of the radio K correction. The sources are selected by their flux density at a fixed observing frequency (1.4 GHz). The emitted frequency is higher by a factor $(1 + z)$. If the ratio of the flux density of the core to the flux density of extended emission at 1.4 GHz in the rest frame of the source is $R(1.4 GHz) = S_{\rm 1.4 GHz}^{\rm core}/S_{\rm 1.4 GHz}^{\rm ext}$, then that ratio at the emitted frequency is $R(1.4 GHz) \times (1+z)^{\alpha_{\rm ext} - \alpha_{\rm core}}$. Here $\alpha_{\rm core}$ and $\alpha_{\rm ext}$ are the spectral indices  of the core and the extended emission, respectively. Typically, $\alpha_{core} \approx 0$ and $\alpha_{ext}$ is in the range $0.6$ to $1.0$ or more, so sources appear increasingly core dominated at higher redshifts. The high redshift sample therefore contains an increasing proportion of sources whose core flux density exceeds 70 mJy at 1.4 GHz (in our frame) but whose extended emission is too weak to detect.

Because the core emission is generally believed to be beamed, a corollary of this effect is that a flux limited sample, such as this one, is increasingly biased towards small angles to the line of sight at increasing redshift. However, we can always construct a sub-sample that is unbiased in orientation, by selecting on the unbeamed extended emission rather than the total flux density. We will do this in future papers.

\section{Summary}
\label{s:conclude}

We have presented VLA observations of a new and complete sample of 123 high redshift ($ z \ge 2.5 $) quasars. The images are from re-imaging archival data from the NRAO archives, where available (43 sources), and from new observations for previously unobserved sources. We show images of the 61 sources that are significantly resolved, and list structural properties for all of them. 

The distribution of morphological types for this sample is strikingly different from what is seen in low redshift samples, but this is readily understood in terms of the increasing emission frequency at increasing redshift. Whether or not there are morphological properties that do evolve with cosmological epoch will be investigated in subsequent papers.

\acknowledgments
\section{Acknowledgments}
\label{s:Acknowledgments}
This work has been supported by NSF grants 0607453 and 1009261. C.C.C. was supported at NRL by NASA DPR S-15633-Y.

The National Radio Astronomy Observatory is a facility of the National Science Foundation operated under cooperative agreement by Associated Universities, Inc.

This research has made use of the NASA/IPAC Extragalactic Database (NED) which is operated by the Jet Propulsion Laboratory, California Institute of Technology, under contract with the National Aeronautics and Space Administration. 

Funding for SDSS-III has been provided by the Alfred P. Sloan Foundation, the Participating Institutions, the National Science Foundation, and the U.S. Department of Energy Office of Science. The SDSS-III web site is http://www.sdss3.org/.

SDSS-III is managed by the Astrophysical Research Consortium for the Participating Institutions of the SDSS-III Collaboration including the University of Arizona, the Brazilian Participation Group, Brookhaven National Laboratory, Carnegie Mellon University, University of Florida, the French Participation Group, the German Participation Group, Harvard University, the Instituto de Astrofisica de Canarias, the Michigan State/Notre Dame/JINA Participation Group, Johns Hopkins University, Lawrence Berkeley National Laboratory, Max Planck Institute for Astrophysics, Max Planck Institute for Extraterrestrial Physics, New Mexico State University, New York University, Ohio State University, Pennsylvania State University, University of Portsmouth, Princeton University, the Spanish Participation Group, University of Tokyo, University of Utah, Vanderbilt University, University of Virginia, University of Washington, and Yale University.

\newpage
\clearpage

\begin{table*}
\scriptsize
\begin{center}
\caption{Measured Quantities}
    \tabcolsep 5.0pt
\begin{tabular}{rrccccrrrrrr}
\hline\hline
\multicolumn{1}{c}{IAU name} &
\multicolumn{1}{c}{$z$} &
\multicolumn{1}{c}{R.A.} &
\multicolumn{1}{c}{Decl.} &
\multicolumn{1}{c}{mag} &
\multicolumn{1}{c}{morph-} &
\multicolumn{1}{c}{$S_{\rm 1.4}^{\rm total}$} &
\multicolumn{1}{c}{$S_{\rm 1.4}^{core}$} &
\multicolumn{1}{c}{$\theta_{B}$} &
\multicolumn{1}{c}{$\theta_{F}$} &
\multicolumn{1}{c}{BA} &
\multicolumn{1}{c}{Refs.} \\
\multicolumn{1}{c}{} &
\multicolumn{1}{c}{} &
\multicolumn{1}{c}{(J2000)} &
\multicolumn{1}{c}{(J2000)} &
\multicolumn{1}{c}{} & 
\multicolumn{1}{c}{ology} & 
\multicolumn{1}{c}{arcsec} &
\multicolumn{1}{c}{arcsec} &
\multicolumn{1}{c}{deg} &
\multicolumn{1}{c}{} \\
\hline
J0730+4049 & 2.501 	& 07 30 51.346 & +40 49 50.83	 & 18.55 	& D & 381 & 374 & 1.5$\pm$0.1 &  &  &   4 \\  
J0733+2536 & 2.686 	& 07 33 08.784 & +25 36 25.06	 & 19.38 	& T & 578 & 26 & 4.7$\pm$0.1 & 3.5$\pm$0.1 & 33 & 1, 2, 3, 6 \\  
J0733+2721 & 2.941 	& 07 33 20.491 & +27 21 03.53	 & 19.50 	& D & 216 & 214 & 5.4$\pm$0.1 &  &  &  \\ 
J0736+6513 & 3.035 	& 07 36 21.300 & +65 13 12.00	 & 18.18 	& T & 81 & 62 & 22.1$\pm$0.4 & 8.6$\pm$0.4 & 5 & 2, 6 \\  
J0744+2119 & 2.505 	& 07 44 47.272 & +21 20 00.48	 & 20.08 	& C & 52 & 52 &  &  &  &  \\  
J0746+2549 & 2.987 	& 07 46 25.874 & +25 49 02.13	 & 19.43 	& C & 428 & 428 &  &  &  & 4 \\  
J0752+5808 & 2.940 	& 07 52 09.679 & +58 08 52.26	 & 19.62*	& C & 152 & 152 & &  &  & 4 \\  
J0753+4231 & 3.595 	& 07 53 03.337 & +42 31 30.77	 & 17.77 	& C & 709 & 709 &  &  &  & 4 \\  
J0756+3714 & 2.515 	& 07 56 28.250 & +37 14 55.65	 & 19.85 	& C & 235 & 235 &  &  &  &  4 \\  
J0801+1153 & 2.672 	& 08 01 00.484 & +11 53 23.59	 & 19.13 	& T & 117 & 53 & 4.7$\pm$0.1 & 6.0$\pm$0.1 & 9 & 1 \\  
J0801+4725 & 3.267 	& 08 01 37.682 & +47 25 28.24	 & 19.17 	& D & 71 & 68 & 3.5$\pm$0.1 &  &  &  \\  
J0805+6144 & 3.033 	& 08 05 18.180 & +61 44 23.70	 & 19.64*	& D & 813 & 798 & 3.8$\pm$0.1 &  &  & 4 \\  
J0807+0432 & 2.877 	& 08 07 57.538 & +04 32 34.53	 & 17.78*	& T & 530 & 326 & 3.9$\pm$0.1 & 2.6$\pm$0.1 & 20 & 2 \\  
J0821+3107 & 2.613 	& 08 21 07.616 & +31 07 51.17	 & 16.87 	& C & 91 & 91 &  &  &  & 4, 5 \\  
J0823+0611 & 2.792 	& 08 23 28.620 & +06 11 46.10	 & 17.81 	& C & 68 & 68 &  &  &  &  \\  
J0824+2341 & 2.611 	& 08 24 28.022 & +23 41 07.95	 & 18.75 	& D & 146 & 145 &  5.4$\pm$ 0.2 &  &  &  \\  
J0833+0959 & 3.731 	& 08 33 22.514 & +09 59 41.14	 & 18.70 	& D & 122 & 121 & 10.6$\pm$0.6 &  &  &  9\\  
J0833+1124 & 2.985 	& 08 33 14.367 & +11 23 36.23	 & 18.30 	& D & 392 & 361 & 0.8$\pm$0.1 &  &  & 2, 6 \\ 
J0833+3531 & 2.703 	& 08 33 01.658 & +35 31 33.68	 & 19.96 	& T & 75 & 22 & 3.1$\pm$0.4 & 4.9$\pm$0.4 & 4 &  \\  
J0847+3831 & 3.186 	& 08 47 15.169 & +38 31 09.98	 & 18.31 	& C & 142 & 142 &  &  &  &  \\  
J0905+3555 & 2.840 	& 09 05 36.065 & +35 55 51.68	 & 18.27 	& C & 87 & 87 &  &  &  &  \\  
J0905+4850 & 2.697 	& 09 05 27.464 & +48 50 49.97	 & 17.39 	& D & 646 & 505 & 3.7$\pm$0.1 &  &  & 4, 7 \\  
J0909+0354 & 3.288 	& 09 09 15.915 & +03 54 42.98	 & 19.73 	& D & 132 & 126 & 2.4$\pm$0.3 &  &  &  9 \\  
J0910+2539 & 2.753 	& 09 10 55.237 & +25 39 21.50	 & 18.38 	& C & 179 & 179 &  &  &  &  \\  
J0915+0007 & 3.074 	& 09 15 51.695 & +00 07 13.31	 & 20.54 	& T & 399 & 386 & 3.3$\pm$0.4 & 3.5$\pm$0.4 & 9 &  \\  
J0918+5332 & 3.014 	& 09 18 57.676 & +53 32 20.02	 & 19.68 	& T & 90 & 20 & 3.4$\pm$0.2 & 5.9$\pm$0.2 & 8 &  \\  
J0933+2845 & 3.421 	& 09 33 37.298 & +28 45 32.24	 & 17.83 	& E & 121 & 121 &  &  &  &  \\  
J0934+3050 & 2.895 	& 09 34 47.241 & +30 50 55.77	 & 20.35 	& T & 291 & 88 & 1.7$\pm$0.1 & 2.7$\pm$0.1 & 52 & \\  
J0934+4908 & 2.584 	& 09 34 15.762 & +49 08 21.73	 & 19.13 	& D & 802 & 785 & 5.0$\pm$0.1 &  &  & 4, 8\\  
J0935+3633 & 2.859 	& 09 35 31.842 & +36 33 17.55	 & 18.39 	& C & 291 & 291 &  &  &  & 4 \\  
J0941+1145 & 3.194 	& 09 41 13.558 & +11 45 32.33	 & 19.34 	& T & 218 & 210 & 1.9$\pm$0.2 & 5.4$\pm$0.2 & 16 & 2, 6, 9 \\  
J0944+2554 & 2.916 	& 09 44 42.320 & +25 54 43.31	 & 18.38 	& D & 827 & 586 & 1.5$\pm$0.3 &  &  & 2 \\  
J0947+6327 & 2.612 	& 09 47 59.416 & +63 28 03.11	 & 19.04 	& T & 78 & 33 & 8.3$\pm$1.4 & 6.6$\pm$1.4 & 24 &  \\  
J0958+3922 & 2.941 	& 09 58 56.780 & +39 22 32.40	 & 21.25 	& T & 160 & 46 & 2.7$\pm$0.1 & 9.5$\pm$0.1 & 38 & 10 \\  
J1007+1356 & 2.721 	& 10 07 41.498 & +13 56 29.60	 & 18.47 	& T & 914 & 818 & 4.9$\pm$0.1 & 2.4$\pm$0.1 & 52 & 11 \\  
J1013+3526 & 2.638 	& 10 13 02.299 & +35 26 05.70	 & 17.81 	& C & 200 & 200 &  &  &  & 4, 5 \\  
J1016+2037 & 3.115 	& 10 16 44.322 & +20 37 47.30	 & 19.11 	& D & 730 & 695 & 1.1$\pm$0.3 &  &  & 4 \\  
J1017+6116 & 2.801 	& 10 17 25.887 & +61 16 27.50	 & 18.10 	& C & 474 & 474 &  &  &  & 4, 11 \\  
J1026+2542 & 5.284 	& 10 26 23.621 & +25 42 59.40	 & 19.96 	& C & 239 & 239 &  &  &  &  4 \\  
J1026+3658 & 3.252 	& 10 26 45.340 & +36 58 25.60	 & 19.43 	& C & 195 & 195 &  &  &  &  \\  
J1036+1326 & 3.097 	& 10 36 26.880 & +13 26 51.70	 & 17.78 	& T & 100 & 56 & 8.5$\pm$0.1 & 7.4$\pm$0.1 & 11 &  13 \\  
J1042+0749 & 2.661 	& 10 42 57.589 & +07 48 50.55	 & 17.31 	& C & 457 & 457 &  &  &  &  \\  
J1044+2959 & 2.979 	& 10 44 06.342 & +29 59 00.99	 & 18.97 	& C & 59 & 59 &  &  &  & 4 \\  
J1045+3142 & 3.236 	& 10 45 23.481 & +31 42 31.65	 & 18.73 	& C & 132 & 132 &  &  &  &  \\
J1049+1332 & 2.767 	& 10 49 41.098 & +13 32 55.68	 & 18.74 	& T & 110 & 87 & 5.8$\pm$0.1 & 18.0$\pm$0.1 & 44 &  \\  
J1050+3430 & 2.520 	& 10 50 58.123 & +34 30 10.94	 & 20.60 	& C & 583 & 583 &  &  &  & 4 \\  
J1057+0325 & 2.832 	& 10 57 26.622 & +03 24 48.09	 & 19.32 	& T & 152 & 88 & 2.1$\pm$0.1 & 2.7$\pm$0.1 & 2 & \\  
J1058+0443 & 2.622 	& 10 58 58.580 & +04 43 47.80	 & 20.07 	& T & 55 & 47 & 3.3$\pm$0.1 & 4.7$\pm$0.1 & 40 &  \\  
J1101+0010 & 3.694 	& 11 01 47.890 & +00 10 39.44	 & 20.14 	& C & 188 & 188 &  &  &  &  \\  
J1103+0232 & 2.517 	& 11 03 44.540 & +02 32 10.00	 & 18.33 	& C & 160 & 160 &  &  &  &  \\  
\hline\hline
\end{tabular}
\end{center}
\end{table*}

\addtocounter{table}{-1}
\begin{table*}
\scriptsize
\begin{center}
\caption{Measured Quantities ({\it continued})}
    \tabcolsep 5.0pt
\begin{tabular}{rrccccrrrrrr}
\hline\hline
\multicolumn{1}{c}{IAU name} &
\multicolumn{1}{c}{$z$} &
\multicolumn{1}{c}{R.A.} &
\multicolumn{1}{c}{Decl.} &
\multicolumn{1}{c}{mag} &
\multicolumn{1}{c}{morph-} &
\multicolumn{1}{c}{$S_{\rm 1.4}^{\rm total}$} &
\multicolumn{1}{c}{$S_{\rm 1.4}^{core}$} &
\multicolumn{1}{c}{$\theta_{B}$} &
\multicolumn{1}{c}{$\theta_{F}$} &
\multicolumn{1}{c}{BA} &
\multicolumn{1}{c}{Refs.} \\
\multicolumn{1}{c}{} &
\multicolumn{1}{c}{} &
\multicolumn{1}{c}{(J2000)} &
\multicolumn{1}{c}{(J2000)} &
\multicolumn{1}{c}{} &
\multicolumn{1}{c}{ology} &
\multicolumn{1}{c}{arcsec} &
\multicolumn{1}{c}{arcsec} &
\multicolumn{1}{c}{deg} &
\multicolumn{1}{c}{} \\
\hline
J1119+6004 & 2.641      & 11 19 14.345 & +60 04 57.20    & 17.06        & C & 109 & 109 &  &  &  & 4 \\
J1127+5650 & 2.890      & 11 27 40.135 & +56 50 14.79    & 19.53        & D & 529 & 510 & 0.6$\pm$0.1 &  &  & 4, 6 \\
J1128+2326 & 3.049 	& 11 28 51.701 & +23 26 17.35	 & 18.36 	& E & 141 & 130 &  &  &  &  \\  
J1130+0731 & 2.659 	& 11 30 17.380 & +07 32 12.90	 & 17.17 	& C & 35 & 35 &  &  &  & 5 \\  
J1137+2935 & 2.644 	& 11 37 20.880 & +29 35 39.10	 & 19.54 	& C & 76 & 76 &  &  &  &  \\  
J1150+4332 & 3.037 	& 11 50 16.603 & +43 32 05.91	 & 19.69 	& C & 160 & 160 &  &  &  & 4 \\  
J1151+4008 & 2.740 	& 11 51 16.930 & +40 08 22.20	 & 19.36 	& C & 83 & 83 &  &  &  &  \\  
J1204+5228 & 2.736 	& 12 04 36.800 & +52 28 41.78	 & 18.36 	& T & 940 & 632 & 4.2$\pm$0.1 & 0.7$\pm$0.1 & 33 & 4, 12 \\  
J1213+3248 & 2.516 	& 12 13 03.824 & +32 47 36.97	 & 18.90 	& T & 129 & 116 & 1.2$\pm$0.1 & 0.9$\pm$0.1 & 23 &  \\  
J1215+6422 & 3.239 	& 12 15 48.906 & +64 22 28.46	 & 18.56 	& C & 65 & 65 &  &  &  & 4 \\  
J1217+3306 & 2.606 	& 12 17 32.540 & +33 05 38.00	 & 18.37 	& D & 208 & 200 & 2.3$\pm$0.1 &  &  & 6 \\  
J1217+3435 & 2.651 	& 12 17 15.208 & +34 35 37.87	 & 18.44 	& T & 181 & 101 & 2.1$\pm$0.4 & 5.5$\pm$0.4 & 6 & 6 \\  
J1217+5835 & 2.551 	& 12 17 11.019 & +58 35 26.25	 & 19.09 	& T & 412 & 412 & 1.1$\pm$0.1 & 1.6$\pm$0.1 & 131 & 4 \\  
J1223+5038 & 3.501 	& 12 23 43.169 & +50 37 53.40	 & 17.41 	& E & 178 & 171 &  &  &  & 4 \\  
J1242+3720 & 3.827 	& 12 42 09.812 & +37 20 05.69	 & 19.39 	& T & 816 & 799 & 6.4$\pm$0.5 & 10.9$\pm$0.5 & 21 & 4 \\  
J1246+0104 & 2.510 	& 12 46 23.727 & +01 04 02.35	 & 19.54 	& D & 117 & 3 & 5.5$\pm$0.1 &  &  & \\  
J1301+1905 & 3.069 	& 13 01 21.010 & +19 04 21.40	 & 18.18 	& C & 146 & 146 &  &  &  &  \\  
J1310+6229 & 2.614 	& 13 10 46.559 & +62 29 09.05	 & 20.02 	& C & 292 & 292 &  &  &  &  \\  
J1319+6217 & 3.073 	& 13 19 07.484 & +62 17 21.34	 & 19.27 	& C & 214 & 214 &  &  &  & 4 \\  
J1322+3912 & 2.992 	& 13 22 55.664 & +39 12 07.95	 & 17.58 	& C & 124 & 124 &  &  &  & 4 \\  
J1325+1123 & 4.415 	& 13 25 12.493 & +11 23 29.77	 & 19.17 	& C & 74 & 74 &  &  &  &  \\  
J1329+5009 & 2.654 	& 13 29 05.803 & +50 09 26.40	 & 20.17 	& C & 164 & 164 &  &  &  & 4 \\  
J1330+4954 & 2.881 	& 13 30 29.389 & +49 54 27.48	 & 18.64 	& C & 66 & 66 &  &  &  & 4 \\  
J1337+3152 & 3.182 	& 13 37 24.699 & +31 52 54.52	 & 18.47 	& C & 84 & 84 &  &  &  &  \\  
J1339+6328 & 2.562 	& 13 39 23.783 & +63 28 58.42	 & 19.25 	& C & 480 & 480 &  &  &  & 4 \\  
J1340+3754 & 3.110 	& 13 40 22.952 & +37 54 43.83	 & 18.54 	& C & 276 & 276 &  &  &  & 4, 9\\  
J1342+5110 & 2.599 	& 13 42 24.317 & +51 10 12.45	 & 19.20 	& C & 154 & 154 &  &  &  & 4 \\  
J1346+2900 & 2.721 	& 13 46 37.440 & +29 00 42.40	 & 20.92 	& D & 96 & 90 & 2.1$\pm$0.4 &  &  &  \\  
J1353+5725 & 3.476 	& 13 53 26.017 & +57 25 52.80	 & 19.31 	& T & 319 & 207 & 7.8$\pm$0.2 & 4.3$\pm$0.2 & 8 & 13 \\  
J1356+2918 & 3.244 	& 13 56 52.540 & +29 18 18.80	 & 20.36 	& T & 125 & 90 & 4.3$\pm$0.4 & 1.0$\pm$0.4 & 2 &  \\  
J1400+0425 & 2.550 	& 14 00 48.443 & +04 25 30.87	 & 20.32 	& T & 281 & 170 & 2.9$\pm$0.1 & 1.3$\pm$0.1 & 48 & \\  
J1404+0728 & 2.885 	& 14 04 32.992 & +07 28 46.96	 & 18.83 	& C & 158 & 158 &  &  &  &  \\  
J1405+0415 & 3.209 	& 14 05 01.120 & +04 15 35.82	 & 19.36 	& D & 729 & 580 & 3.3$\pm$0.4 &  &  & 14, 15 \\  
J1406+3433 & 2.563 	& 14 06 53.847 & +34 33 37.31	 & 18.49 	& C & 169 & 169 &  &  &  & 4 \\  
J1413+4505 & 3.118 	& 14 13 18.864 & +45 05 23.01	 & 19.09 	& C & 141 & 141 &  &  &  & 4 \\  
J1429+2607 & 2.914 	& 14 29 50.916 & +26 07 50.32	 & 18.13 	& T & 415 & 371 & 9.1$\pm$0.4 & 2.2$\pm$0.4 & 3 & 4\\  
J1429+5406 & 2.990 	& 14 29 21.879 & +54 06 11.12	 & 19.81 	& D & 1036 & 916 & 0.35$\pm$0.04 &  &  & 4\\  
J1430+4204 & 4.715 	& 14 30 23.742 & +42 04 36.49	 & 19.14 	& D & 153 & 151 & 3.6$\pm$0.1 &  &  & 4, 9 \\  
J1435+5435 & 3.809 	& 14 35 33.779 & +54 35 59.31	 & 19.95 	& D & 96 & 83 & 1.3$\pm$0.3 &  &  &  \\  
J1439+1117 & 2.592 	& 14 39 12.044 & +11 17 40.56	 & 18.33 	& C & 32 & 32 &  &  &  &  \\  
J1445+0958 & 3.552 	& 14 45 16.465 & +09 58 36.07	 & 17.88 	& C & 2614 & 2614 &  &  &  & 6, 11 \\  
J1450+0910 & 2.611 	& 14 50 31.169 & +09 10 27.95	 & 19.44 	& E & 157 & 148 &  &  &  &  \\
J1454+5003 & 2.849 	& 14 54 08.322 & +50 03 30.99	 & 19.32 	& C & 830 & 830 &  &  &  & \\  
J1455+4431 & 2.689 	& 14 55 54.136 & +44 31 37.65	 & 19.49 	& C & 288 & 288 &  &  &  & 4 \\  
J1457+3439 & 2.732 	& 14 57 57.303 & +34 39 50.39	 & 19.50 	& T & 227 & 184 & 1.7$\pm$0.4 & 1.6$\pm$0.4 & 32 & 4, 6 \\  
J1459+3253 & 3.329 	& 14 59 27.051 & +32 53 58.66	 & 19.43 	& T & 119 & 53 & 3.8$\pm$0.4 & 3.3$\pm$0.4 & 22 &  \\  
J1459+4442 & 3.401 	& 14 59 35.458 & +44 42 07.92	 & 19.99 	& C & 109 & 109 &  &  &  & 4 \\  
J1502+5521 & 3.323 	& 15 02 06.525 & +55 21 46.64	 & 19.80 	& T & 83 & 21 & 0.8$\pm$0.2 & 1.0$\pm$0.2 & 28 & \\  
J1503+0419 & 3.667 	& 15 03 28.888 & +04 19 48.99	 & 18.20 	& C & 137 & 137 & &  &  &  9\\  
J1510+5702 & 4.313 	& 15 10 02.922 & +57 02 43.38	 & 19.89 	& D & 278 & 273 & 2.6$\pm$0.5 &  &  & 4, 9, 11, 15 \\  
\hline\hline
\end{tabular}
\end{center}
\end{table*}

\addtocounter{table}{-1}
\begin{table*}
\scriptsize
\begin{center}
\caption{Measured Quantities ({\it continued})}
    \tabcolsep 5.0pt
\begin{tabular}{rrccccrrrrrr}
\hline\hline
\multicolumn{1}{c}{IAU name} &
\multicolumn{1}{c}{$z$} &
\multicolumn{1}{c}{R.A.} &
\multicolumn{1}{c}{Decl.} &
\multicolumn{1}{c}{mag} &
\multicolumn{1}{c}{morph-} &
\multicolumn{1}{c}{$S_{\rm 1.4}^{\rm total}$} &
\multicolumn{1}{c}{$S_{\rm 1.4}^{core}$} &
\multicolumn{1}{c}{$\theta_{B}$} &
\multicolumn{1}{c}{$\theta_{F}$} &
\multicolumn{1}{c}{BA} &
\multicolumn{1}{c}{Refs.} \\
\multicolumn{1}{c}{} &
\multicolumn{1}{c}{} &
\multicolumn{1}{c}{(J2000)} &
\multicolumn{1}{c}{(J2000)} &
\multicolumn{1}{c}{} &
\multicolumn{1}{c}{ology} &
\multicolumn{1}{c}{arcsec} &
\multicolumn{1}{c}{arcsec} &
\multicolumn{1}{c}{deg} &
\multicolumn{1}{c}{} \\
\hline
J1520+4732 & 2.815 	& 15 20 43.602 & +47 32 49.14	 & 18.56 	& C & 94 & 94 &  &  &  &  \\  
J1521+1756 & 3.053 	& 15 21 17.583 & +17 56 01.09	 & 19.27 	& C & 181 & 181 &  &  &  & 4 \\  
J1528+5310 & 2.816 	& 15 28 21.659 & +53 10 30.48	 & 19.79 	& D & 178 & 163 & 1.8$\pm$0.4 &  &  &  \\  
J1535+4836 & 2.561 	& 15 35 14.653 & +48 36 59.70	 & 17.81 	& C & 108 & 108 &  &  &  &  4 \\  
J1540+4739 & 2.566 	& 15 40 58.710 & +47 38 27.60	 & 18.97 	& T & 221 & 36 & 7.9$\pm$0.4 & 5.0$\pm$0.4 & 16 & \\  
J1541+5348 & 2.539 	& 15 41 25.461 & +53 48 13.03	 & 18.92 	& C & 253 & 253 &  &  &  & 4 \\  
J1547+4208 & 2.739 	& 15 47 59.043 & +42 08 55.57	 & 19.73 	& C & 70 & 70 &  &  &  &  \\  
J1559+0304 & 3.891 	& 15 59 30.973 & +03 04 48.26	 & 19.77 	& T & 384 & 377 & 17.4$\pm$0.4 & 2.6$\pm$0.4 & 89 & 9, 15\\  
J1602+2410 & 2.531 	& 16 02 12.601 & +24 10 10.61	 & 18.58 	& T & 341 & 18 & 1.7$\pm$0.1 & 4.1$\pm$0.1 & 3 & \\  
J1603+5730 & 2.850 	& 16 03 55.931 & +57 30 54.41	 & 17.26 	& C & 362 & 362 &  &  &  & 2, 4 \\  
J1605+3038 & 2.670 	& 16 05 23.701 & +30 38 37.15	 & 19.29 	& C & 51 & 51 &  &  &  &  \\  
J1610+1811 & 3.118 	& 16 10 05.289 & +18 11 43.47	 & 18.25 	& D & 258 & 223 & 4.8$\pm$0.1 &  &  & 5 \\  
J1612+2758 & 3.549 	& 16 12 53.417 & +27 58 42.57	 & 19.60 	& T & 70 & 49 & 1.9$\pm$0.1 & 1.4$\pm$0.1 & 26 &  \\  
J1616+0459 & 3.215 	& 16 16 37.557 & +04 59 32.74	 & 19.22 	& D & 416 & 347 & 0.9$\pm$0.2 &  &  & 6, 11, 15 \\  
J1625+4134 & 2.550 	& 16 25 57.670 & +41 34 40.63	 & 21.58 	& T & 1715 & 1661 & 0.8$\pm$0.1 & 1.1$\pm$0.1 & 12 &  4, 11 \\  
J1632+2643 & 2.683 	& 16 32 21.051 & +26 43 53.46	 & 17.64 	& C & 267 & 267 &  &  &  & 4 \\  
J1655+1948 & 3.260 	& 16 55 43.568 & +19 48 47.12	 & 19.85 	& E & 194 & 188 &  &  &  & 5 \\  
J1655+3242 & 3.189 	& 16 55 19.225 & +32 42 41.13	 & 19.45 	& D & 180 & 14 & 3.2$\pm$0.2 &  &  &  \\  
J1656+1826 & 2.546 	& 16 56 34.089 & +18 26 26.35	 & 19.99 	& C & 221 & 221 &  &  &  & 4 \\  
J1704+0134 & 2.842 	& 17 04 07.489 & +01 34 08.47	 & 19.62* & C? & 293 & 290 &  &  &  &  \\  
J1707+0148 & 2.568 	& 17 07 34.415 & +01 48 45.70	 & 18.52*	& C & 640 & 640 &  &  &  & 6, 11 \\  
J1707+1846 & 2.518 	& 17 07 53.748 & +18 46 39.02	 & 19.36 	& D & 334 & 332 & 10.7$\pm$0.1 &  &  & 6 \\  
J1715+2145 & 4.011 	& 17 15 21.250 & +21 45 31.83	 & 21.48 	& T & 387 & 383 & 0.5$\pm$0.1 & 0.4$\pm$0.1 & 19 &  \\  
\hline\hline
\end{tabular}
\end{center}
Magnitudes are SDSS $i$-band unless noted with asterisks*, in which case, they are $R2$-band magnitudes from the USNO-B1.0 (Monet al al.\ 2003).\\
Flux densities, $S_{\rm 1.4}$, at 1.4 GHz.\\
References: (1) Lonsdale et al. 1993; (2) Barthel et al. 1988; (3) Hintzen et al. 1983; (4) Helmboldt et al. 2007; (5) Bourda et al. 2011; (6) Neff \& Hutchings 1990; (7) Stanghellini et al. 1990; (8) Snellen et al. 1995; (9) Paragi 
et al. 1999; (10) Naundorf et al. 1992; (11) Pushkarev \& Kovalev 2012; (12) Pentericci et al. 2000; (13) Cseh et al. 2010; (14) Yang et al. 2008; (15) O’Sullivan et al. 2011.
\end{table*}

\begin{table*}
\footnotesize
\begin{center}
\caption{Image Properties}
\begin{tabular}{cccccc}
\hline\hline
\multicolumn{1}{c}{Source} &
\multicolumn{1}{c}{Frequency} &
\multicolumn{1}{c}{Array} &
\multicolumn{1}{c}{Bottom Contour} &
\multicolumn{1}{c}{Map Peak} &
\multicolumn{1}{c}{Time on Source} \\
\multicolumn{1}{c}{} &
\multicolumn{1}{c}{(GHz)} &
\multicolumn{1}{c}{} &
\multicolumn{1}{c}{$\%$ of map peak} &
\multicolumn{1}{c}{(mJy)} &
\multicolumn{1}{c}{(sec)} \\
\hline
J0730+4049 & 4.860 & A & $0.15\%$ & 331 & 220 \\
J0733+2536 & 4.860 & A & $0.72\%$ & 27 & 470 \\
J0733+2721 & 1.425 & A & $0.06\%$ & 213 & 560 \\
J0736+6513 & 1.490 & A & $0.25\%$ & 61 & 940 \\
J0801+1153 & 1.425 & A & $0.30\%$ & 52 & 570 \\
J0801+4725 & 1.425 & A & $0.26\%$ & 68 & 560 \\
J0805+6144 & 4.860 & B & $0.02\%$ & 1079 & 2250 \\
J0807+0432 & 8.440 & A & $0.05\%$ & 276 & 2390 \\
J0824+2341 & 1.425 & A & $0.05\%$ & 145 & 580 \\
J0833+0959 & 1.425 & A & $0.05\%$ & 119 & 580 \\
J0833+1124 & 8.440 & A & $0.05\%$ & 300 & 620 \\
J0833+3531 & 1.425 & A & $0.70\%$ & 27 & 527 \\
J0905+4850 & 4.860 & A & $0.10\%$ & 433 & 280 \\
J0909+0354 & 4.860 & A & $0.18\%$ & 176 & 130 \\
J0915+0007 & 1.425 & A & $0.02\%$ & 386 & 580 \\
J0918+5332 & 4.860 & A & $0.89\%$ & 13 & 560 \\
J0934+3050 & 8.440 & A & $1.05\%$ & 15 & 280 \\
J0934+4908 & 1.400 & A & $0.08\%$ & 785 & 310 \\
J0941+1145 & 1.490 & A & $0.08\%$ & 210 & 950 \\
J0944+2554 & 8.440 & A & $0.07\%$ & 161 & 630 \\
J0947+6327 & 1.400 & B & $0.30\%$ & 34 & 180 \\
J0958+3922 & 1.490 & A & $0.13\%$ & 105 & 310 \\
J1007+1356 & 4.760 & A & $0.02\%$ & 770 & 850 \\
J1016+2037 & 4.860 & A & $0.03\%$ & 930 & 5300 \\
J1036+1326 & 1.490 & A & $0.23\%$ & 55 & 510 \\
J1049+1332 & 1.425 & A & $0.09\%$ & 87 & 567 \\
J1057+0325 & 4.860 & A & $0.10\%$ & 36 & 530 \\
J1058+0443 & 4.860 & A & $0.16\%$ & 47 & 580 \\
J1127+5650 & 8.460 & A & $0.05\%$ & 231 & 520 \\
J1204+5228 & 4.710 & A & $0.09\%$ & 157 & 530 \\
J1213+3248 & 4.860 & A & $0.15\%$ & 56 & 567 \\
J1217+3306 & 1.490 & A & $0.07\%$ & 200 & 700 \\
J1217+3435 & 1.490 & A & $0.13\%$ & 104 & 670 \\
J1217+5835 & 8.460 & A & $0.02\%$ & 384 & 4533 \\
J1242+3720 & 1.425 & A & $0.01\%$ & 799 & 330 \\
J1246+0104 & 4.860 & A & $0.36\%$ & 30 & 95 \\
J1346+2900 & 1.425 & A & $0.07\%$ & 90 & 580 \\
\hline\hline
\end{tabular}
\end{center}
\end{table*}

\addtocounter{table}{-1}
\begin{table*}
\footnotesize
\begin{center}
\caption{Image Properties ({\it continued})}
\begin{tabular}{cccccc}
\hline\hline
\multicolumn{1}{c}{Source} &
\multicolumn{1}{c}{Frequency} &
\multicolumn{1}{c}{Array} &
\multicolumn{1}{c}{Bottom Contour} &
\multicolumn{1}{c}{Map Peak} &
\multicolumn{1}{c}{Time on Source} \\
\multicolumn{1}{c}{} &
\multicolumn{1}{c}{(GHz)} &
\multicolumn{1}{c}{} &
\multicolumn{1}{c}{$\%$ of map peak} &
\multicolumn{1}{c}{(mJy)} &
\multicolumn{1}{c}{(sec)} \\
\hline
J1353+5725 & 4.860 & A & $0.10\%$ & 139 & 570 \\
J1356+2918 & 1.425 & A & $0.24\%$ & 84 & 560 \\
J1400+0425 & 4.860 & A & $0.13\%$ & 96 & 540 \\
J1405+0415 & 4.860 & A & $0.03\%$ & 922 & 2010 \\
J1429+2607 & 1.425 & A & $0.04\%$ & 367 & 520 \\
J1429+5406 & 1.415 & A & $0.15\%$ & 1010 & 110 \\
J1430+4204 & 1.425 & A & $0.10\%$ & 150 & 2020 \\
J1435+5435 & 4.860 & A & $0.15\%$ & 73 & 570 \\
J1450+0910 & 1.425 & A & $0.10\%$ & 145 & 677 \\
J1457+3439 & 1.490 & A & $0.05\%$ & 196 & 650 \\
J1459+3253 & 1.425 & A & $0.15\%$ & 55 & 560 \\
J1502+5521 & 4.860 & A & $0.50\%$ & 21 & 570 \\
J1510+5702 & 1.425 & A & $0.03\%$ & 273 & 590 \\
J1528+5310 & 1.490 & A & $0.10\%$ & 159 & 280 \\
J1528+5310 & 4.860 & C & $0.20\%$ & 66 & 280 \\
J1540+4739 & 1.425 & A & $0.23\%$ & 150 & 280 \\
J1559+0304 & 1.425 & A & $0.04\%$ & 377 & 2930 \\
J1602+2410 & 4.860 & A & $0.20\%$ & 42 & 520 \\
J1610+1811 & 4.860 & A & $0.08\%$ & 96 & 520 \\
J1612+2758 & 4.860 & A & $0.14\%$ & 32 & 543 \\
J1616+0459 & 8.440 & A & $0.024\%$ & 714 & 870 \\
J1625+4134 & 4.985 & A & $0.05\%$ & 1110 & 13910 \\
J1655+1948 & 4.860 & A & $0.09\%$ & 151 & 520 \\
J1704+0134 & 1.425 & A & $0.06\%$ & 287 & 690 \\
J1715+2145 & 8.440 & A & $0.06\%$ & 100 & 130 \\
\hline\hline
\end{tabular}
\end{center}
\end{table*}

\begin{table*}
\begin{center}
\caption{Rejected Sources}
\begin{tabular}{cc}
\hline\hline
\multicolumn{1}{c}{Source} &
\multicolumn{1}{c}{Reason for Rejection}\\
\hline
J0747+4618 & radio galaxy\\
J0751+2716 & gravitational lens\\
J0817+2918 & radio galaxy\\
J0849+0950 & flux confused by nearby strong source\\
J0850+1529 & flux confused by nearby strong source\\
J0958+2947 & blend of two sources\\
J1020+1039 & radio galaxy\\
J1401+1513 & gravitational lens\\
J1424+2256 & gravitational lens\\
J1436+6319 & radio galaxy\\
J1506+1029 & radio source and quasar are different objects\\
J1538+0019 & radio galaxy\\
J1600+0412 & original redshifts in NED were in error; SDSS DR10 gives $z = 0.794$\\
J1606+3124 & radio galaxy\\
\hline\hline
\end{tabular}
\end{center}
\end{table*}

\begin{table*}
\begin{center}
\caption{Overview of Morphologies}
\begin{tabular}{ccccc}
\hline\hline
\multicolumn{1}{c}{Redshift Range} & 
\multicolumn{4}{c}{Morphological class} \\
\multicolumn{1}{c}{} &
\multicolumn{1}{c}{C} &
\multicolumn{1}{c}{E} &
\multicolumn{1}{c}{D} &
\multicolumn{1}{c}{T} \\
\hline 
$2.5 \geq z < 5.3$ & 58 (47\%) & 5 (4\%) & 26 (21\%) & 34 (28\%) \\ 
$0 < z <2.5 $& 11 (7\%) & 0 & 22 (13\%) & 135 (80\%) \\
\hline\hline
\end{tabular}
\end{center}
Tabulated are the number of sources in each class, and the corresponding percentages.
\end{table*}


\newpage
\clearpage

\begin{figure}
\figurenum{1}
\plotone{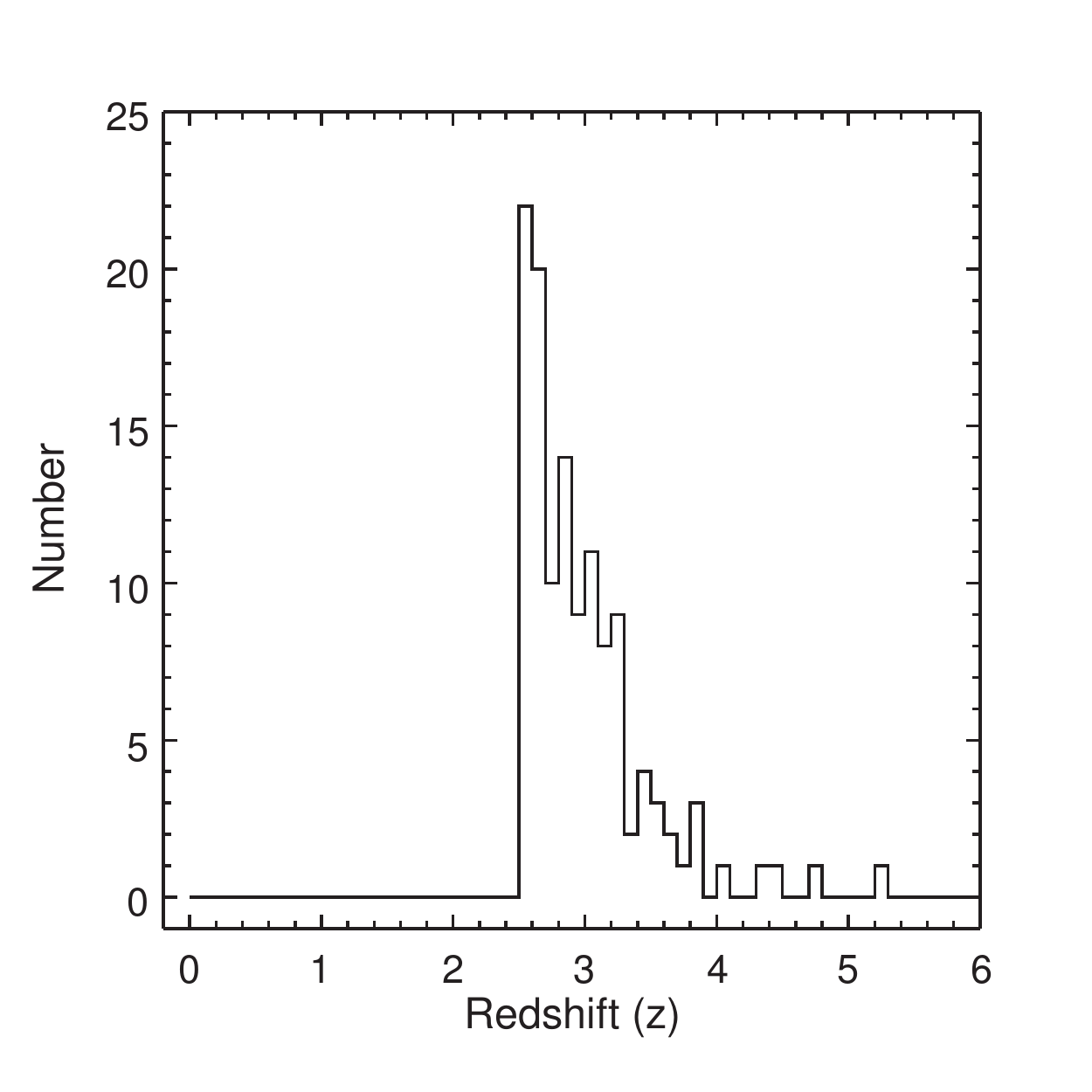}
\caption{Redshift distribution of the 123 source $z \geq 2.5$ sample in bins of $\Delta z=0.1$.}
\end{figure}

\clearpage

\begin{figure}
\figurenum{2}
\centering
\subfloat[Part 1][J0730+4049 at 4.860 GHz]{\includegraphics[width=2.2in]{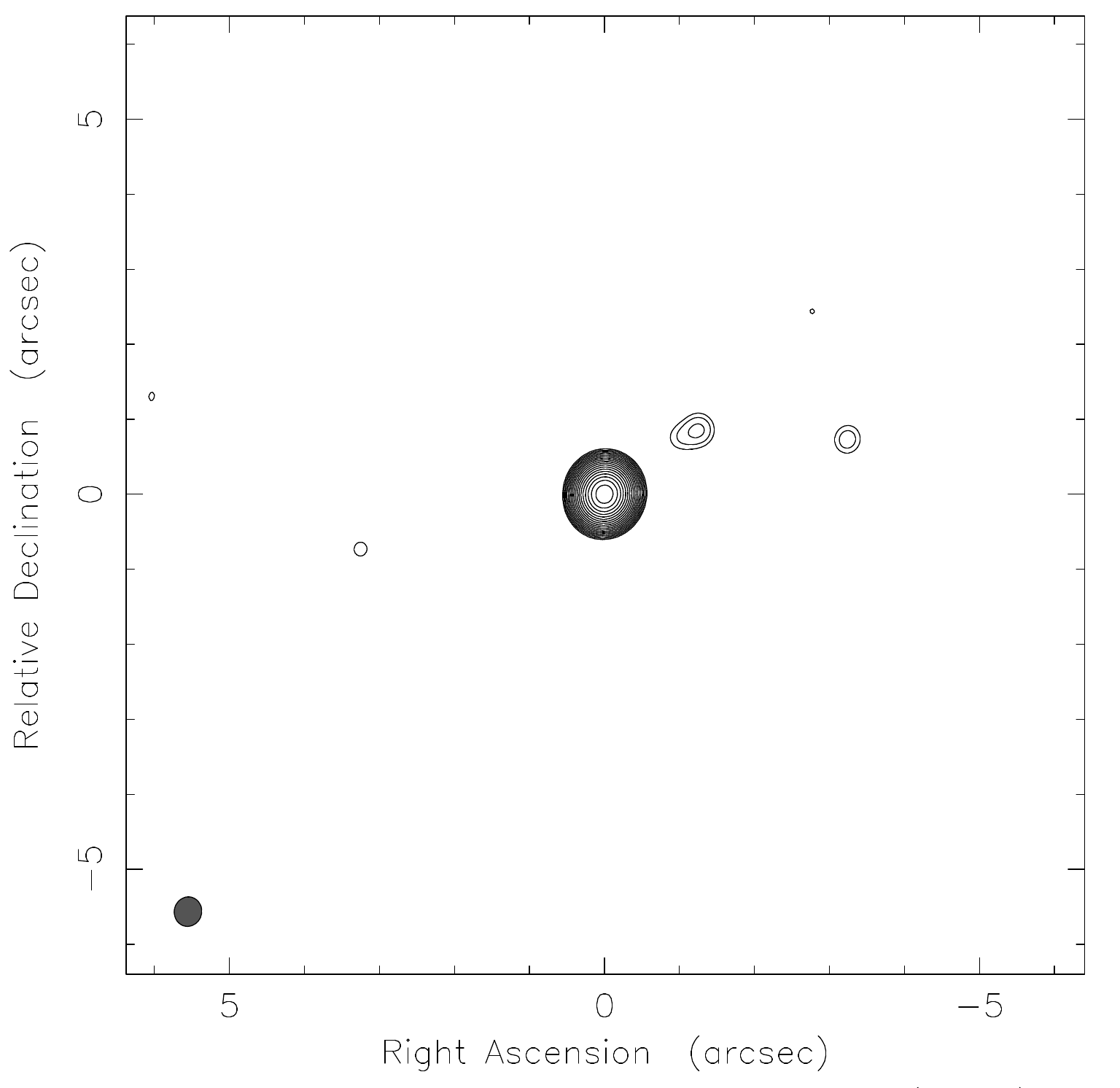} \label{fig:Images1-1}} 
\subfloat[Part 2][J0733+2536 at 4.860 GHz]{\includegraphics[width=2.2in]{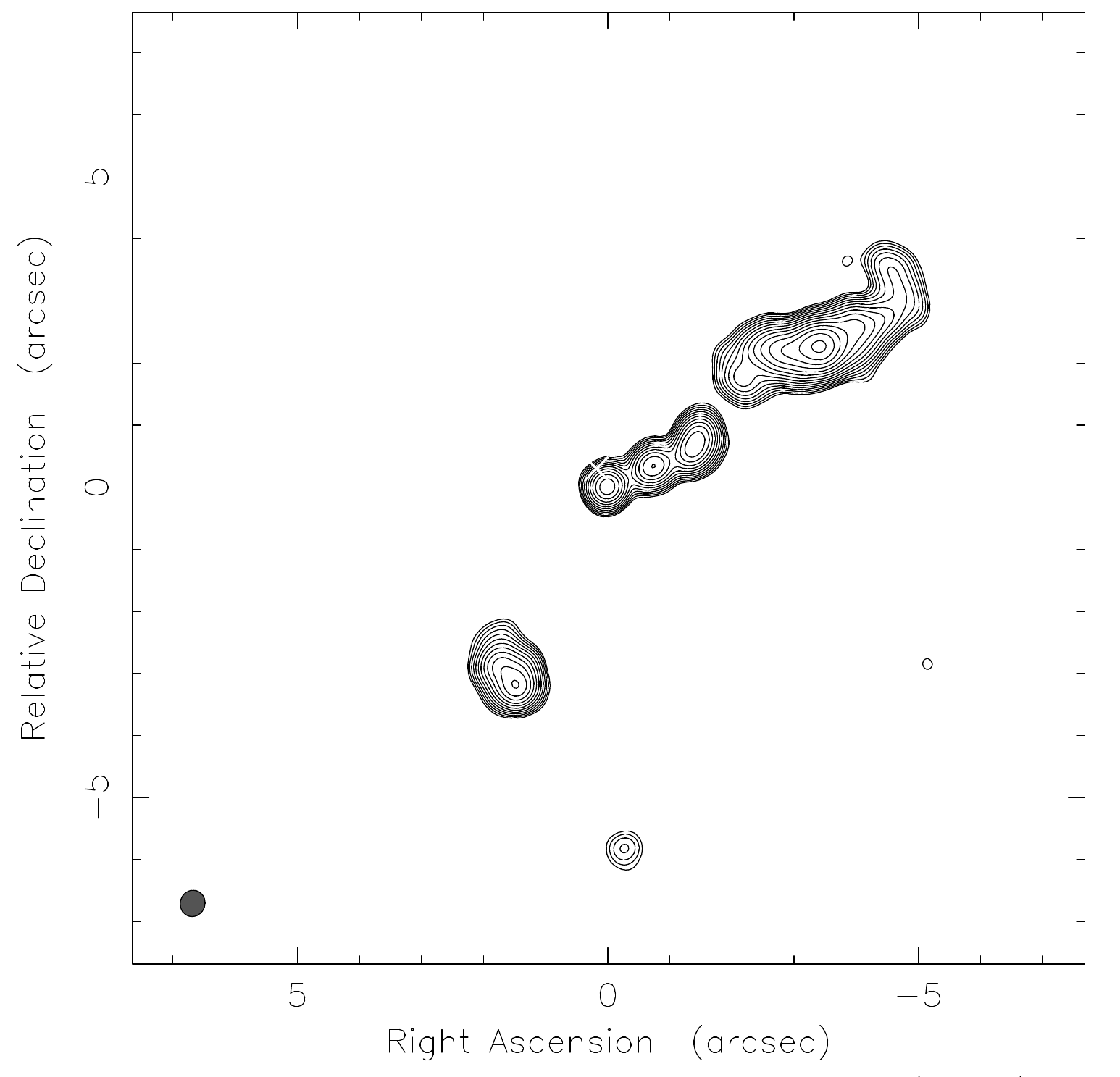} \label{fig:Images1-2}}
\subfloat[Part 3][J0733+2721 at 1.425 GHz]{\includegraphics[width=2.2in]{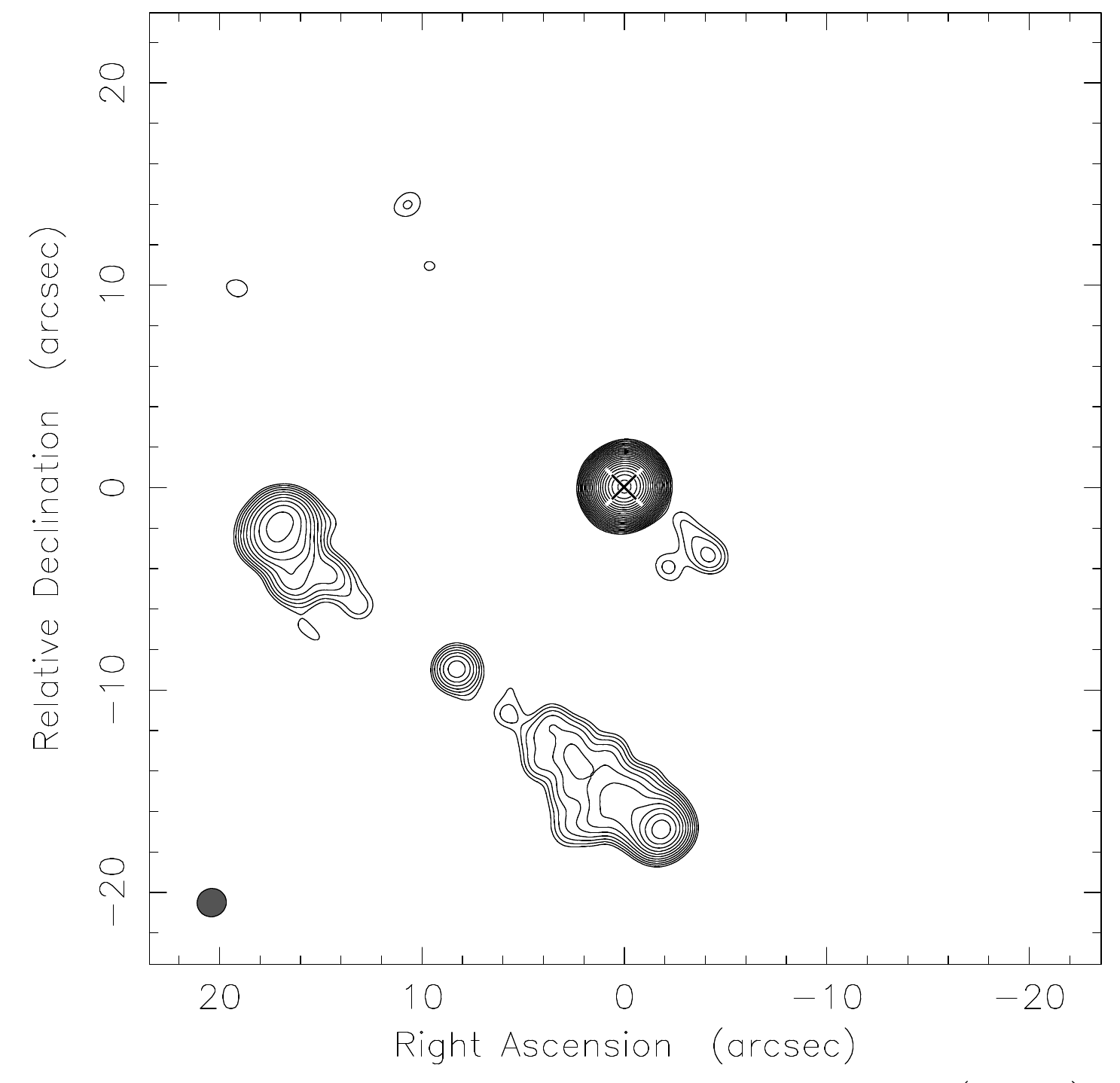} \label{fig:Images1-3}} \\
\subfloat[Part 4][J0736+6513 at 1.490 GHz]{\includegraphics[width=2.2in]{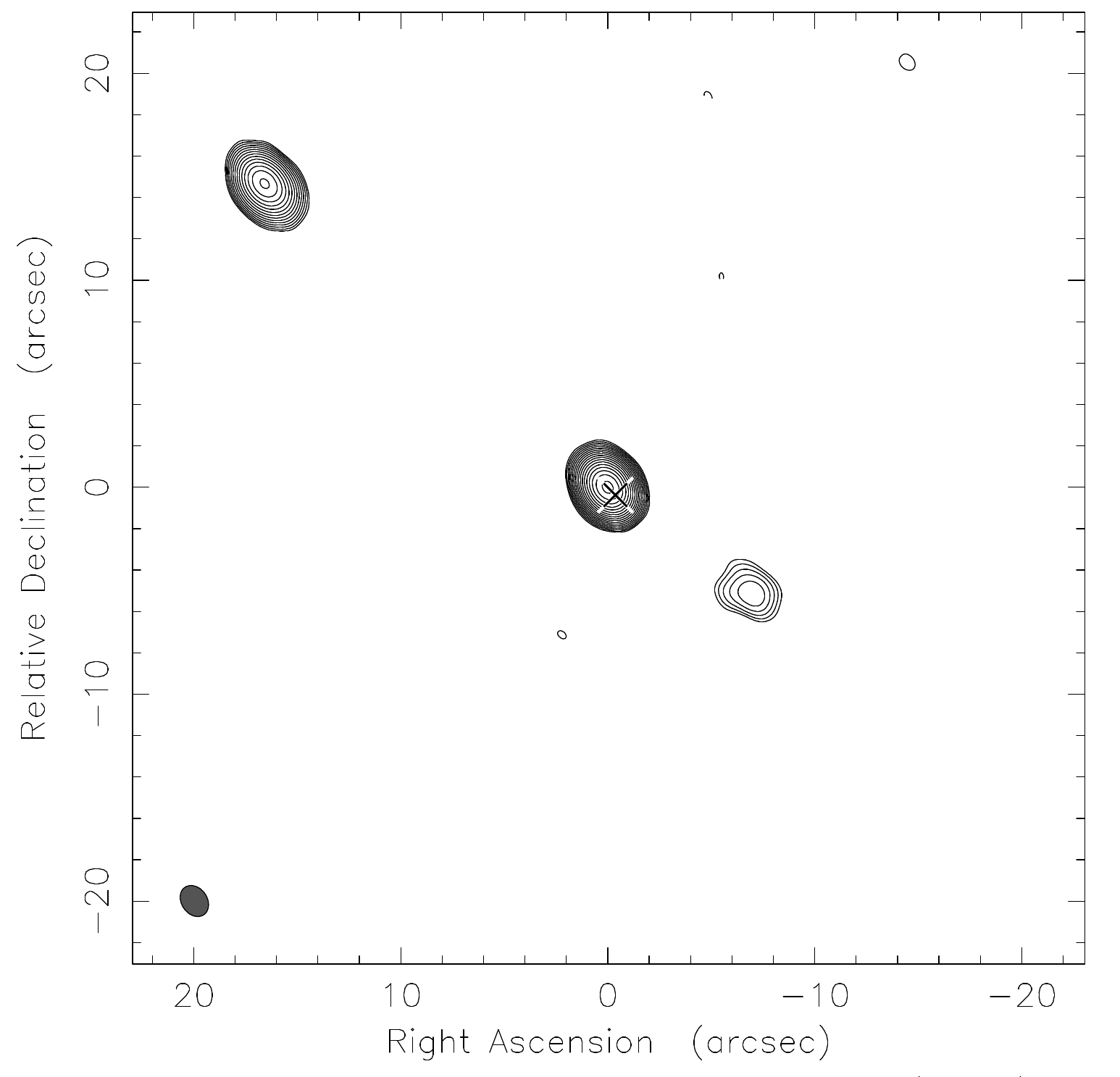} \label{fig:Images1-4}} 
\subfloat[Part 5][J0801+1153 at 1.425 GHz]{\includegraphics[width=2.2in]{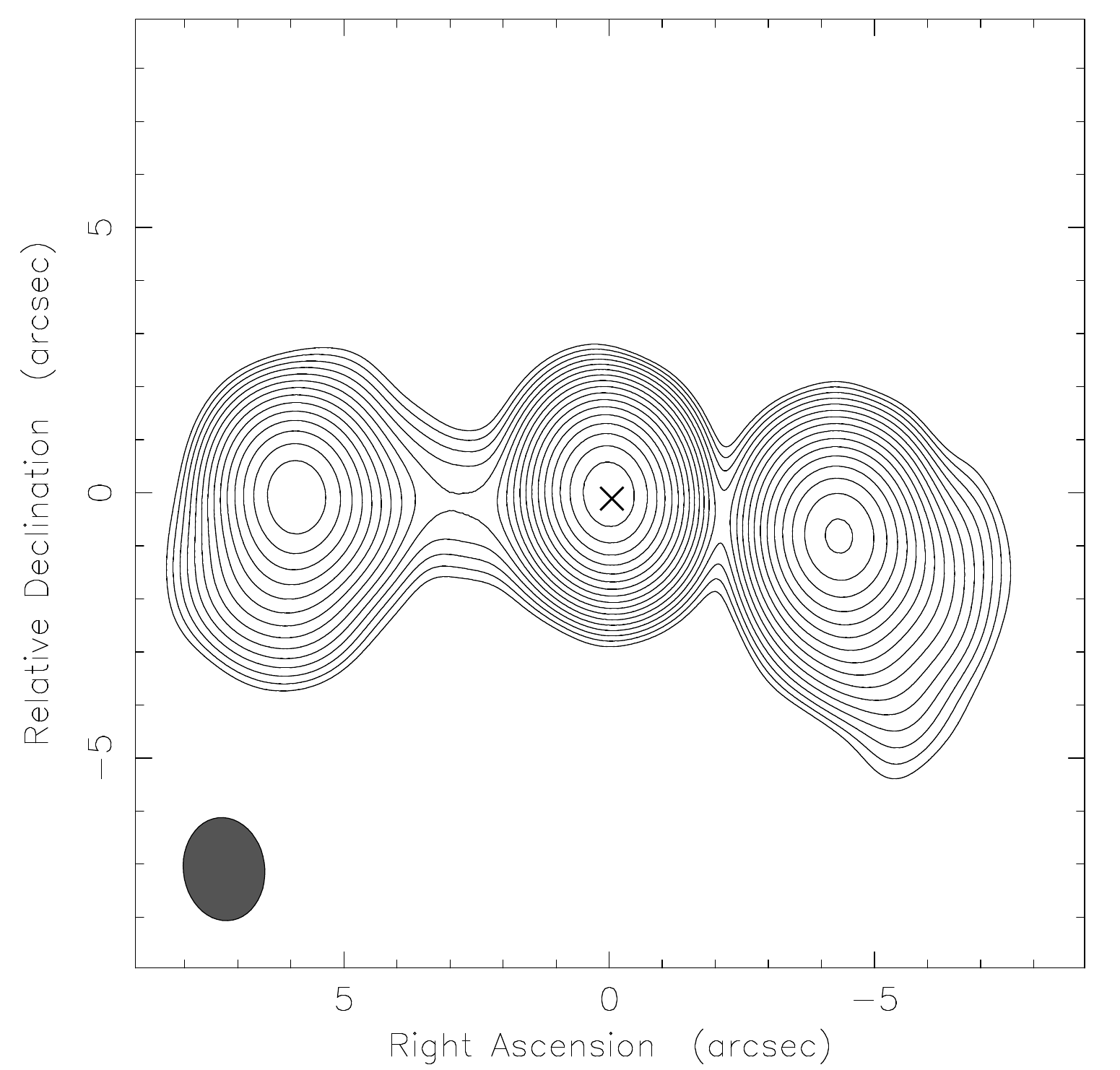} \label{fig:Images2-1}} 
\subfloat[Part 6][J0801+4725 at 1.425 GHz]{\includegraphics[width=2.2in]{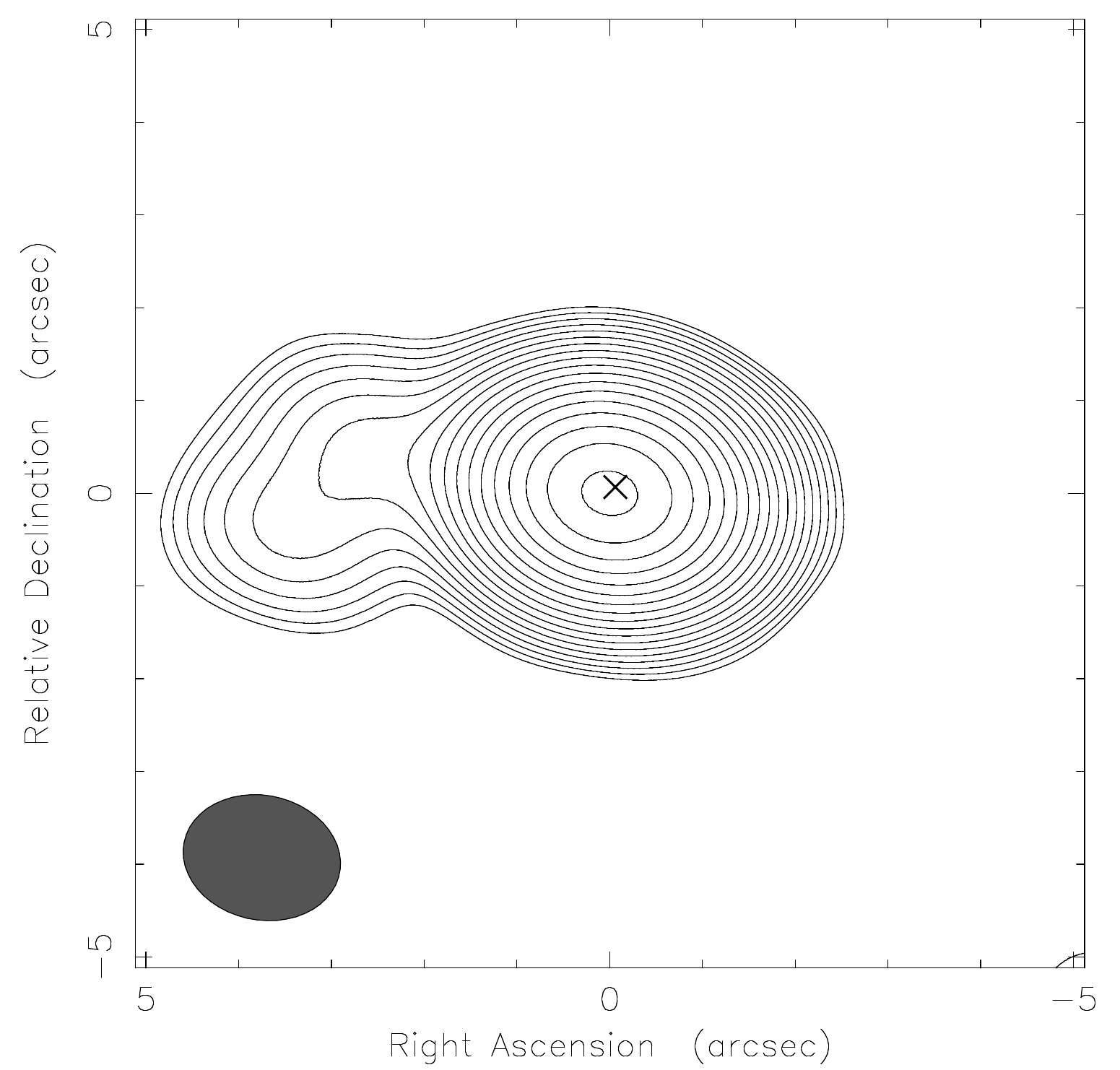} \label{fig:Images2-2}} \\
\subfloat[Part 7][J0805+6144 at 4.860 GHz]{\includegraphics[width=2.2in]{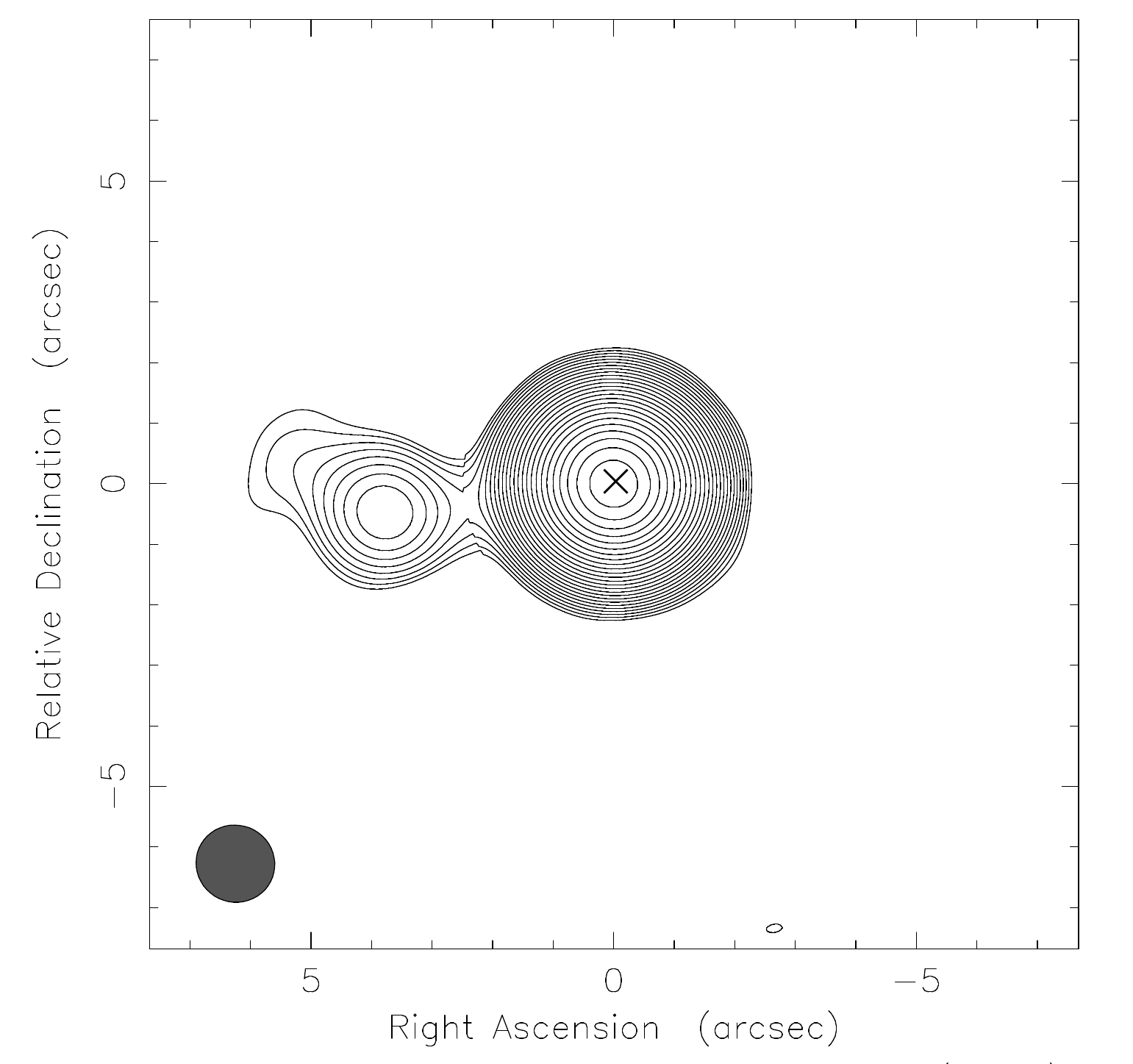} \label{fig:Images2-3}} 
\subfloat[Part 8][J0807+0432 at 8.440 GHz]{\includegraphics[width=2.2in]{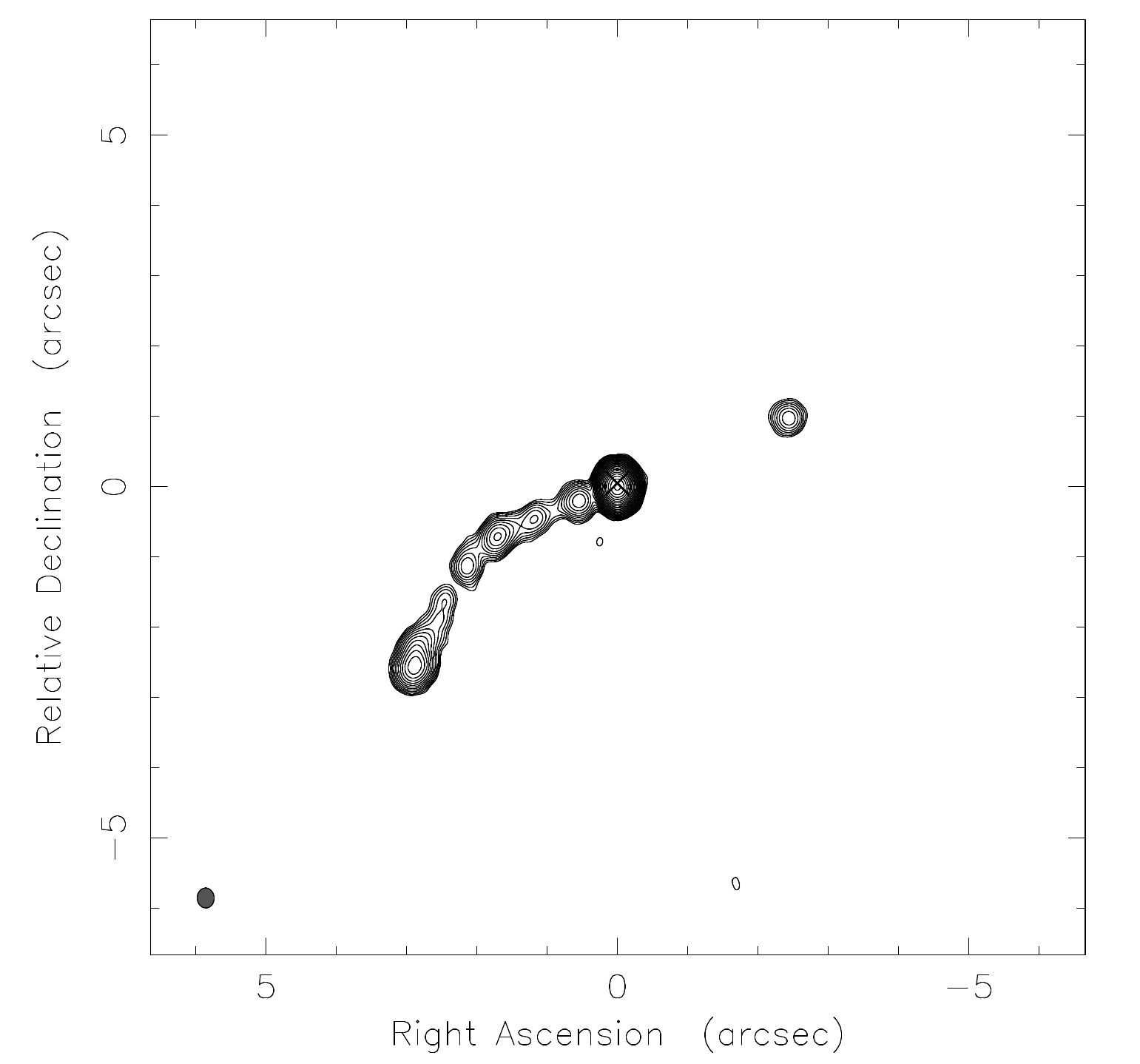} \label{fig:Images2-4}} 
\subfloat[Part 9][J0824+2341 at 1.425 GHz]{\includegraphics[width=2.2in]{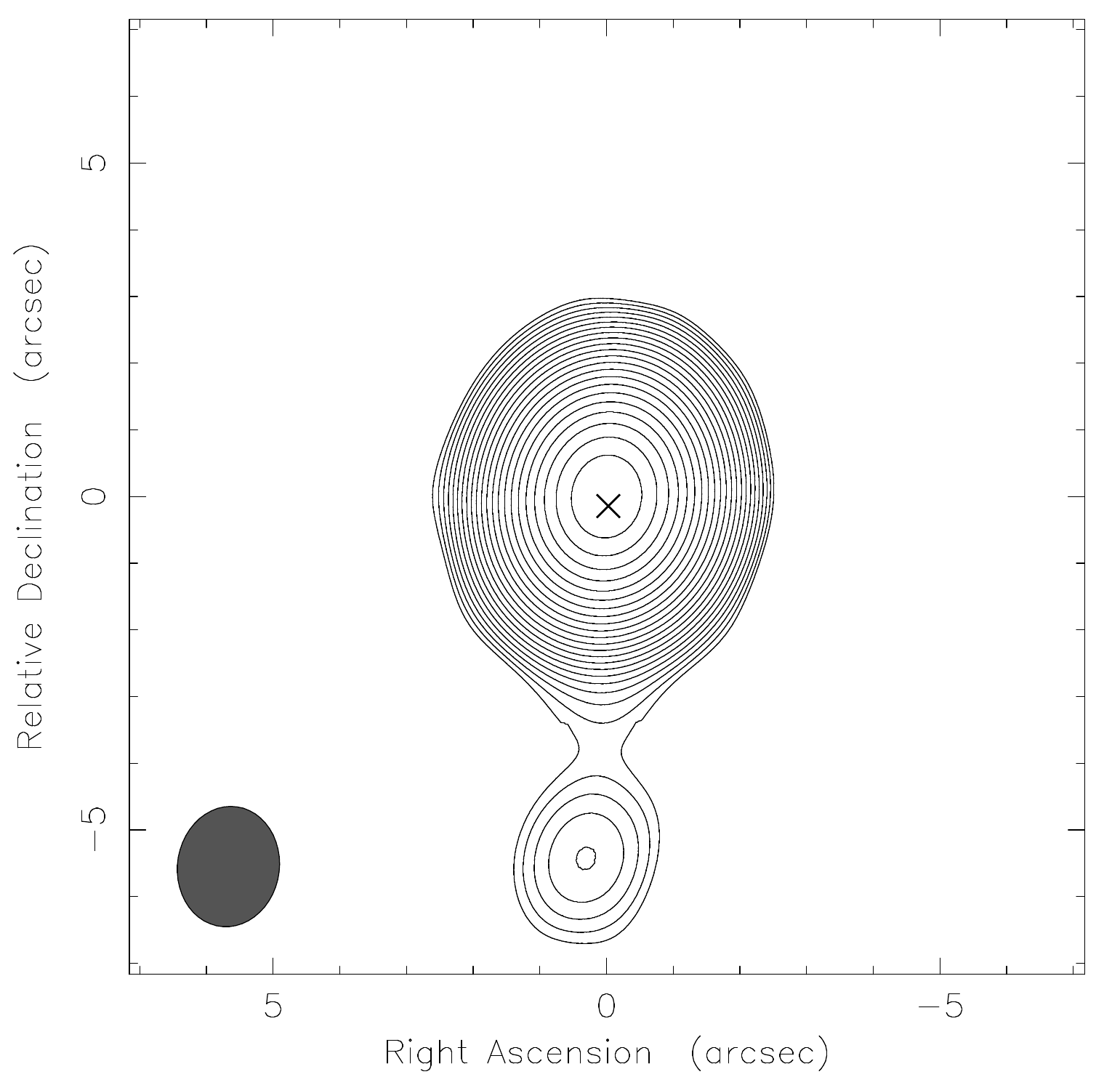} \label{fig:Images3-1}} 
\caption{}
\label{fig:Images2}
\end{figure}

\begin{figure}
\figurenum{3}
\centering
\subfloat[Part 1][J0833+0959 at 1.425 GHz]{\includegraphics[width=2.2in]{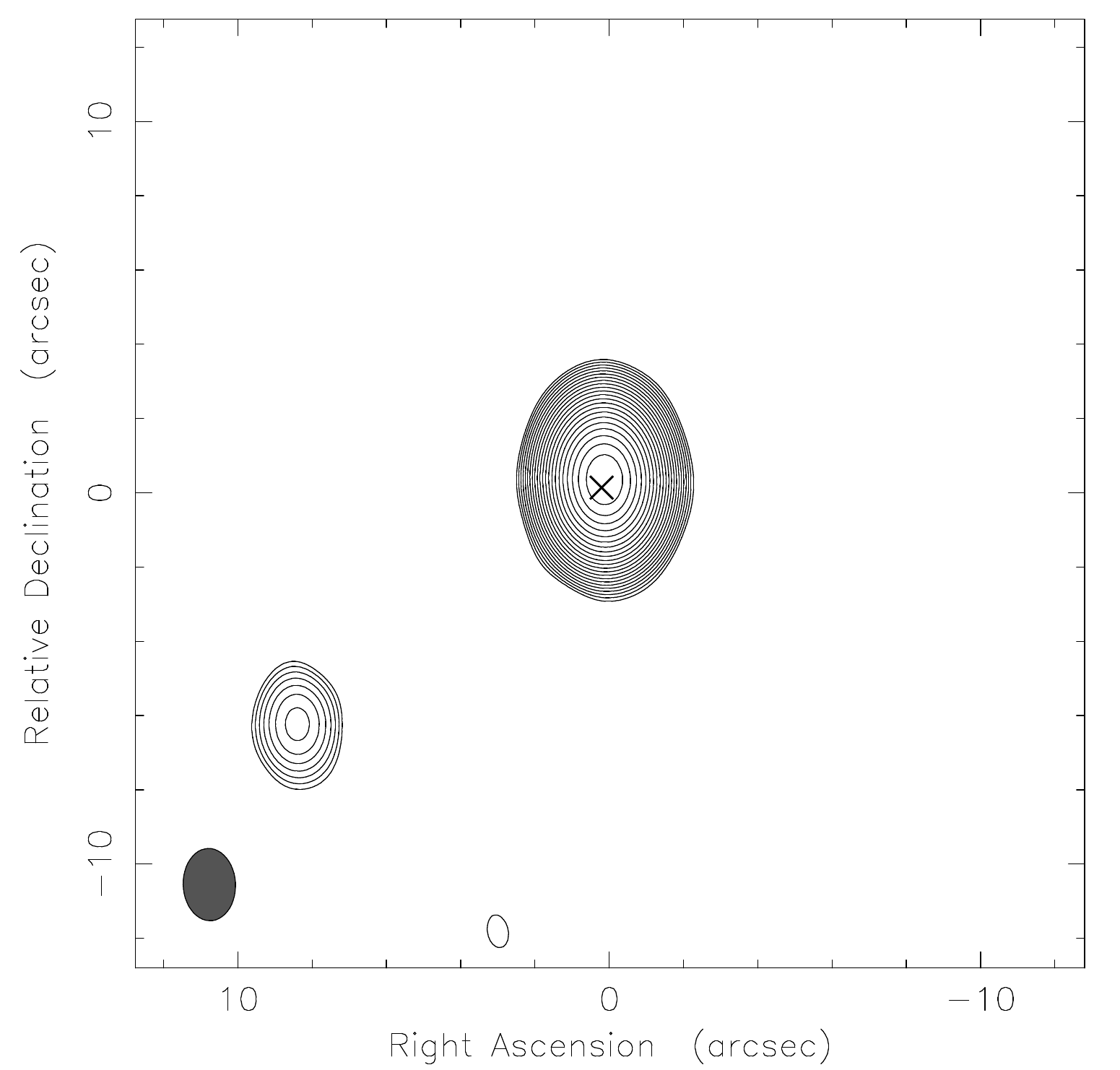} \label{fig:Images3-2}}
\subfloat[Part 2][J0833+1123 at 8.440 GHz]{\includegraphics[width=2.2in]{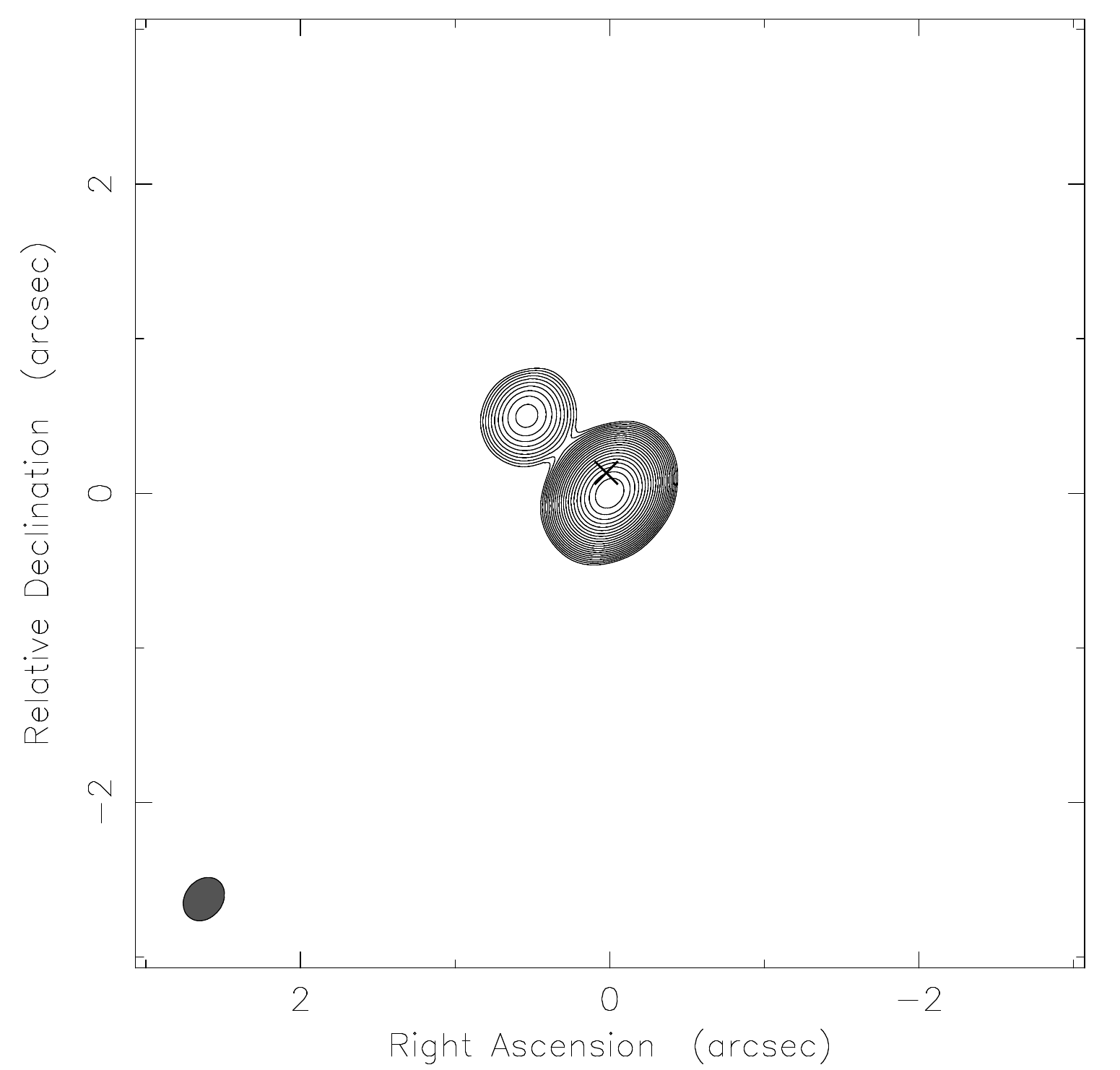} \label{fig:Images3-3}} 
\subfloat[Part 3][J0833+3531 at 1.425 GHz]{\includegraphics[width=2.2in]{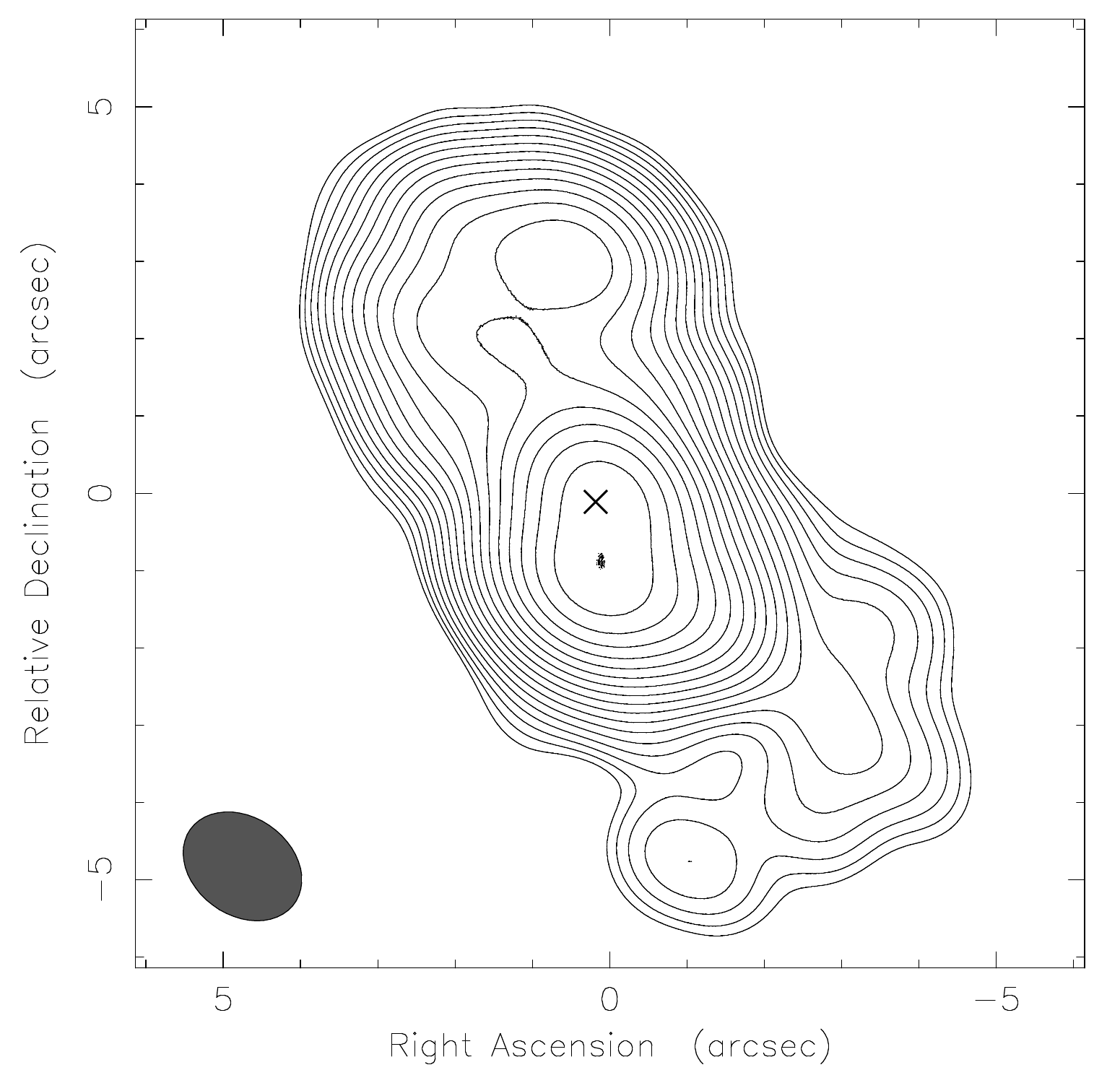} \label{fig:Images3-4}} \\
\subfloat[Part 4][J0905+4850 at 4.860 GHz]{\includegraphics[width=2.2in]{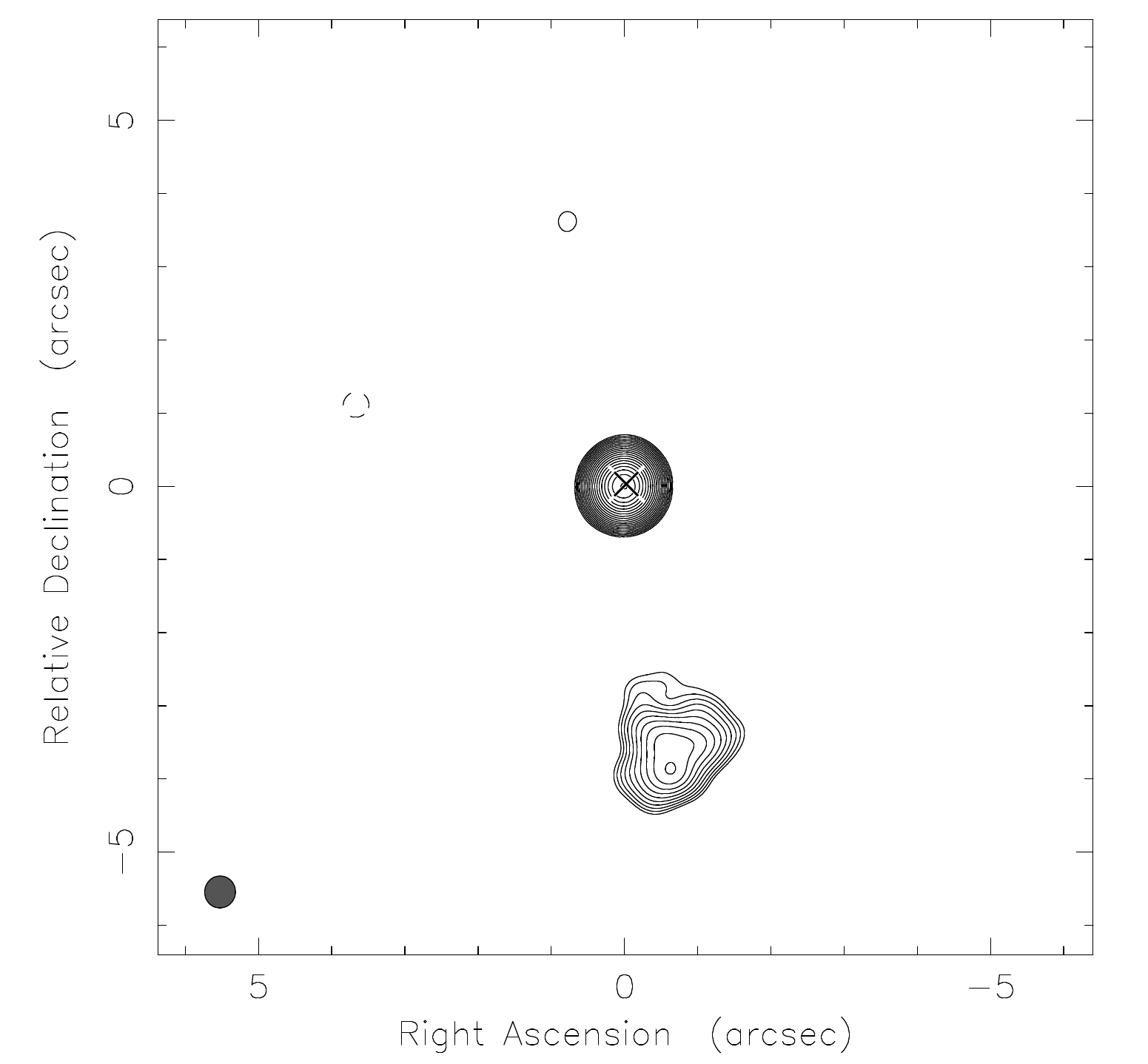} \label{fig:Images4-1}} 
\subfloat[Part 5][J0909+0354 at 4.860 GHz]{\includegraphics[width=2.2in]{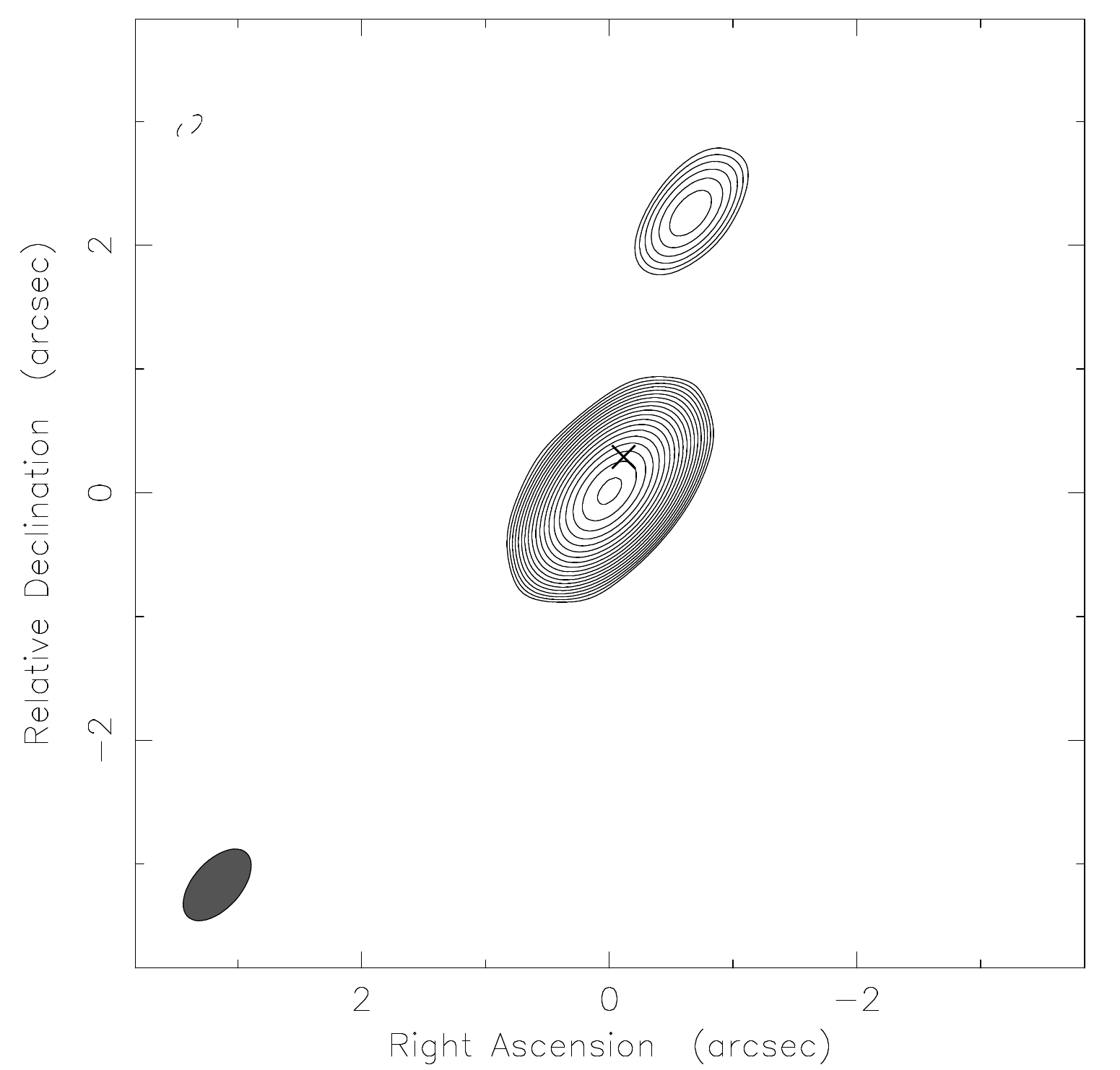} \label{fig:Images4-2}}
\subfloat[Part 6][J0915+0007 at 1.425 GHz]{\includegraphics[width=2.2in]{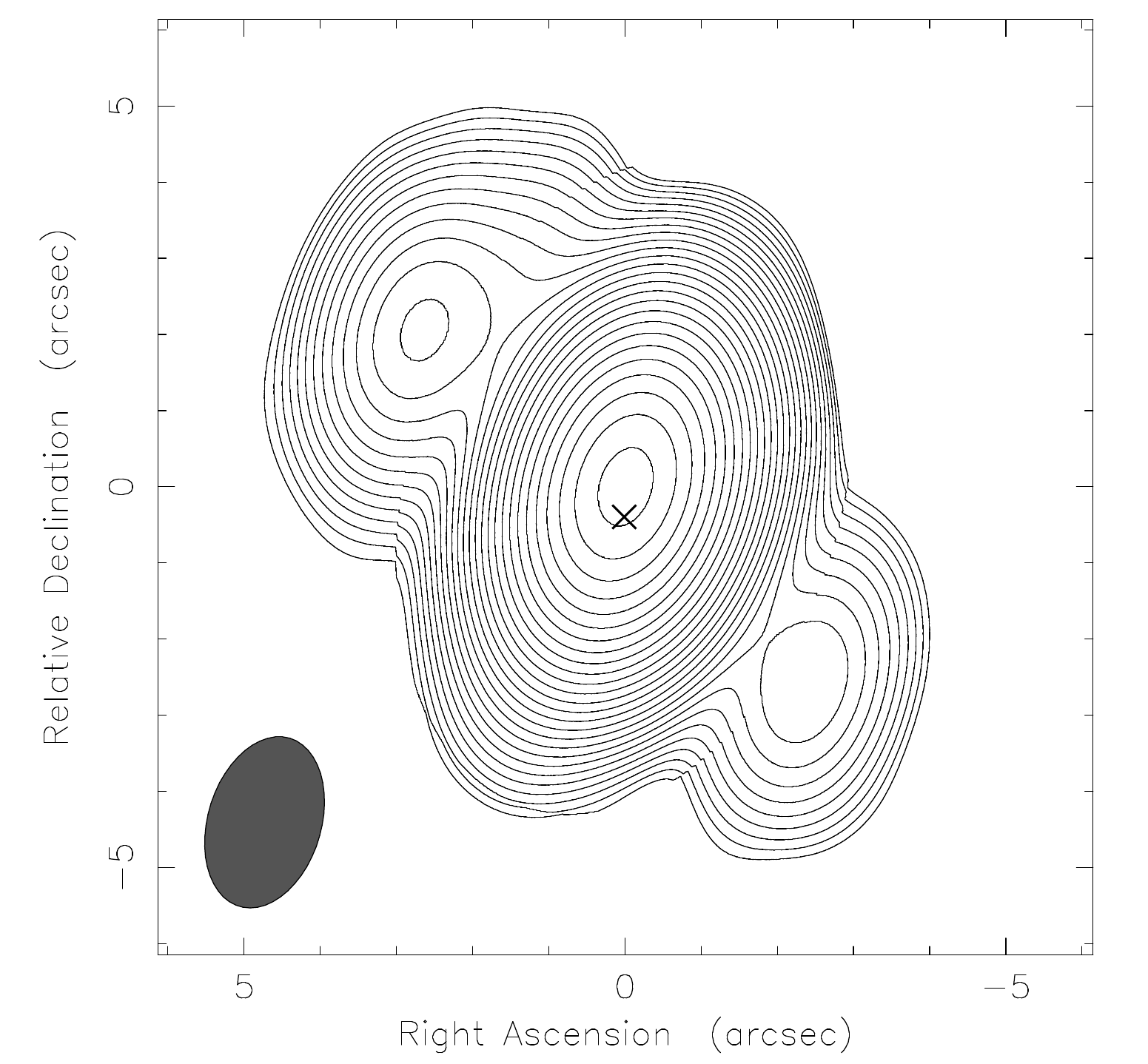} \label{fig:Images4-3}} \\
\subfloat[Part 7][J0918+5332 at 4.860 GHz]{\includegraphics[width=2.2in]{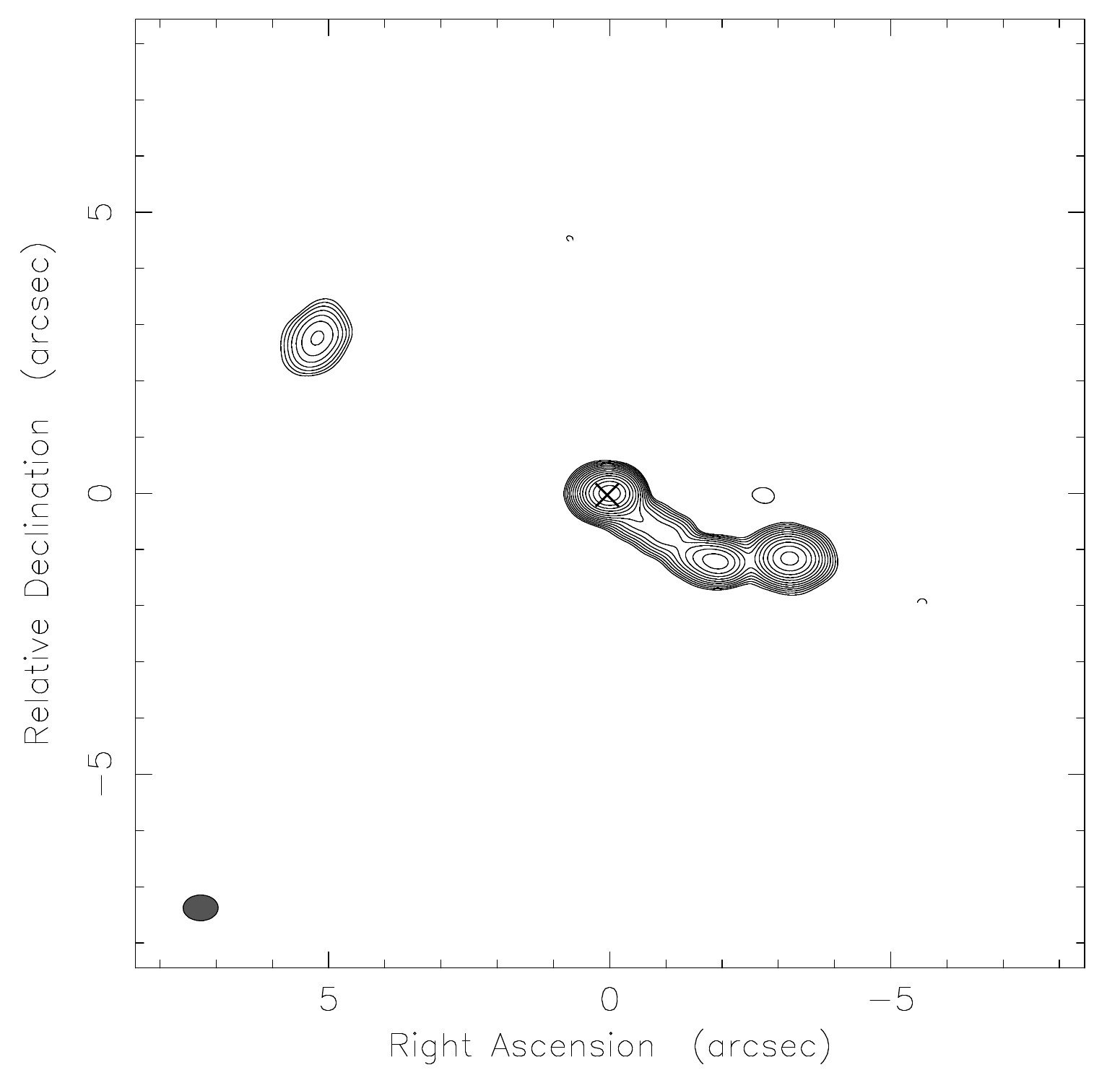} \label{fig:Images4-4}}
\subfloat[Part 8][J0934+3050 at 8.440 GHz]{\includegraphics[width=2.2in]{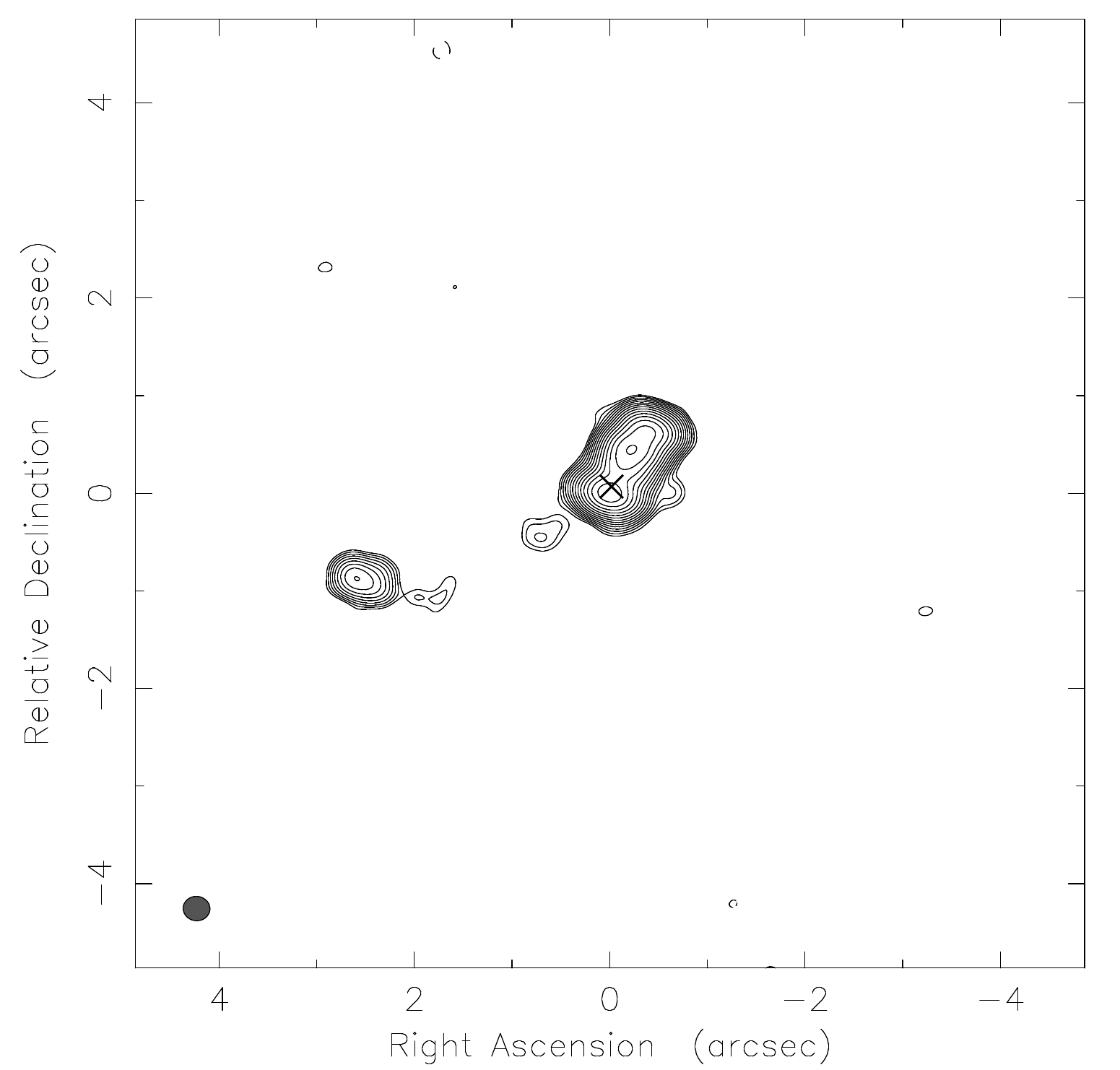} \label{fig:Images5-1}} 
\subfloat[Part 9][J0934+4908 at 1.400 GHz]{\includegraphics[width=2.2in]{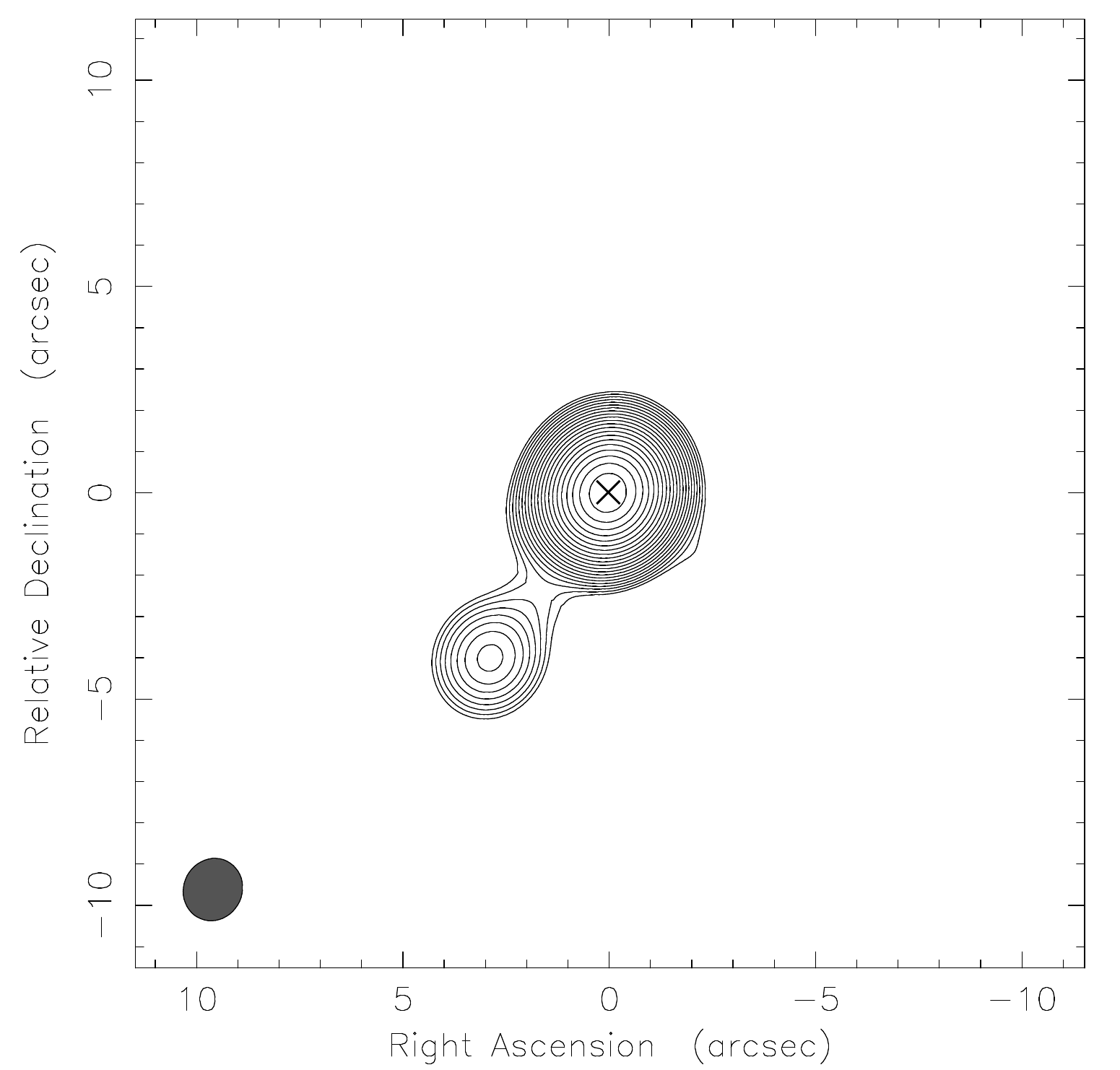} \label{fig:Images5-2}}
\caption{}
\label{fig:Images3}
\end{figure}

\begin{figure}
\figurenum{4}
\centering
\subfloat[Part 1][J0941+1145 at 1.490 GHz]{\includegraphics[width=2.2in]{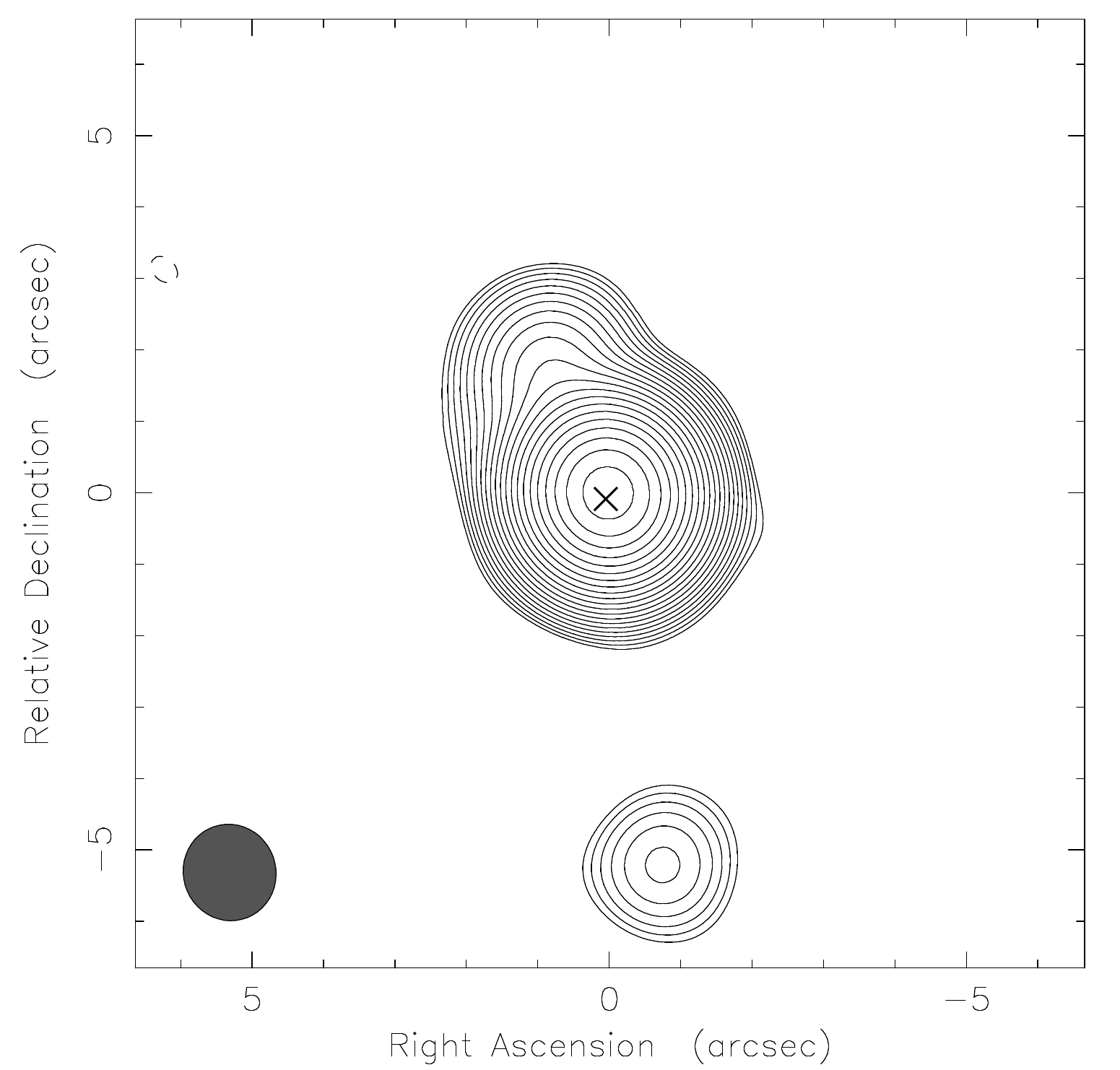} \label{fig:Images5-3}} 
\subfloat[Part 2][J0944+2554 at 8.440 GHz]{\includegraphics[width=2.2in]{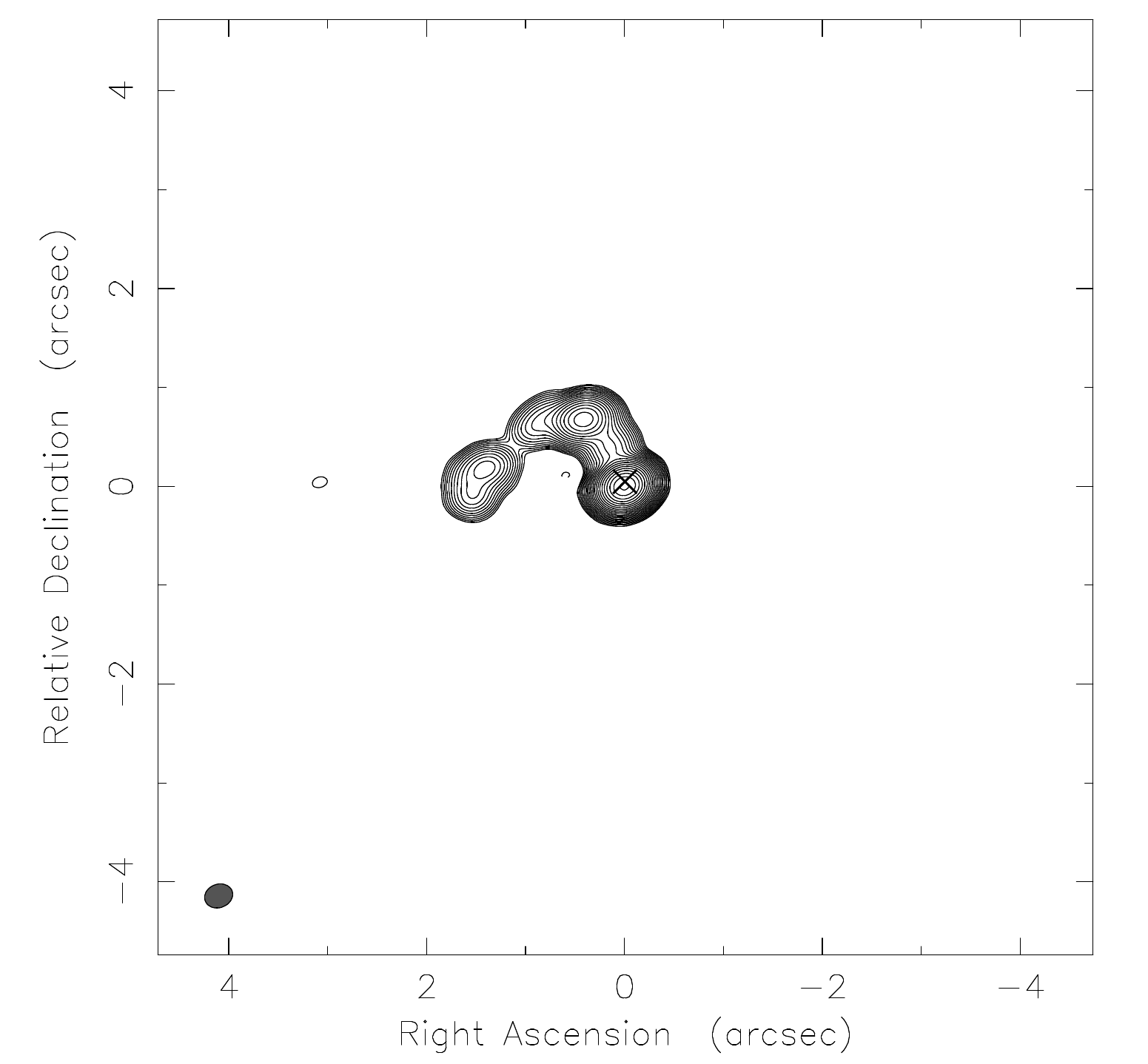} \label{fig:Images5-4}}
\subfloat[Part 3][J0947+6328 at 1.400 GHz]{\includegraphics[width=2.2in]{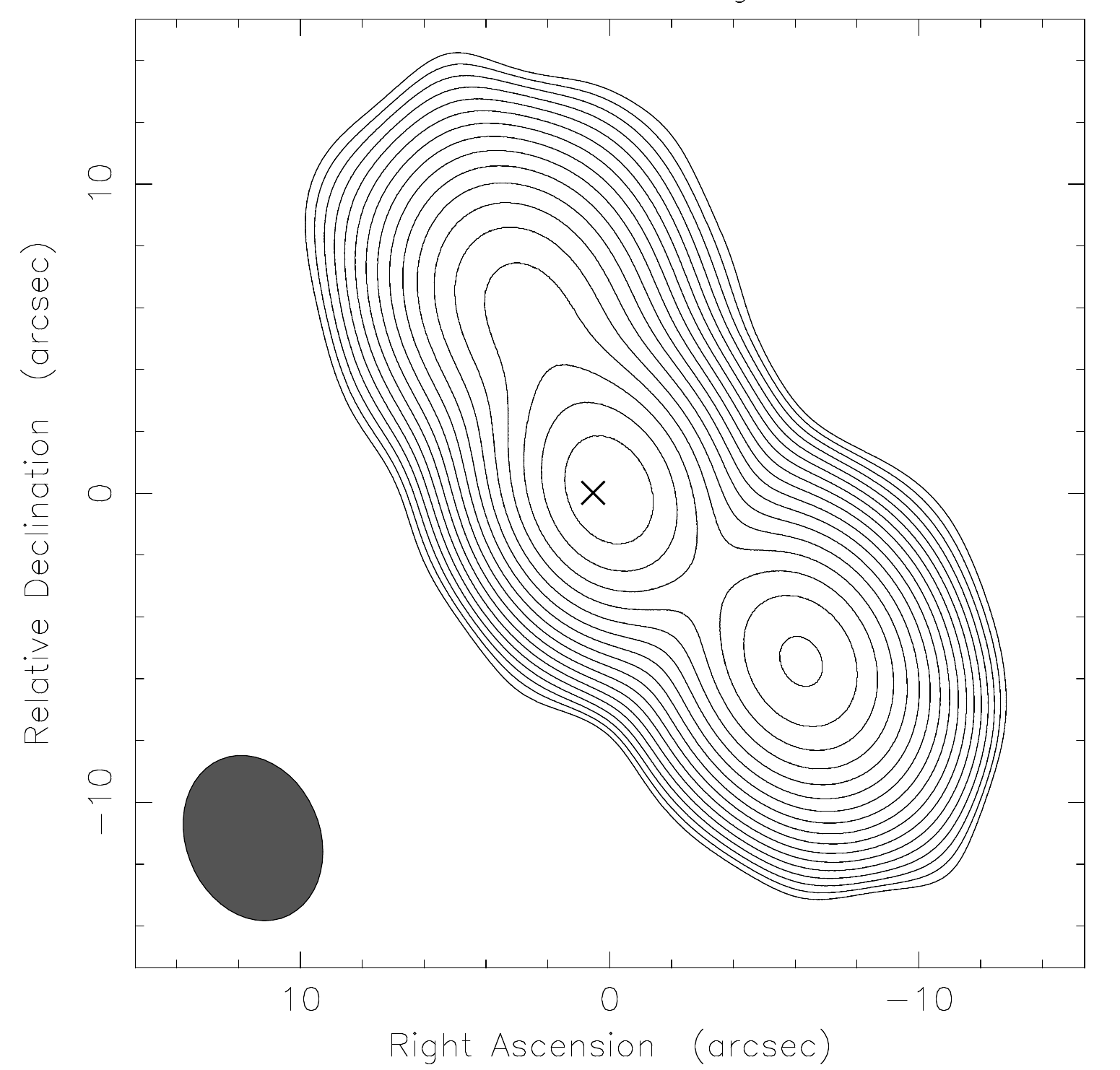} \label{fig:Images6-1}} \\
\subfloat[Part 4][J0958+3922 at 1.490 GHz]{\includegraphics[width=2.2in]{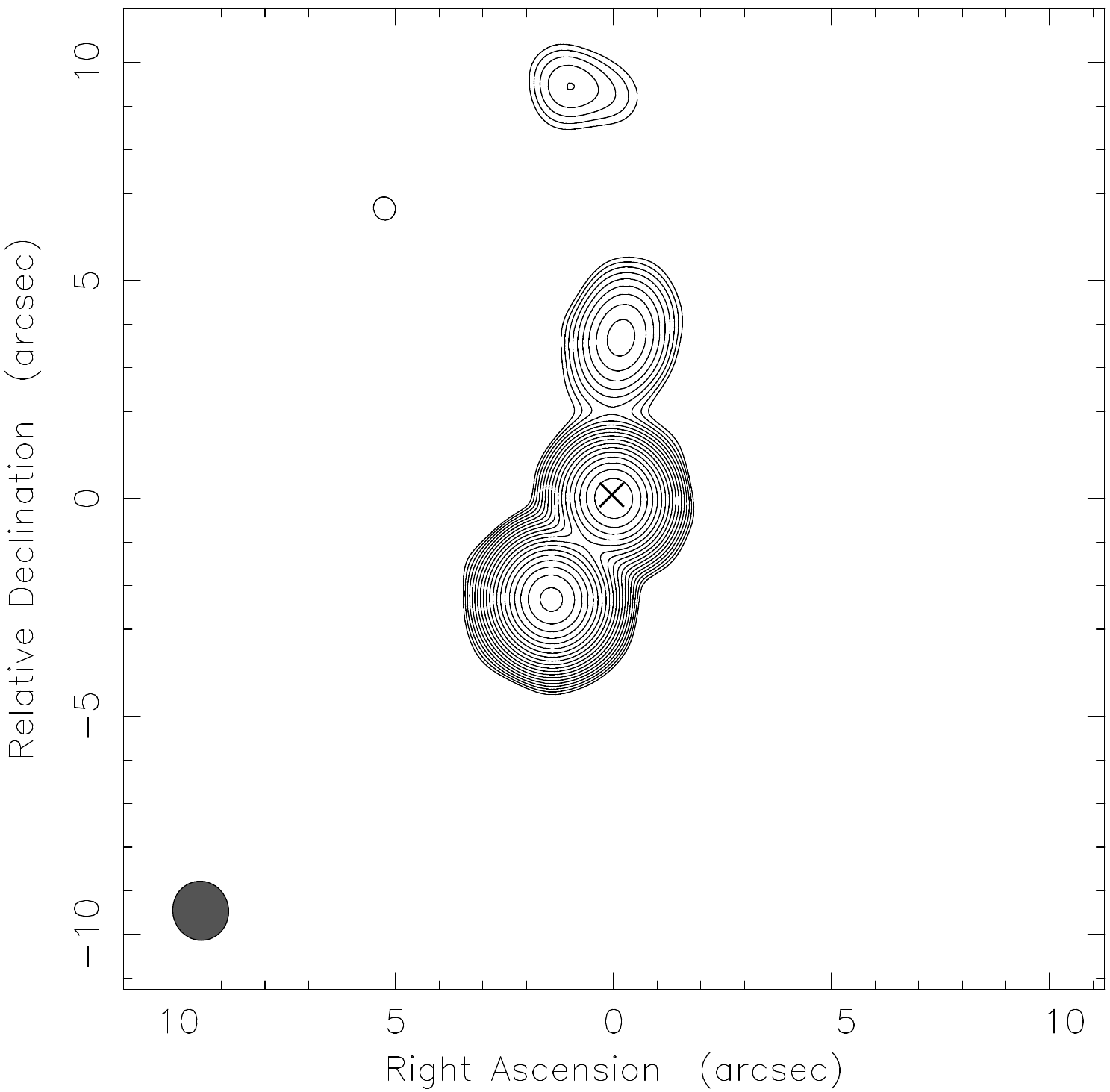} \label{fig:Images6-2}} 
\subfloat[Part 5][J1007+1356 at 4.760 GHz]{\includegraphics[width=2.2in]{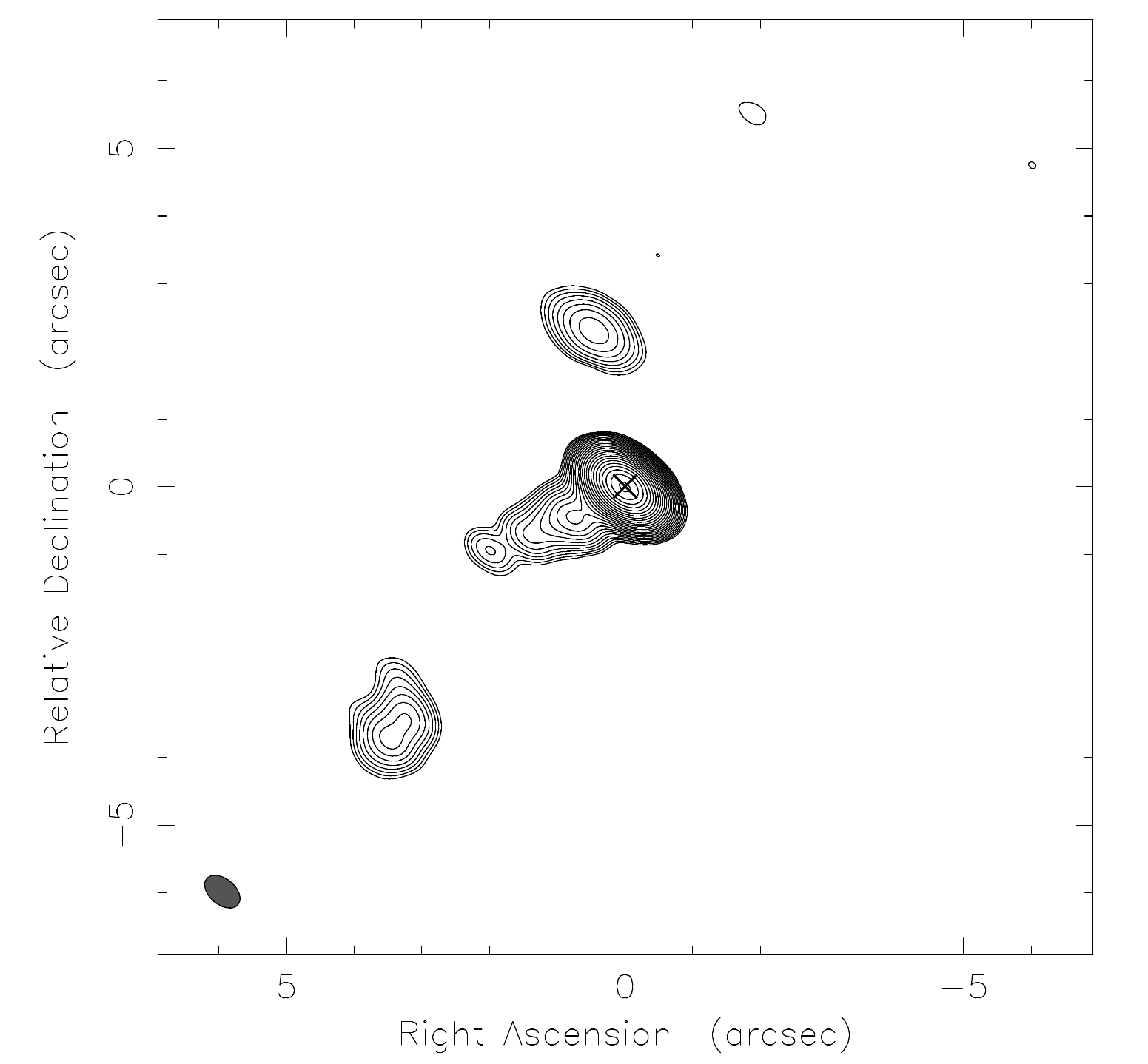} \label{fig:Images6-3}} 
\subfloat[Part 6][J1016+2037 at 4.860 GHz]{\includegraphics[width=2.2in]{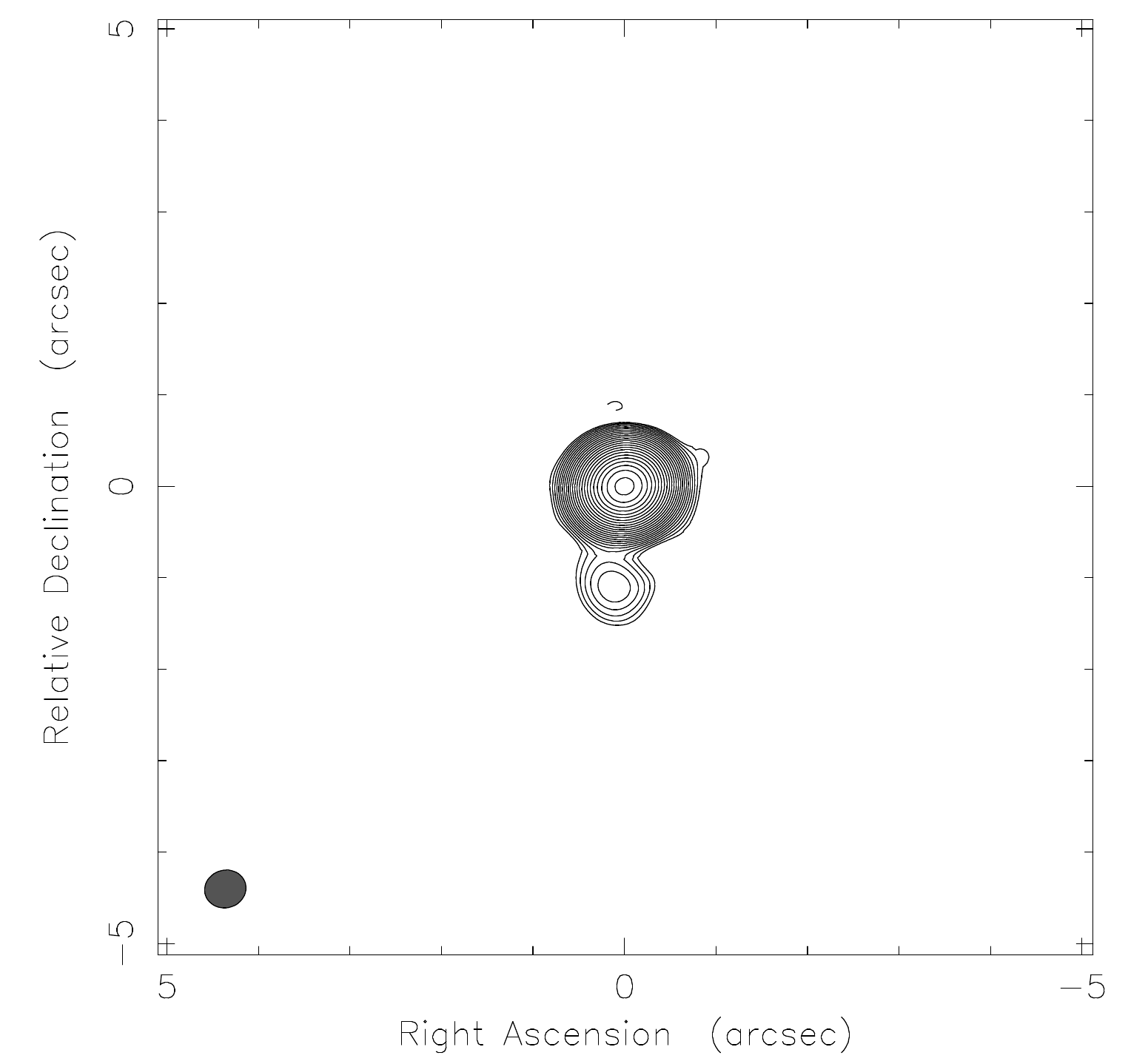} \label{fig:Images6-4}} \\
\subfloat[Part 7][J1036+1326 at 1.490 GHz]{\includegraphics[width=2.2in]{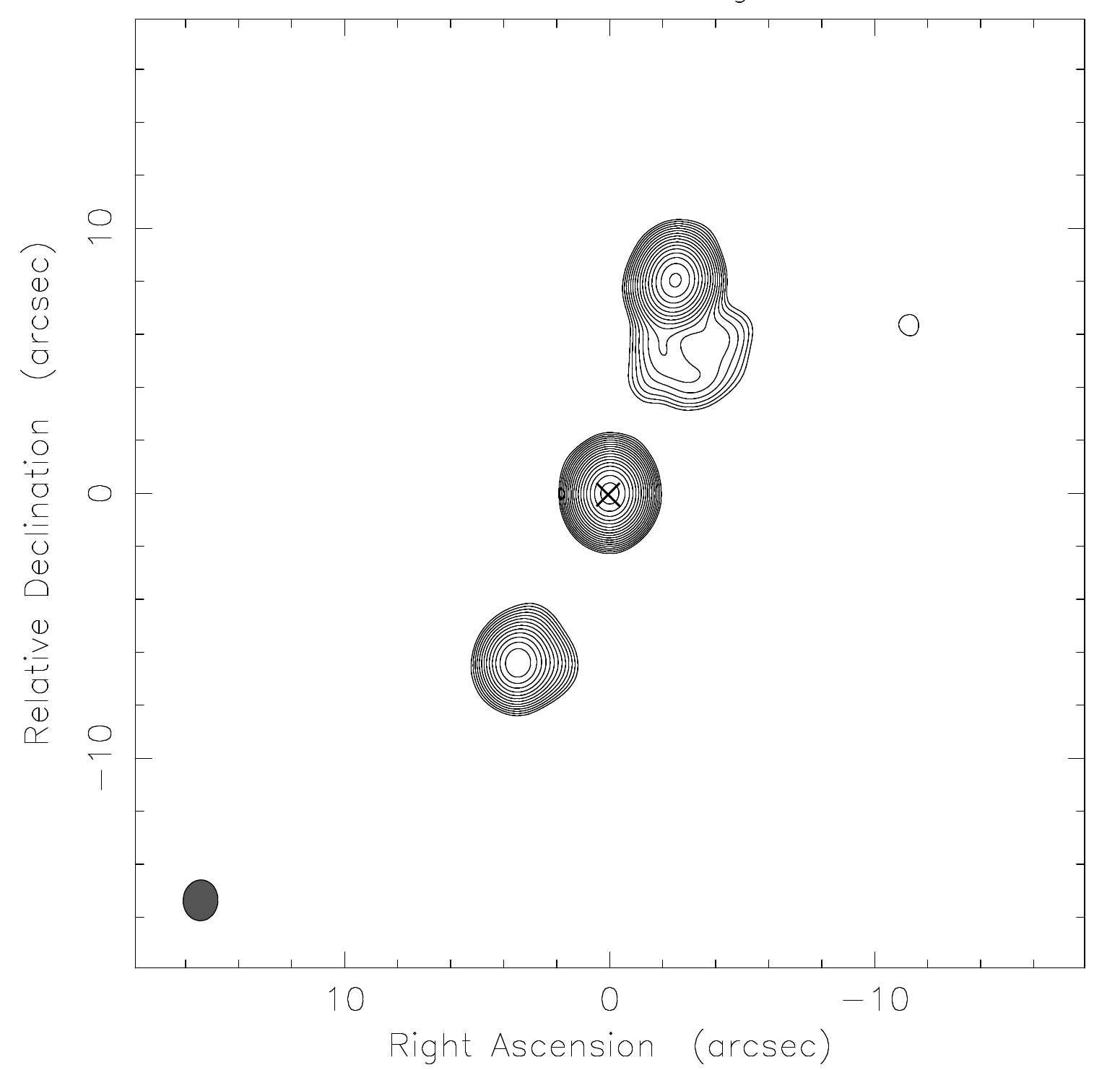} \label{fig:Images7-1}} 
\subfloat[Part 8][J1049+1332 at 1.425 GHz]{\includegraphics[width=2.2in]{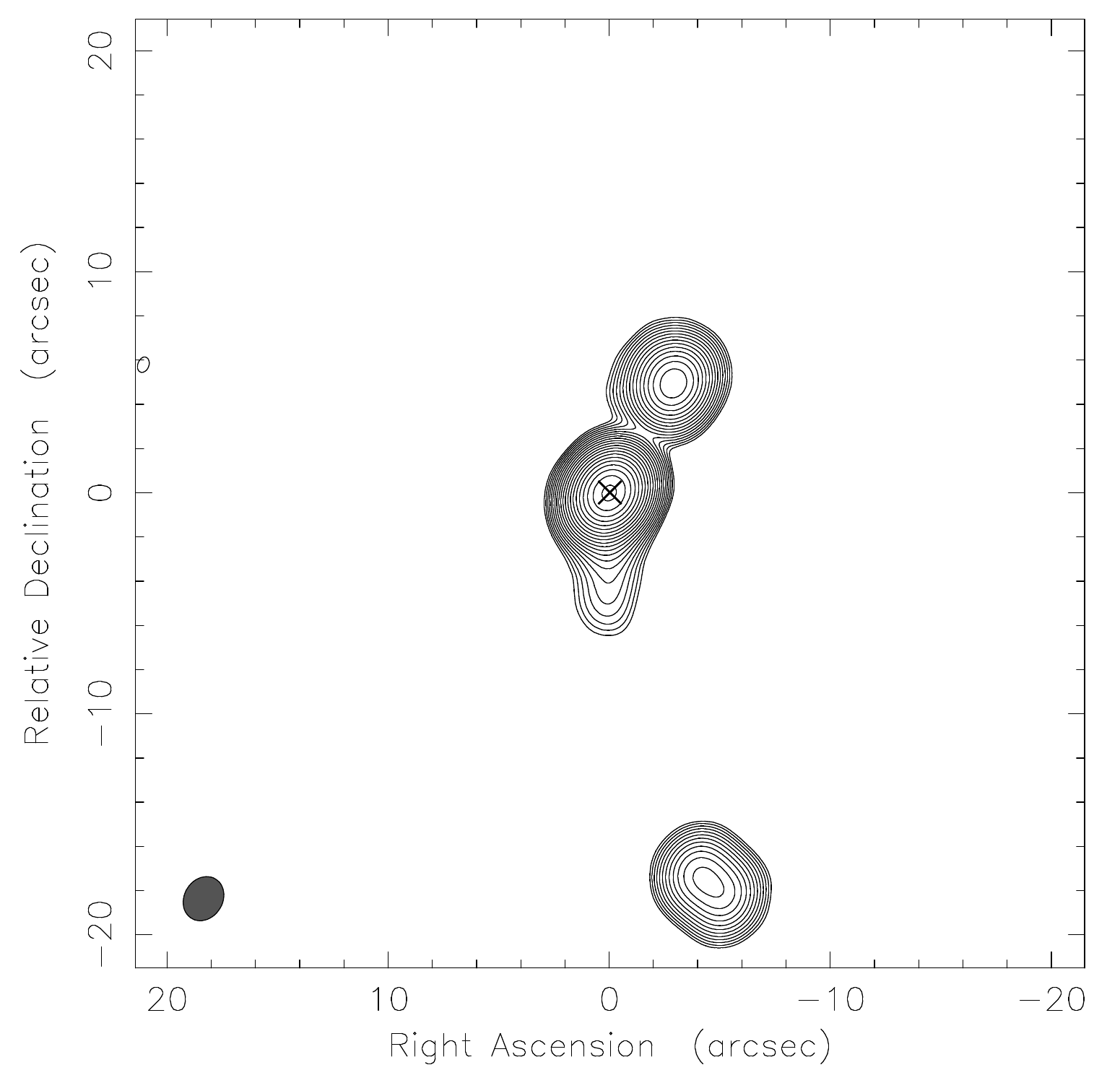} \label{fig:Images7-2}} 
\subfloat[Part 9][J1057+0324 at 4.860 GHz]{\includegraphics[width=2.2in]{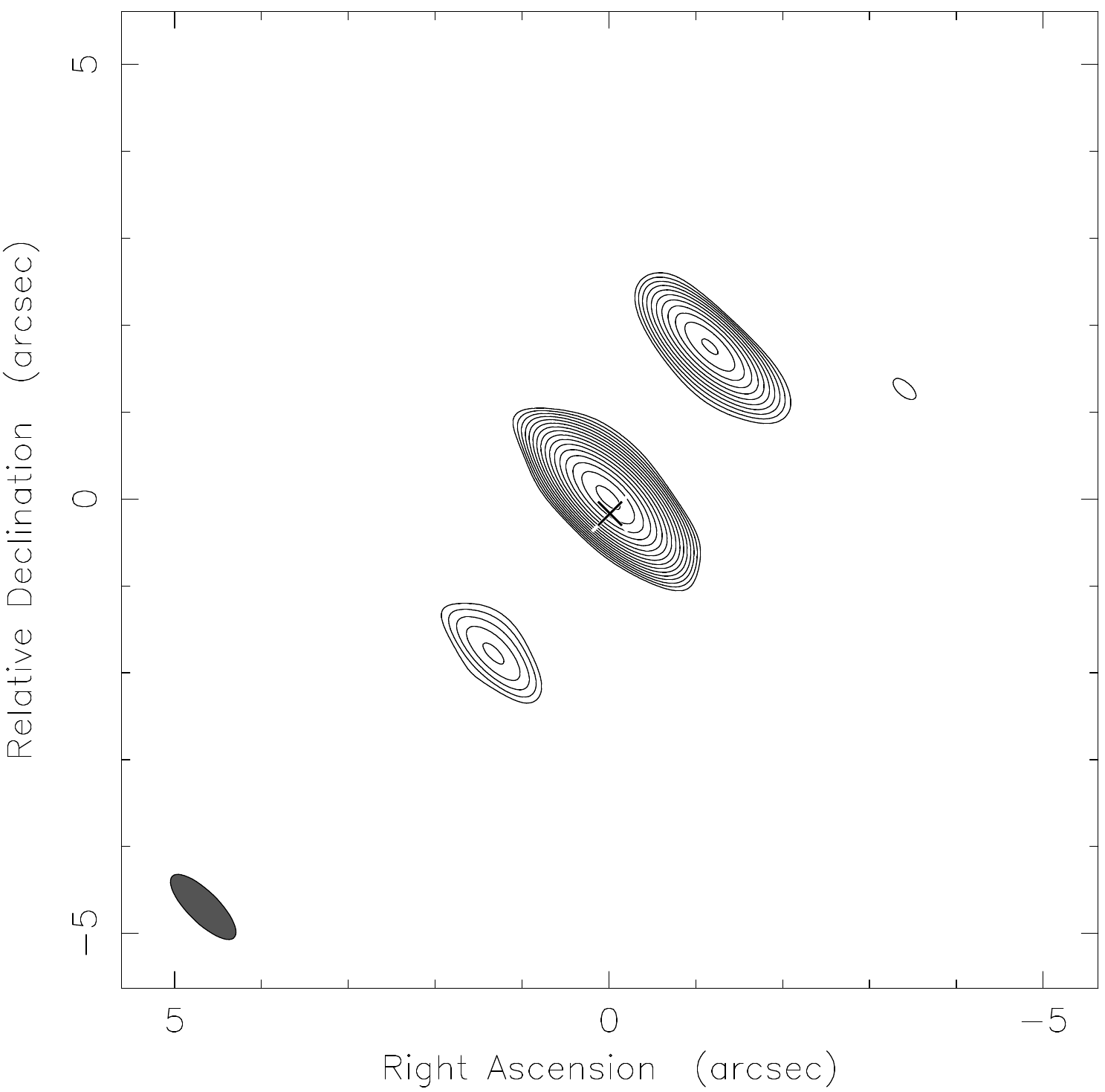} \label{fig:Images7-3}} 
\caption{}
\label{fig:Images4}
\end{figure}

\begin{figure}
\figurenum{5}
\centering
\subfloat[Part 4][J1058+0443 at 4.860 GHz]{\includegraphics[width=2.2in]{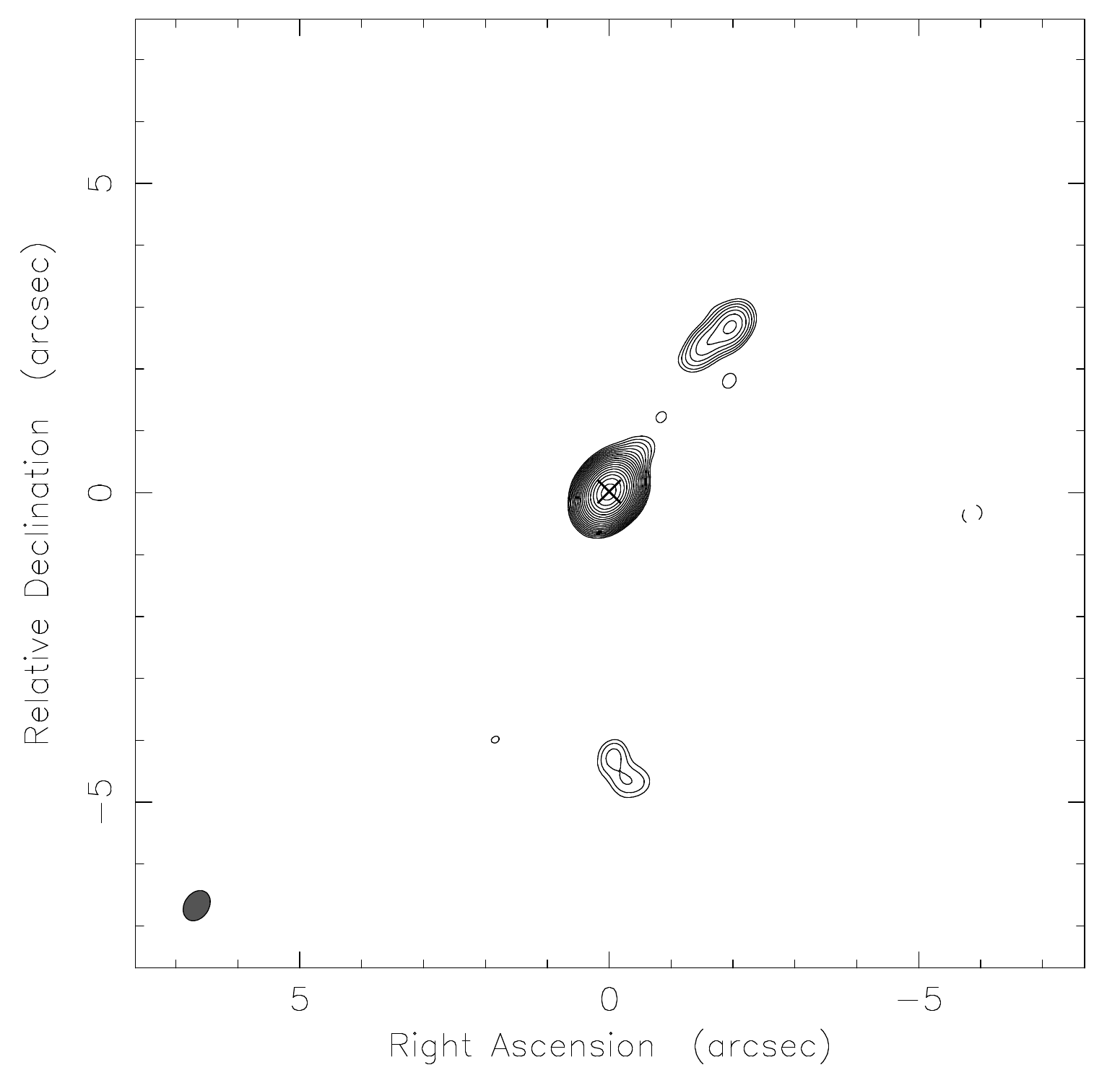} \label{fig:Images7-4}}
\subfloat[Part 1][J1127+5650 at 8.460 GHz]{\includegraphics[width=2.2in]{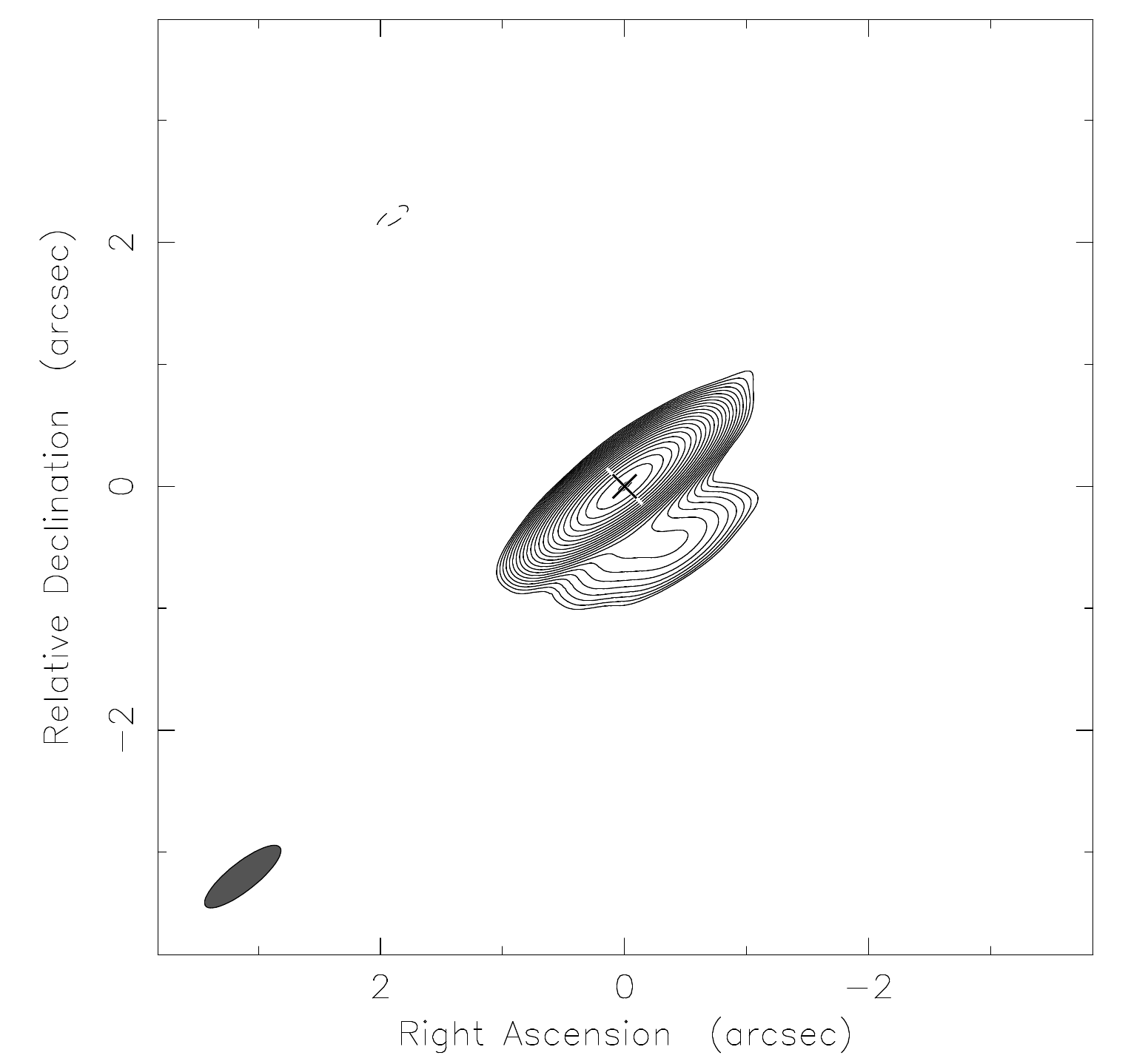} \label{fig:Images8-1}} 
\subfloat[Part 2][J1204+5228 at 4.710 GHz]{\includegraphics[width=2.2in]{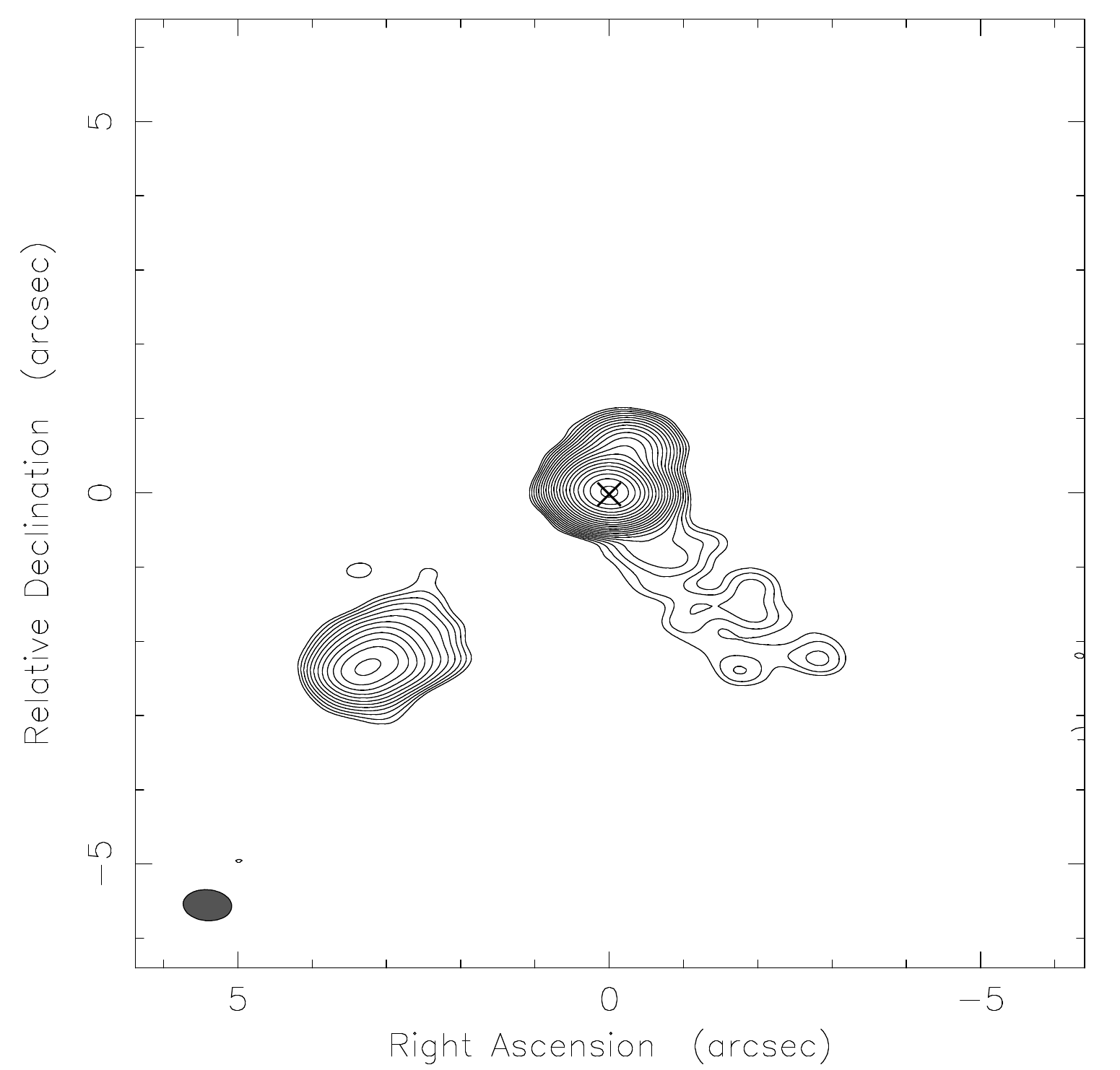} \label{fig:Images8-2}} \\
\subfloat[Part 3][J1213+3247 at 4.860 GHz]{\includegraphics[width=2.2in]{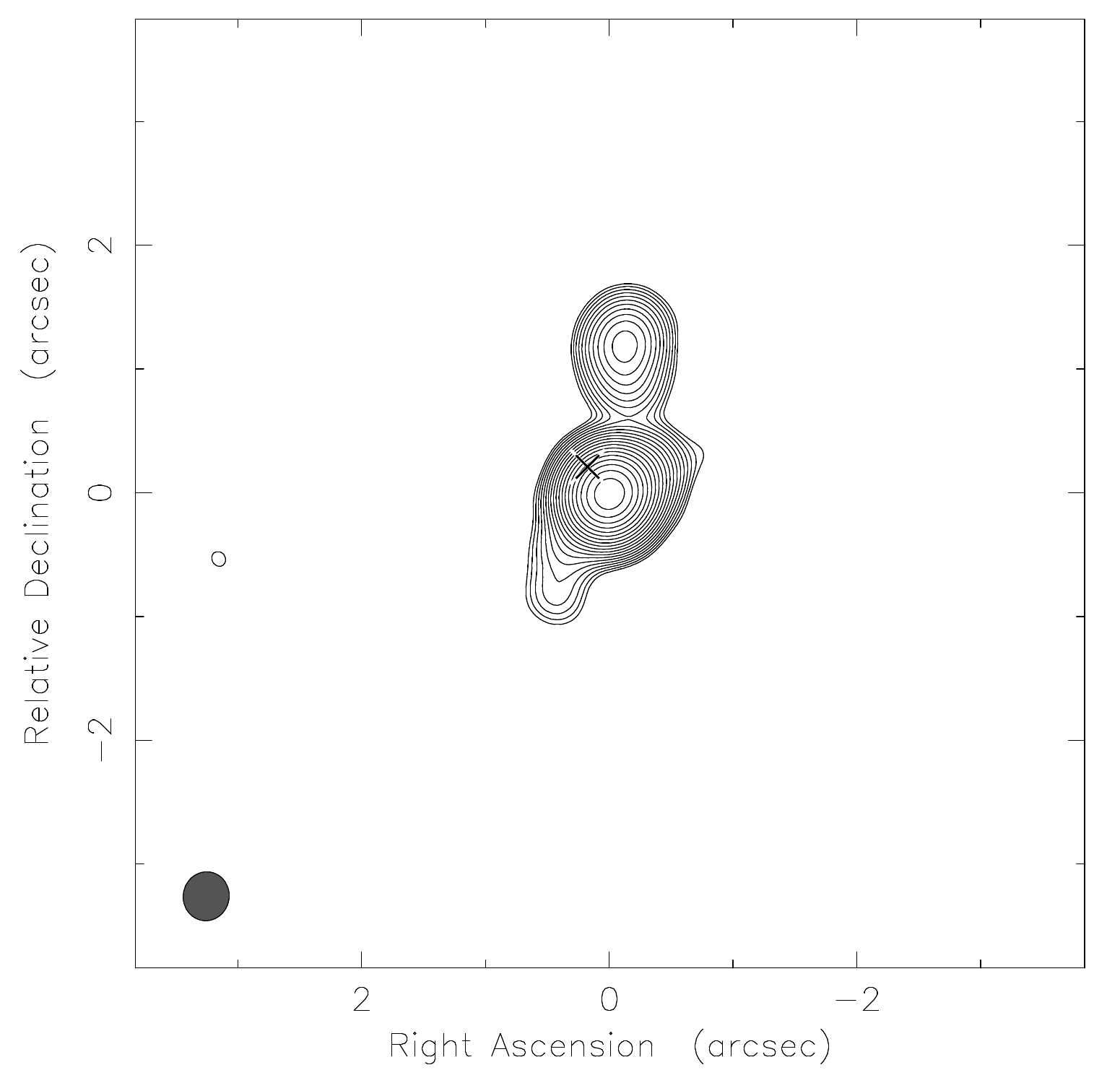} \label{fig:Images8-3}} 
\subfloat[Part 4][J1217+3305 at 1.490 GHz]{\includegraphics[width=2.2in]{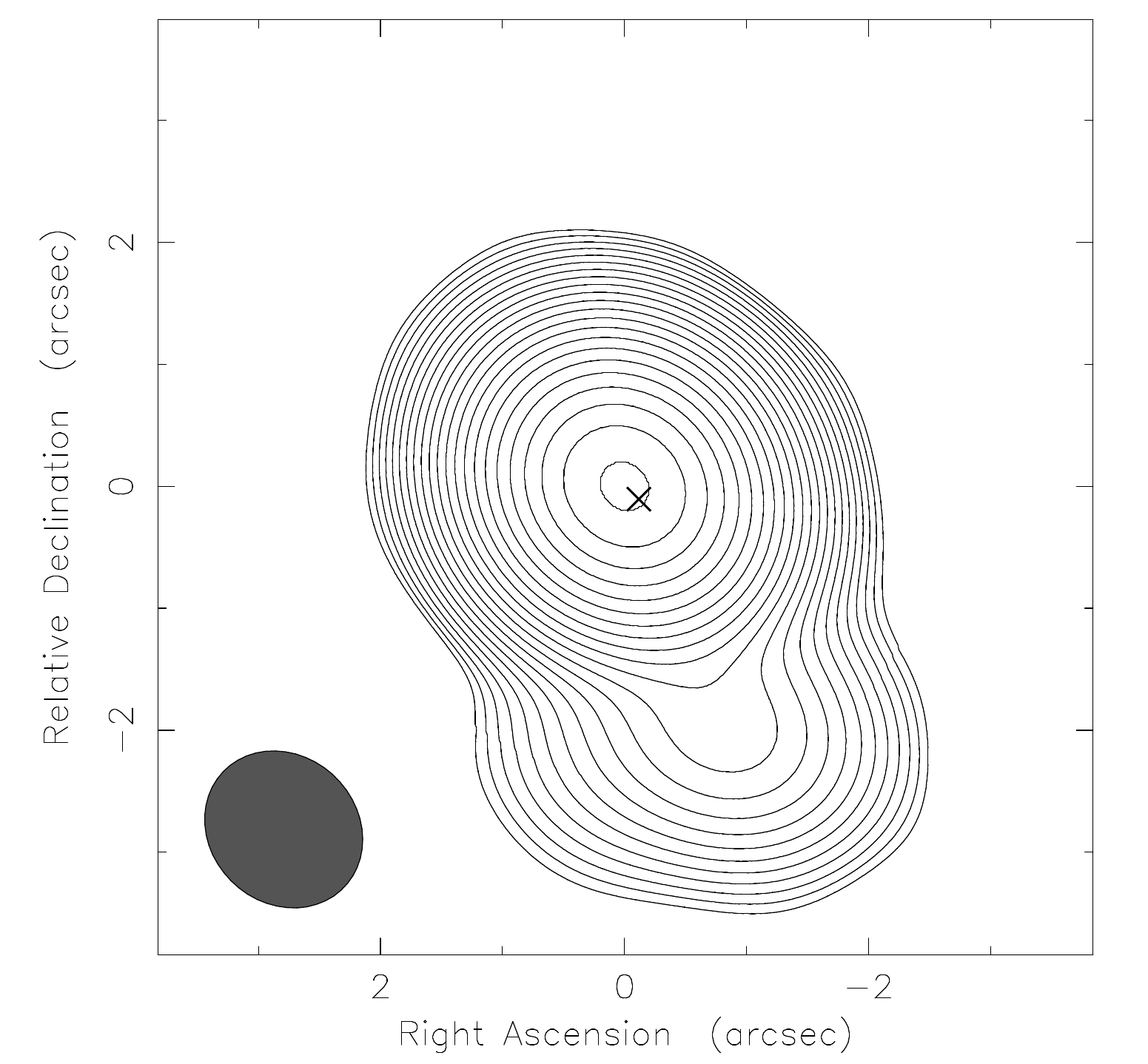} \label{fig:Images8-4}} 
\subfloat[Part 1][J1217+3435 at 1.490 GHz]{\includegraphics[width=2.2in]{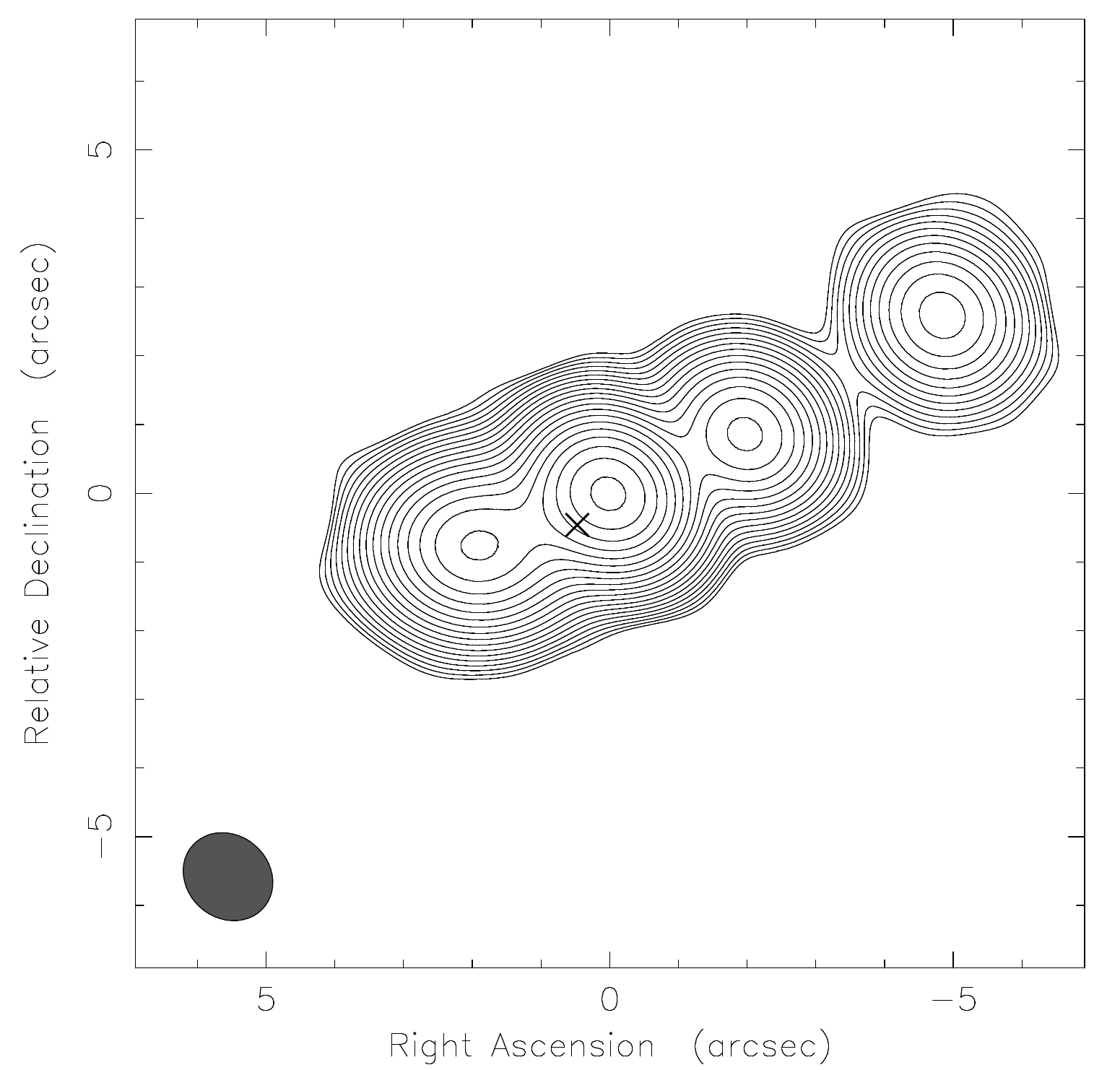} \label{fig:Images9-1}} \\
\subfloat[Part 2][J1217+5835 at 8.460 GHz]{\includegraphics[width=2.2in]{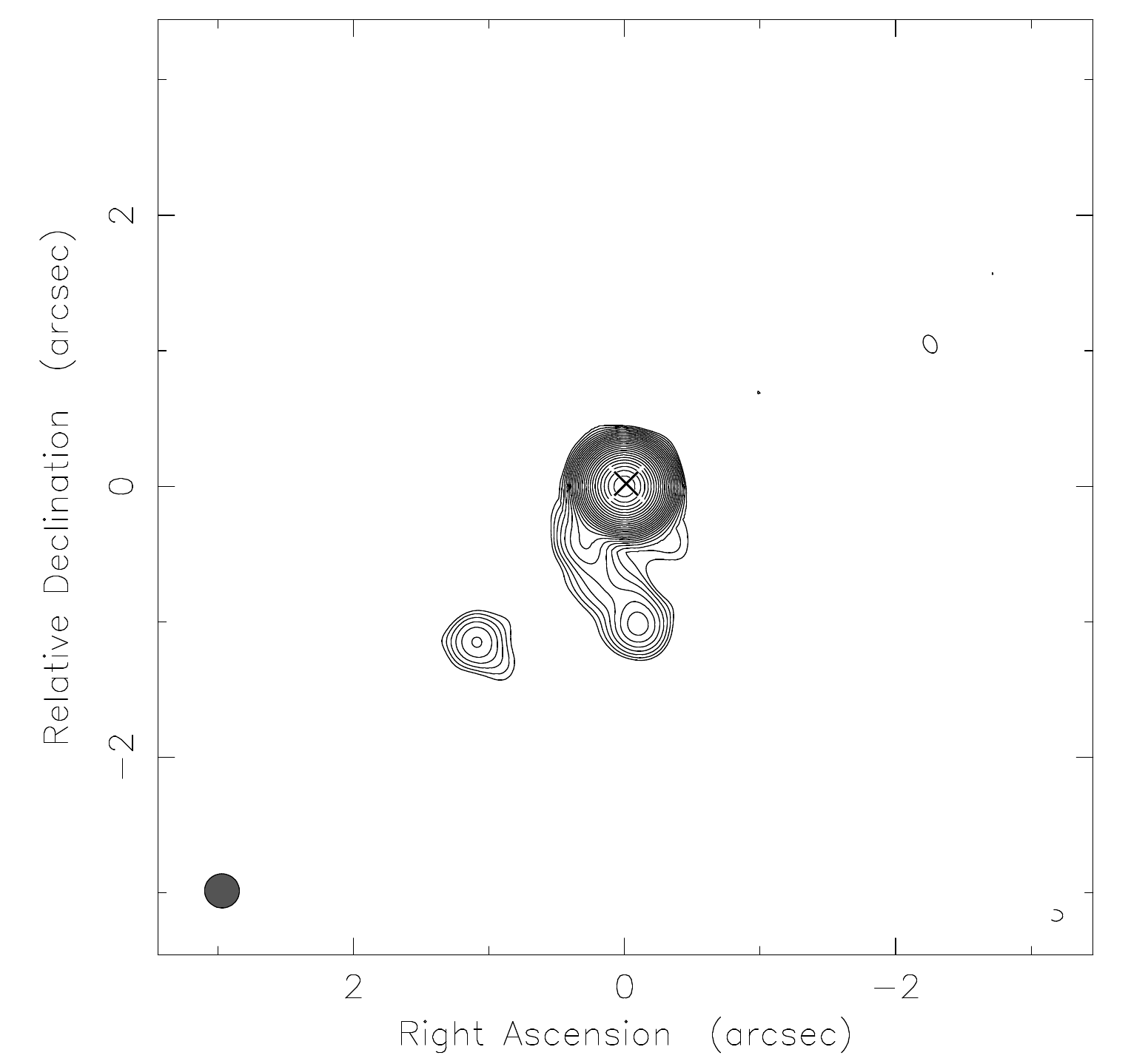} \label{fig:Images9-2}} 
\subfloat[Part 3][J1242+3720 at 1.425 GHz]{\includegraphics[width=2.2in]{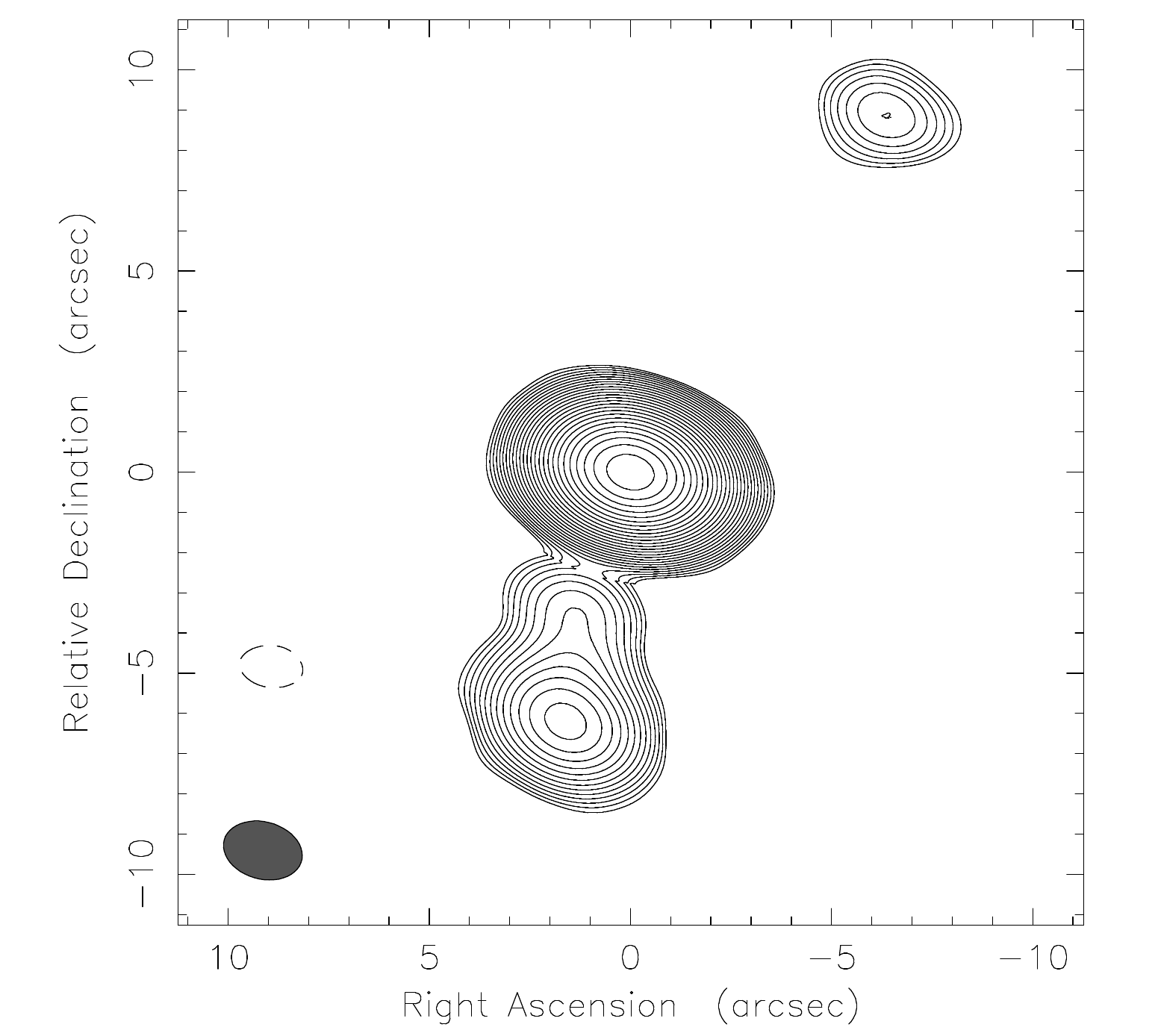} \label{fig:Images9-3}} 
\subfloat[Part 4][J1246+0104 at 4.860 GHz]{\includegraphics[width=2.2in]{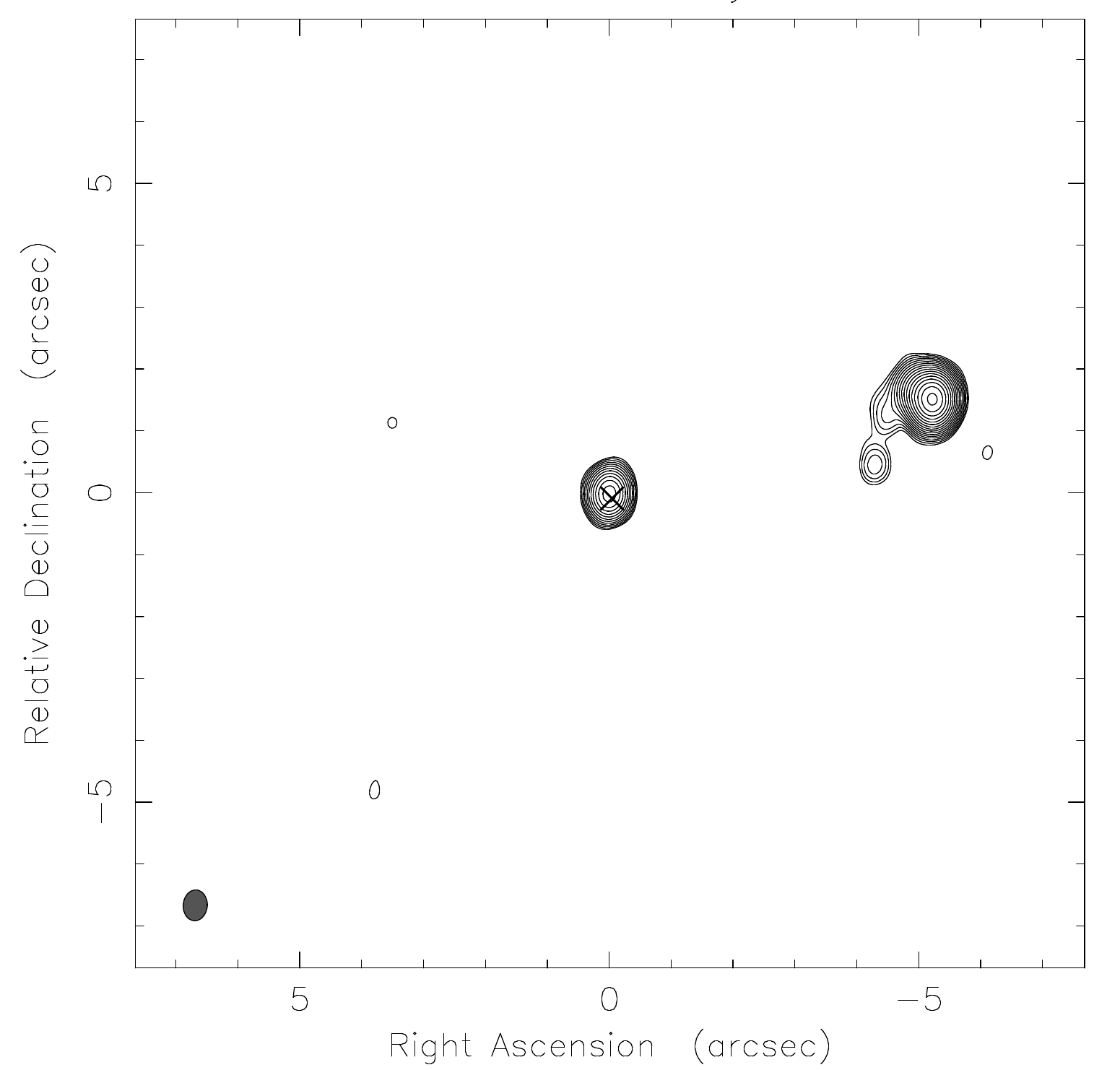} \label{fig:Images9-4}} 
\caption{}
\label{fig:Images5}
\end{figure}

\begin{figure}
\figurenum{6}
\centering
\subfloat[Part 1][J1346+2900 at 4.860 GHz]{\includegraphics[width=2.2in]{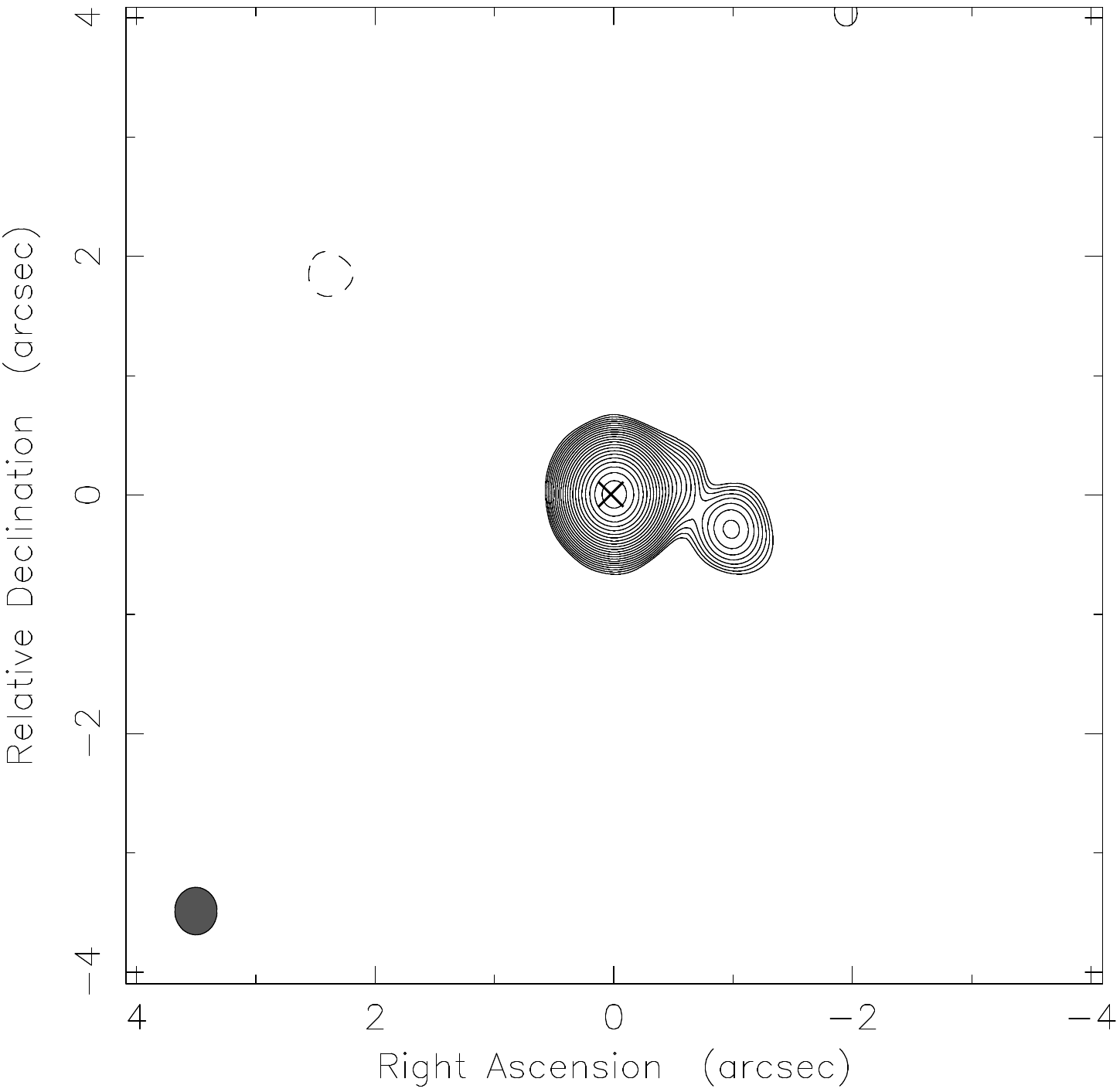} \label{fig:Images10-1}} 
\subfloat[Part 2][J1353+5725 at 4.860 GHz]{\includegraphics[width=2.2in]{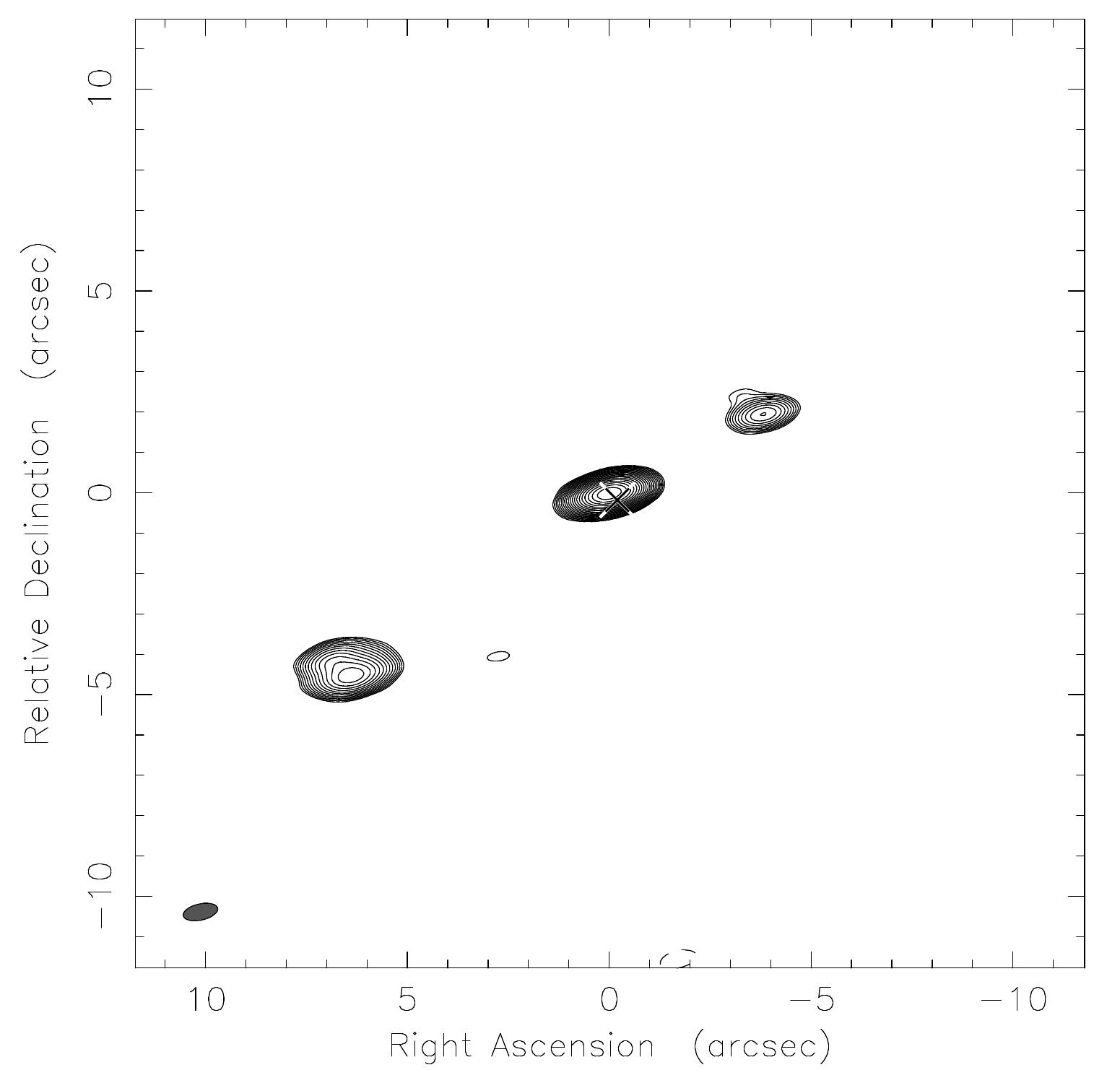} \label{fig:Images10-2}} 
\subfloat[Part 3][J1356+2918 at 1.425 GHz]{\includegraphics[width=2.2in]{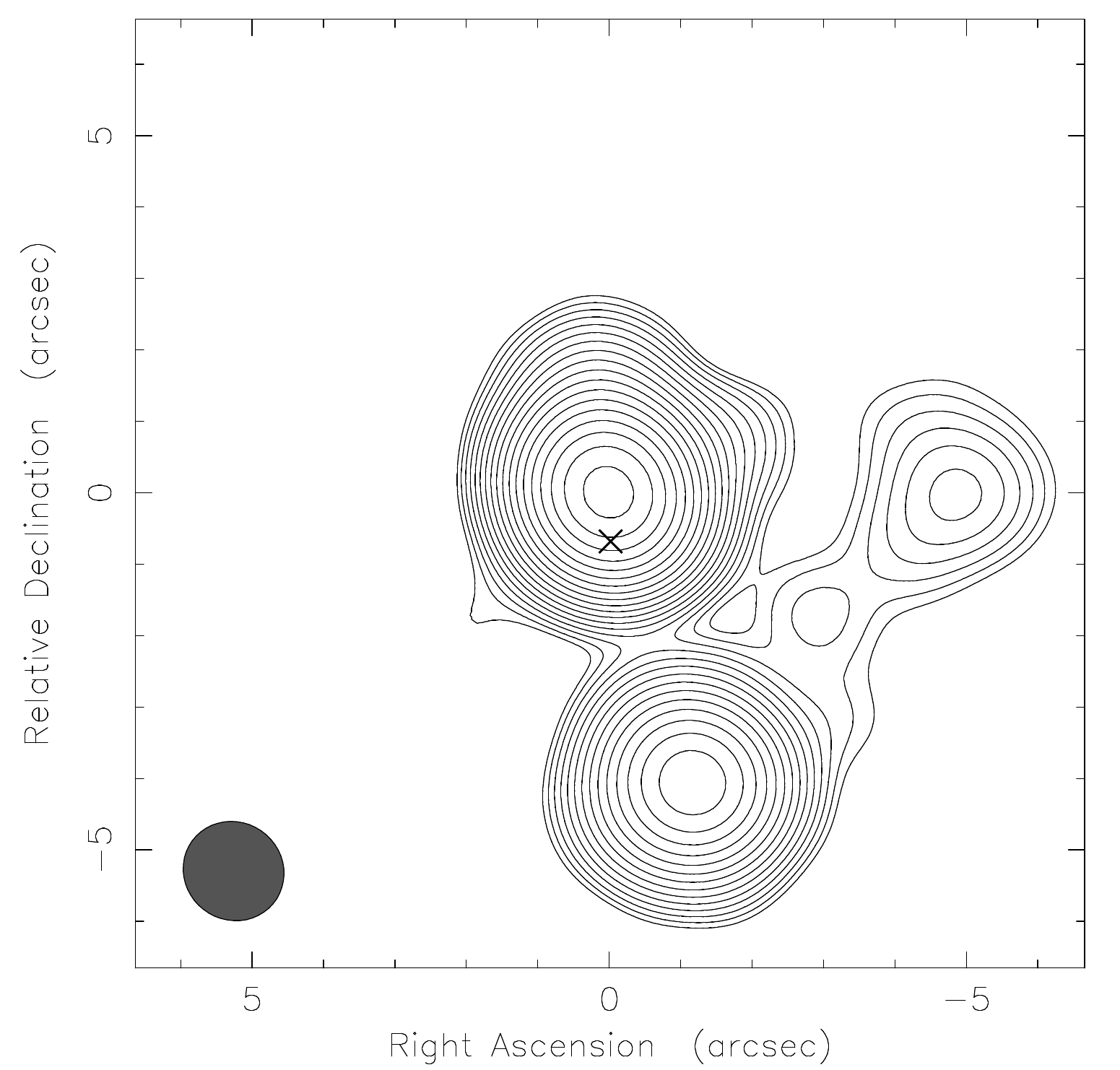} \label{fig:Images10-3}} \\
\subfloat[Part 4][J1400+0425 at 4.860 GHz]{\includegraphics[width=2.2in]{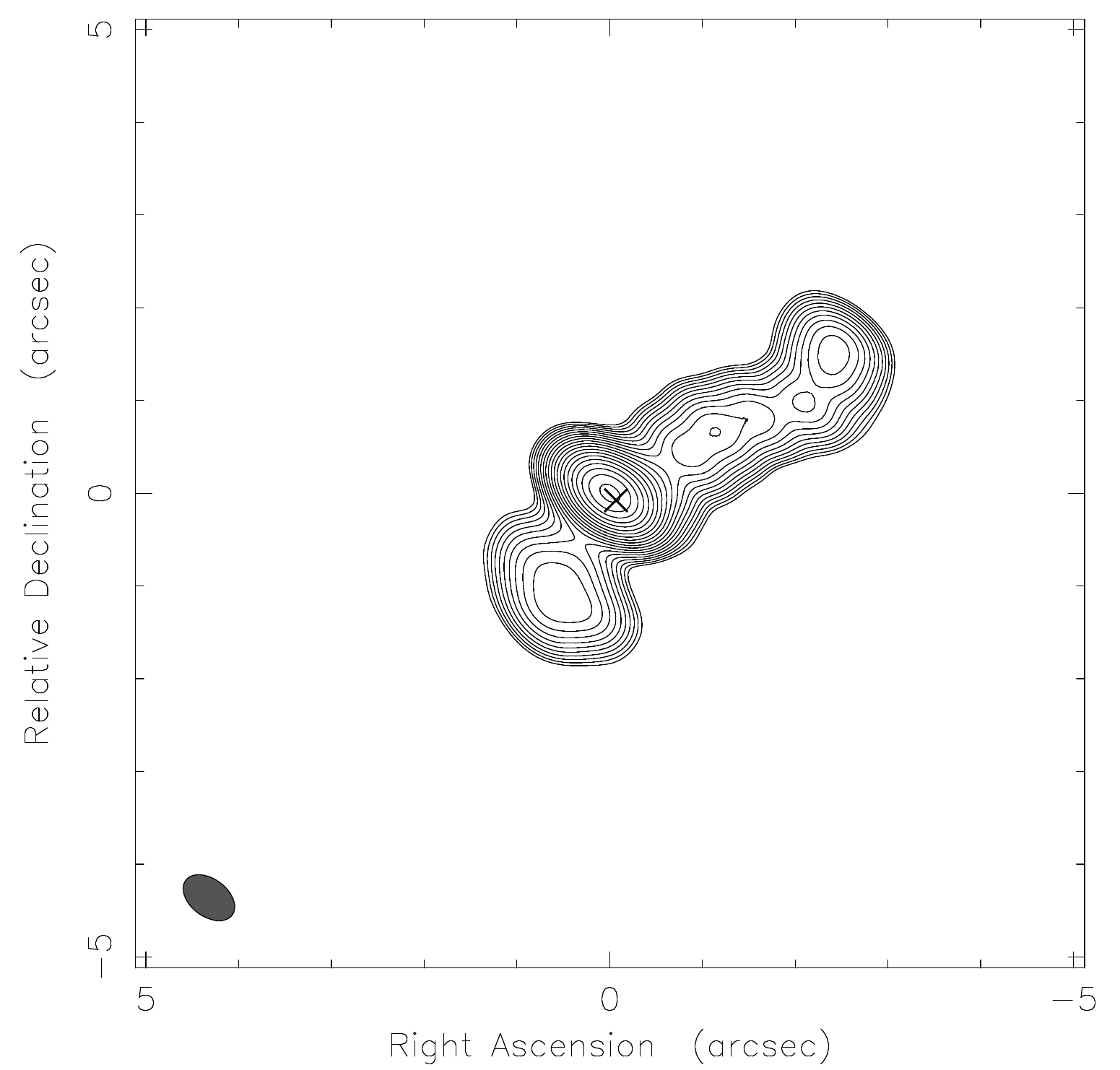} \label{fig:Images10-4}} 
\subfloat[Part 5][J1405+0415 at 4.860 GHz]{\includegraphics[width=2.2in]{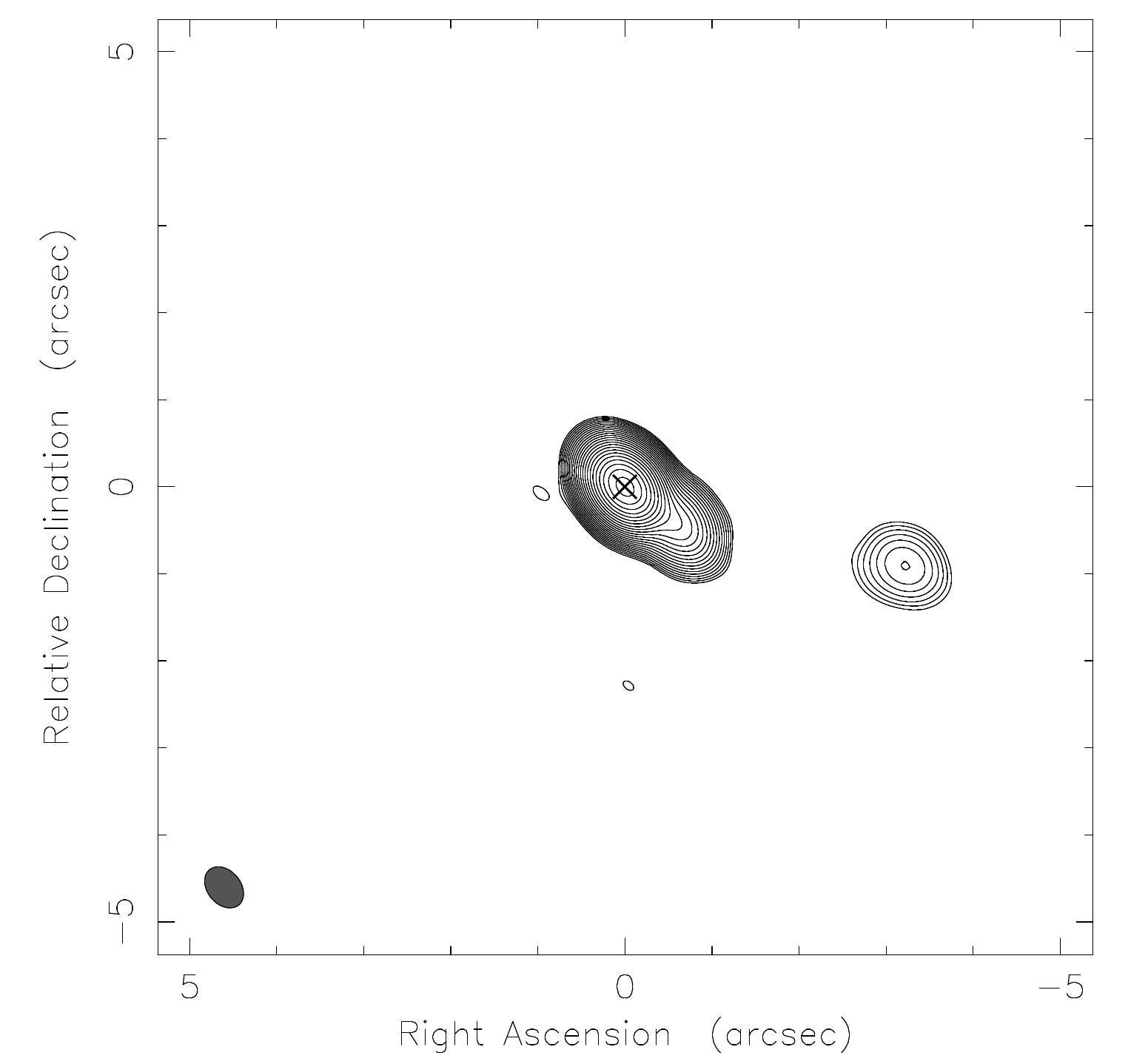} \label{fig:Images11-1}} 
\subfloat[Part 6][J1429+2607 at 1.425 GHz]{\includegraphics[width=2.2in]{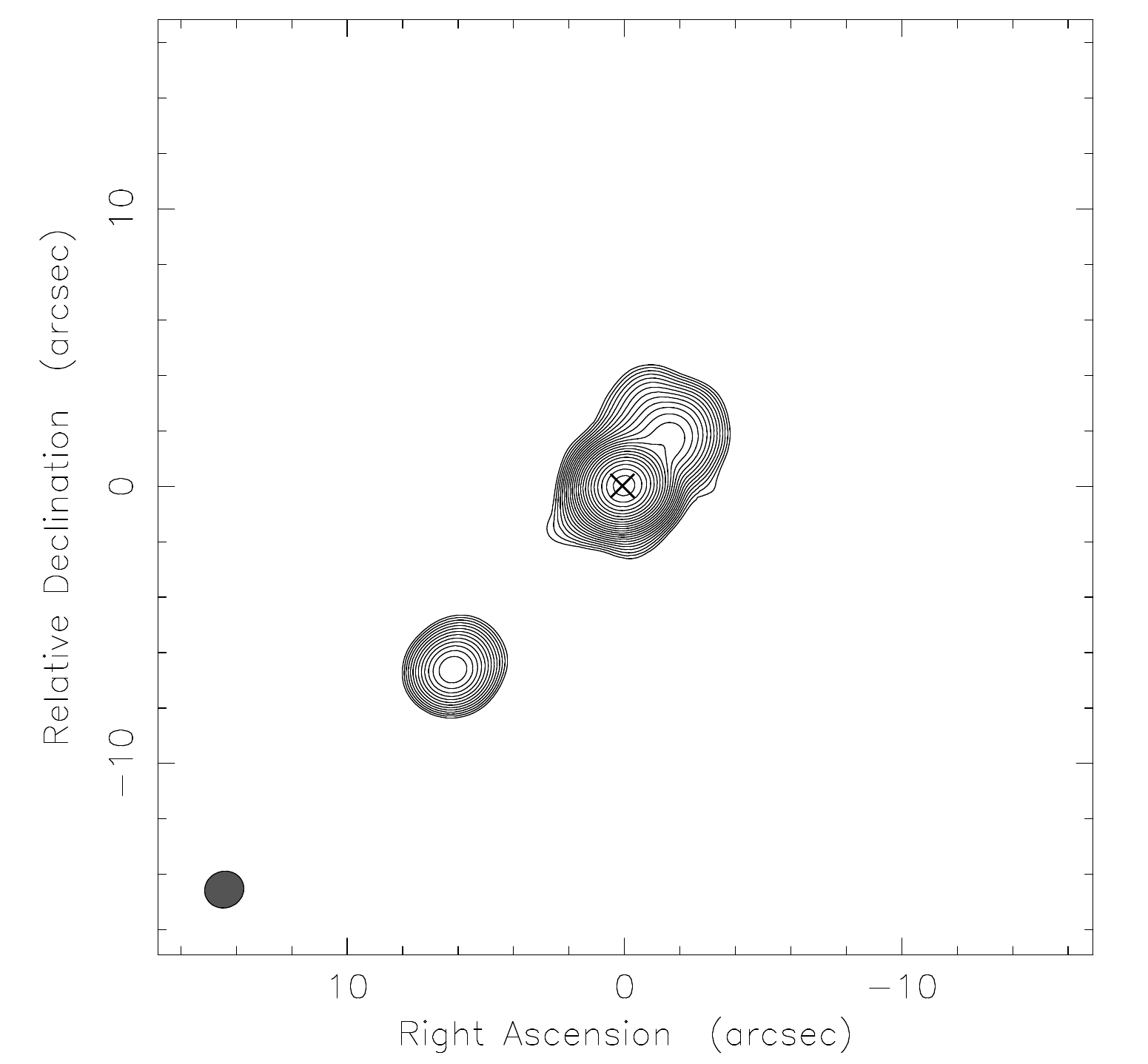} \label{fig:Images11-2}} \\
\subfloat[Part 7][J1429+5406 at 14.940 GHz]{\includegraphics[width=2.2in]{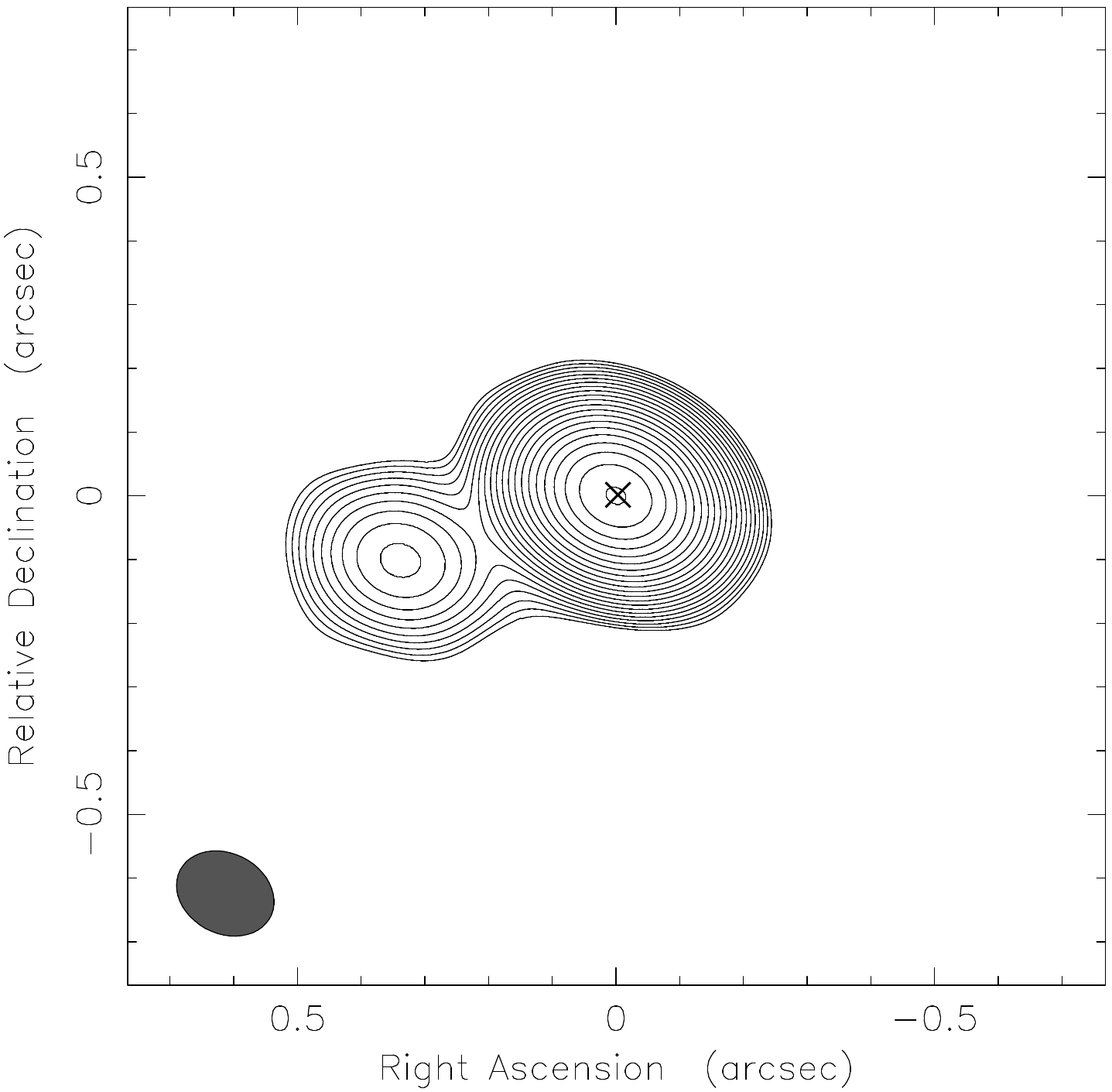} \label{fig:Images11-3}} 
\subfloat[Part 8][J1430+4204 at 1.425 GHz]{\includegraphics[width=2.2in]{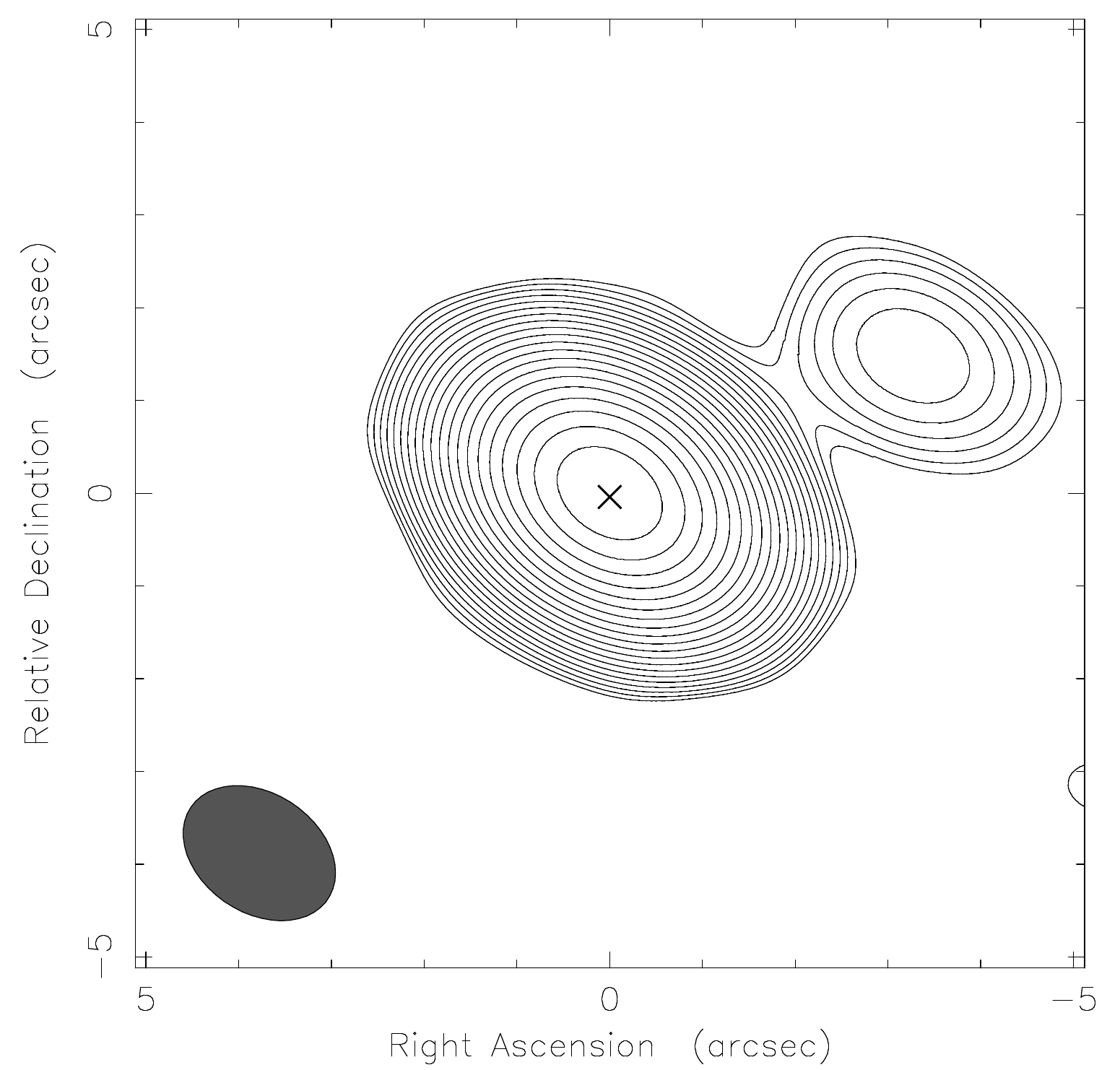} \label{fig:Images11-4}} 
\subfloat[Part 9][J1435+5435 at 4.860 GHz]{\includegraphics[width=2.2in]{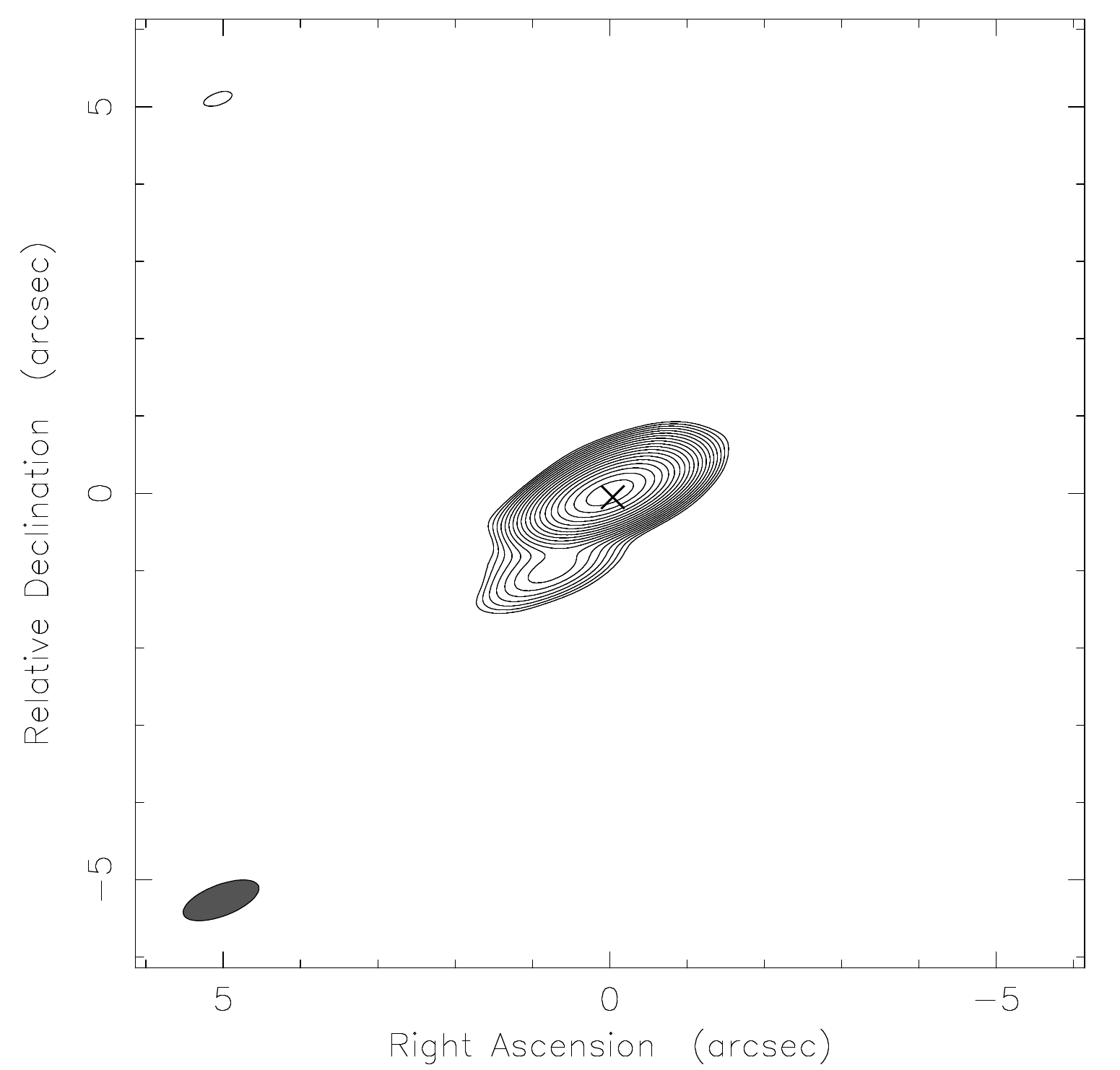} \label{fig:Images12-1}} 
\caption{}
\label{fig:Images6}
\end{figure}

\begin{figure}
\figurenum{7}
\centering
\subfloat[Part 1][J1450+0910 at 1.425 GHz]{\includegraphics[width=2.2in]{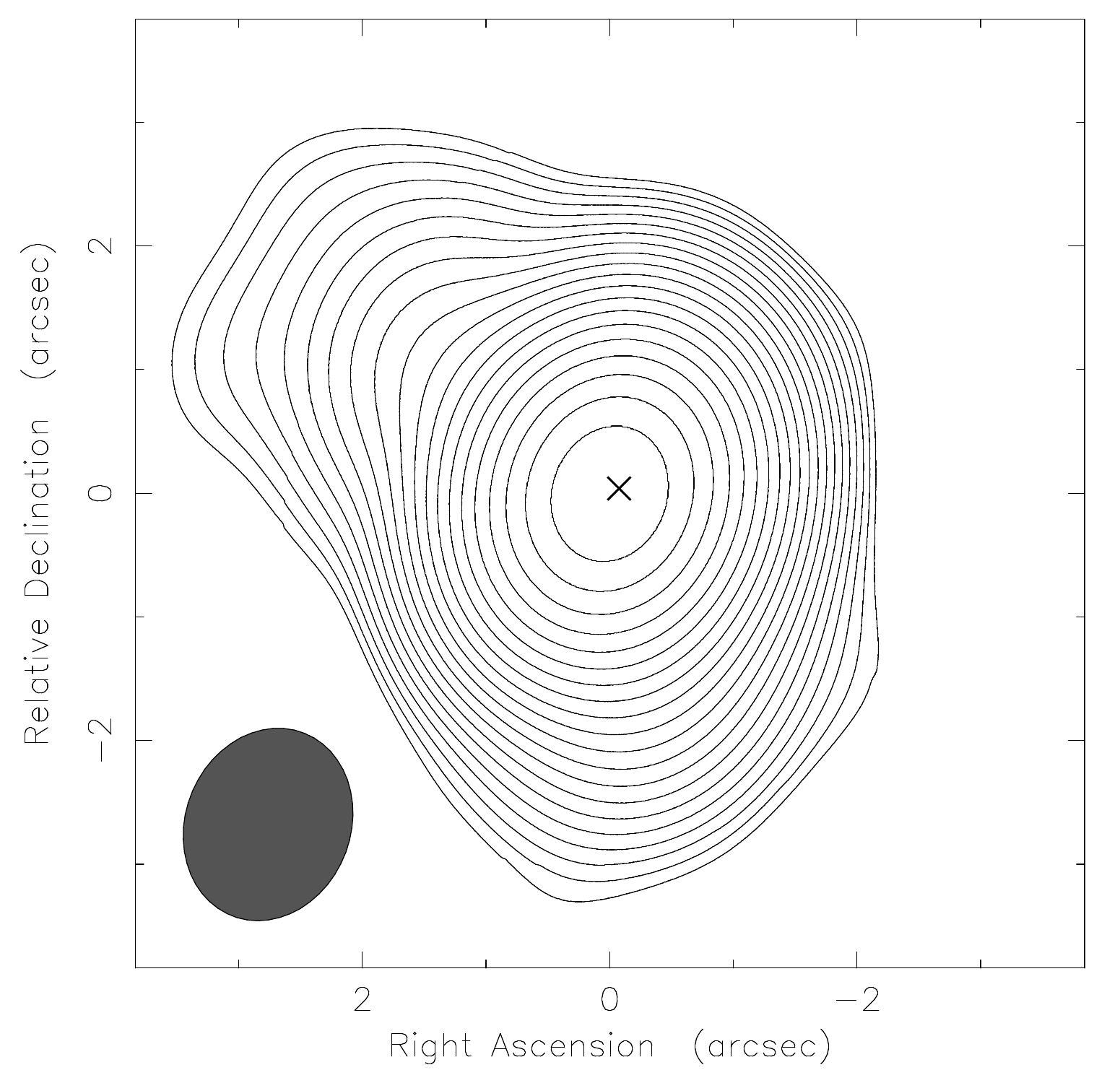} \label{fig:Images12-2}} 
\subfloat[Part 2][J1457+3439 at 4.860 GHz]{\includegraphics[width=2.2in]{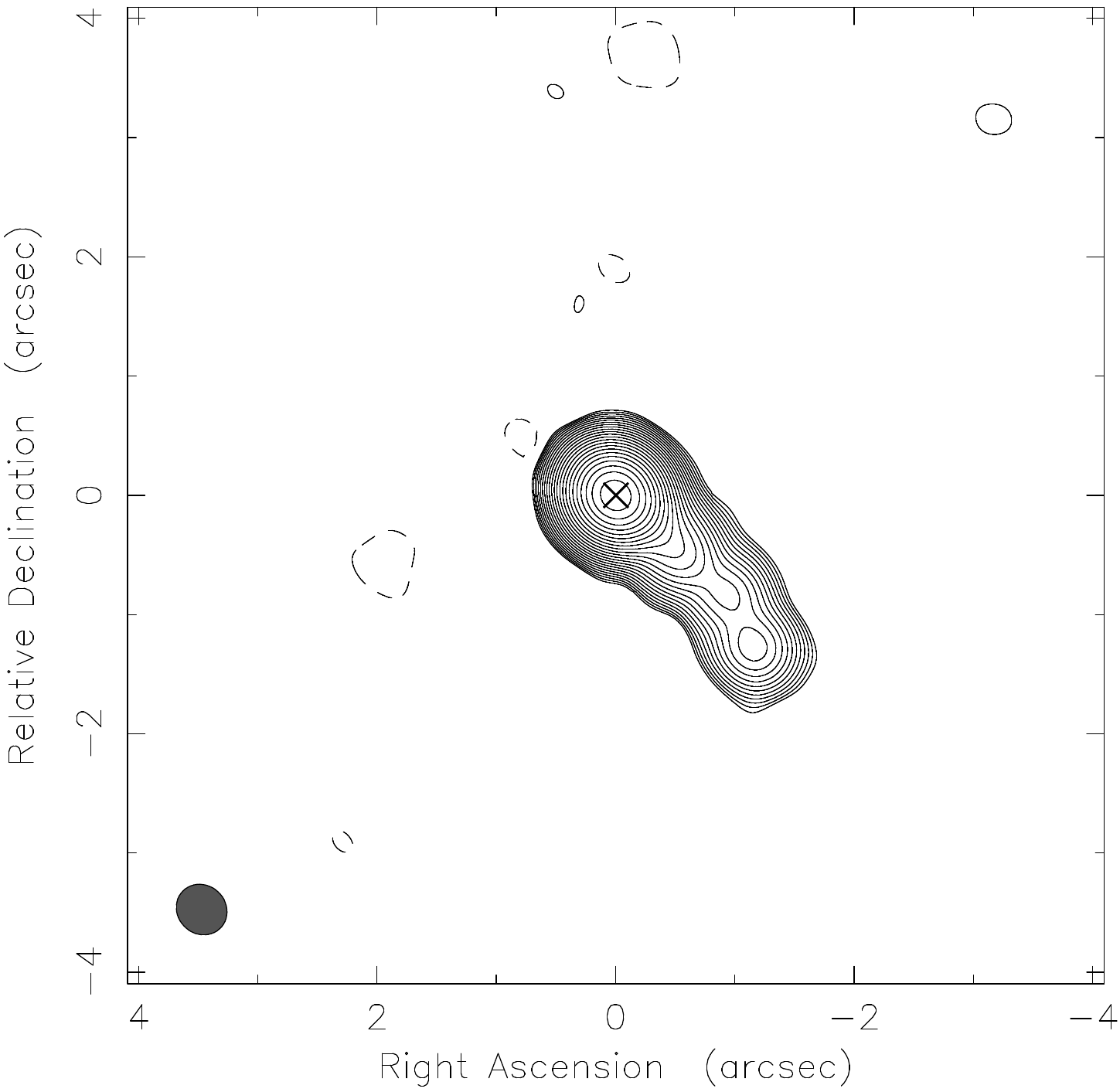} \label{fig:Images12-3}} 
\subfloat[Part 3][J1459+3253 at 4.860 GHz]{\includegraphics[width=2.2in]{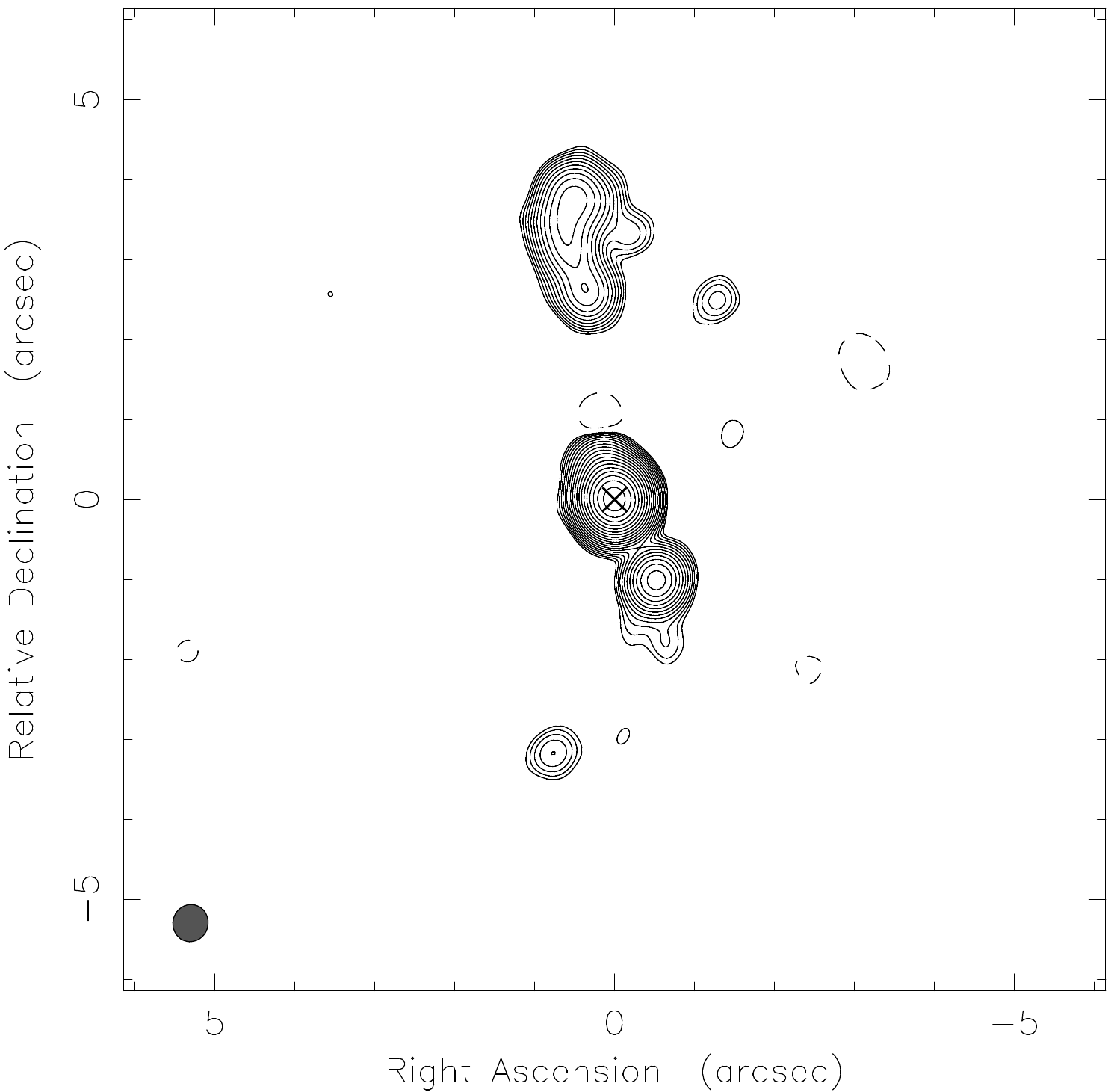} \label{fig:Images12-4}} \\
\subfloat[Part 4][J1502+5521 at 4.860 GHz]{\includegraphics[width=2.2in]{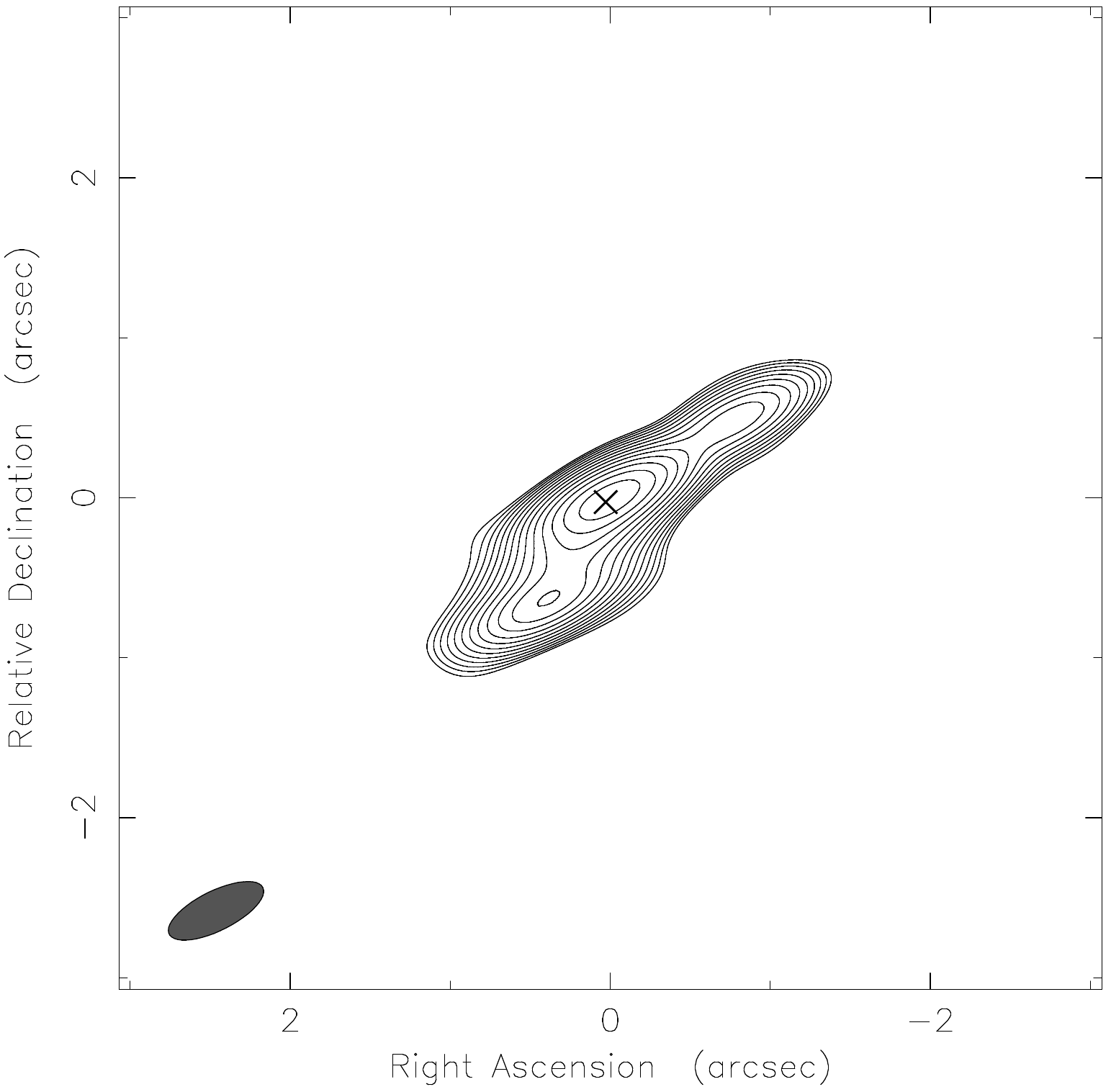} \label{fig:Images13-1}} 
\subfloat[Part 5][J1510+5702 at 1.425 GHz]{\includegraphics[width=2.2in]{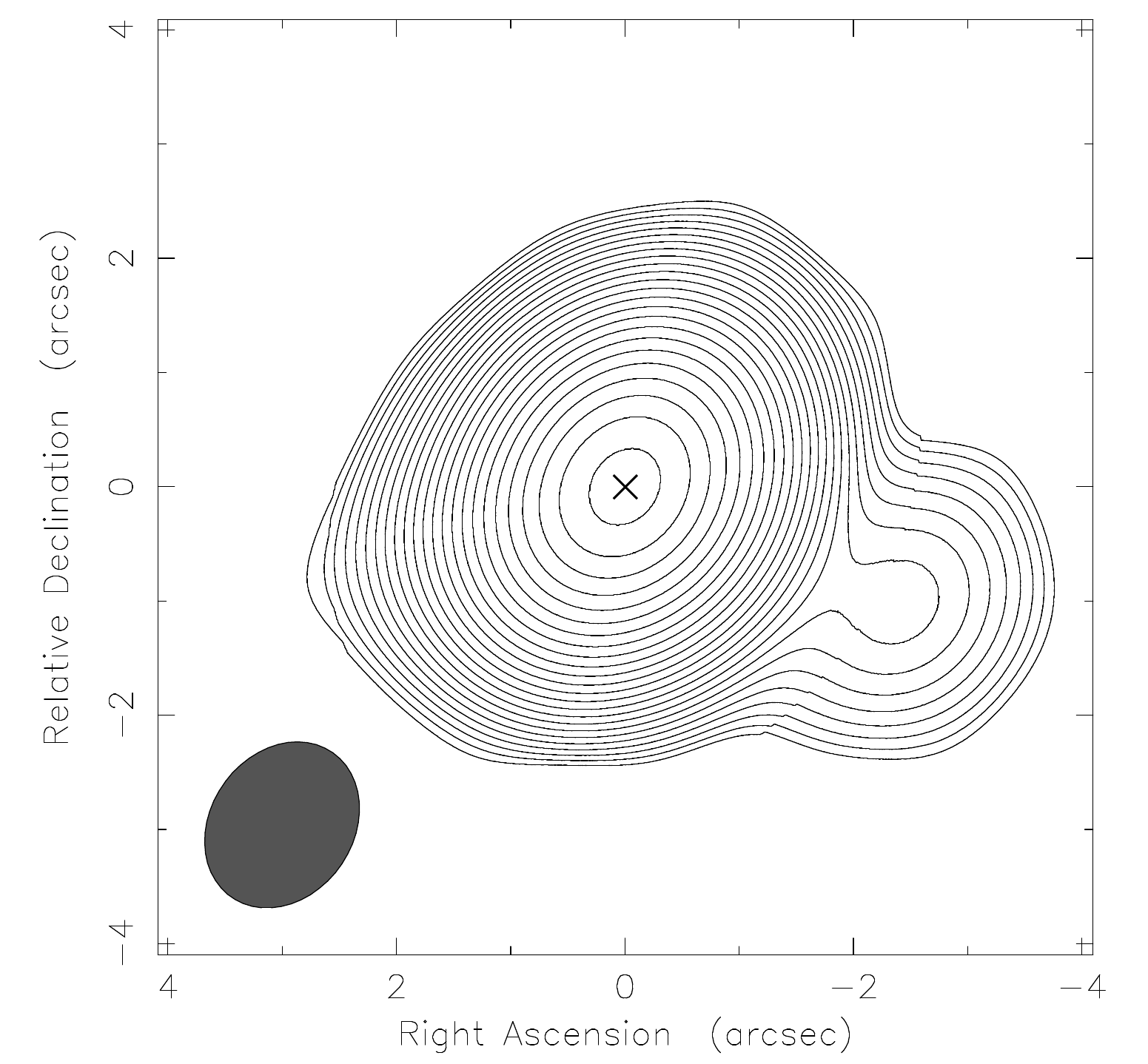} \label{fig:Images13-2}} 
\subfloat[Part 6][J1528+5310 at 1.490 GHz]{\includegraphics[width=2.2in]{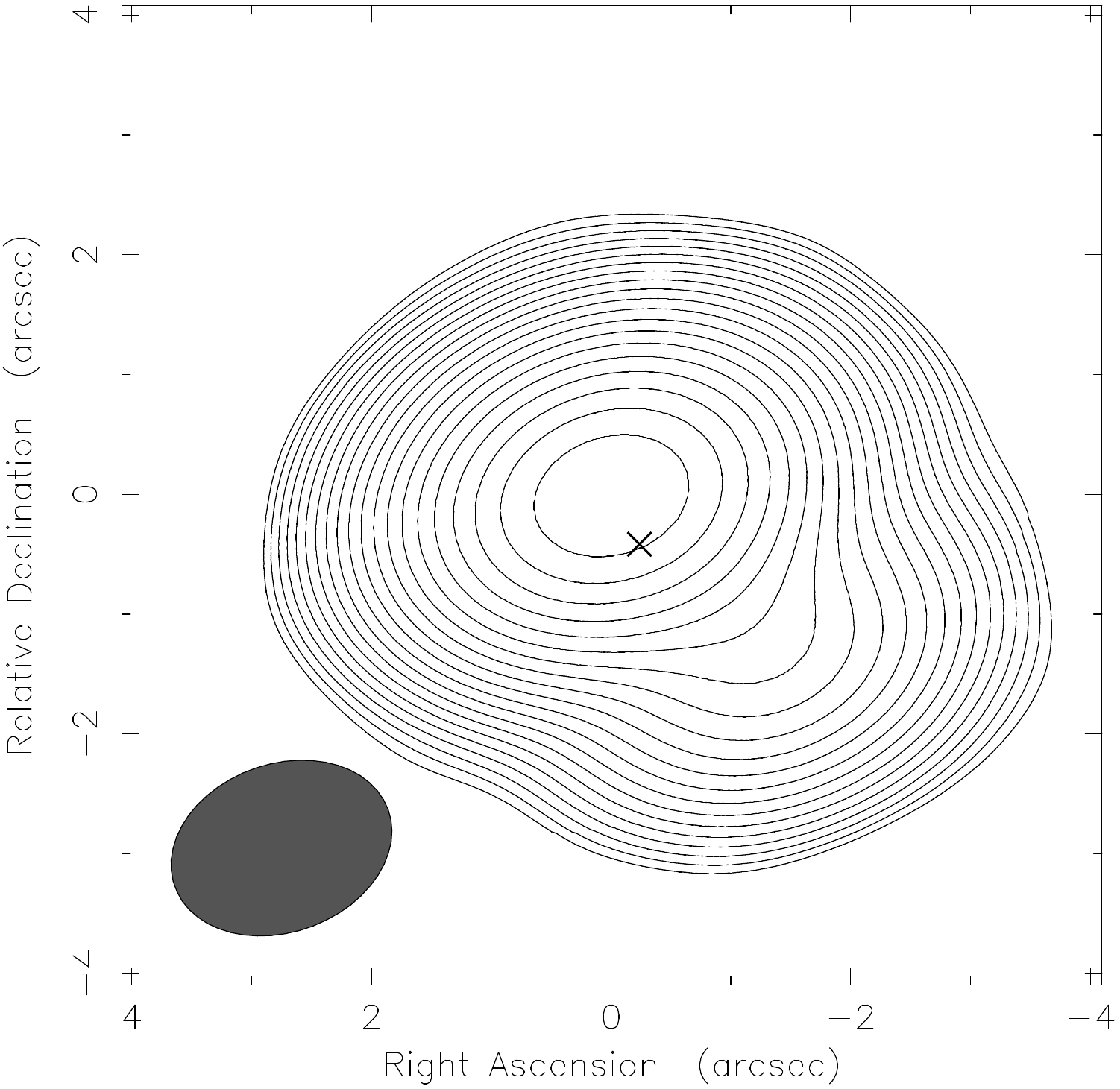} \label{fig:Images13-3}} \\
\subfloat[Part 7][J1528 (wide) at 4.860 GHz]{\includegraphics[width=2.2in]{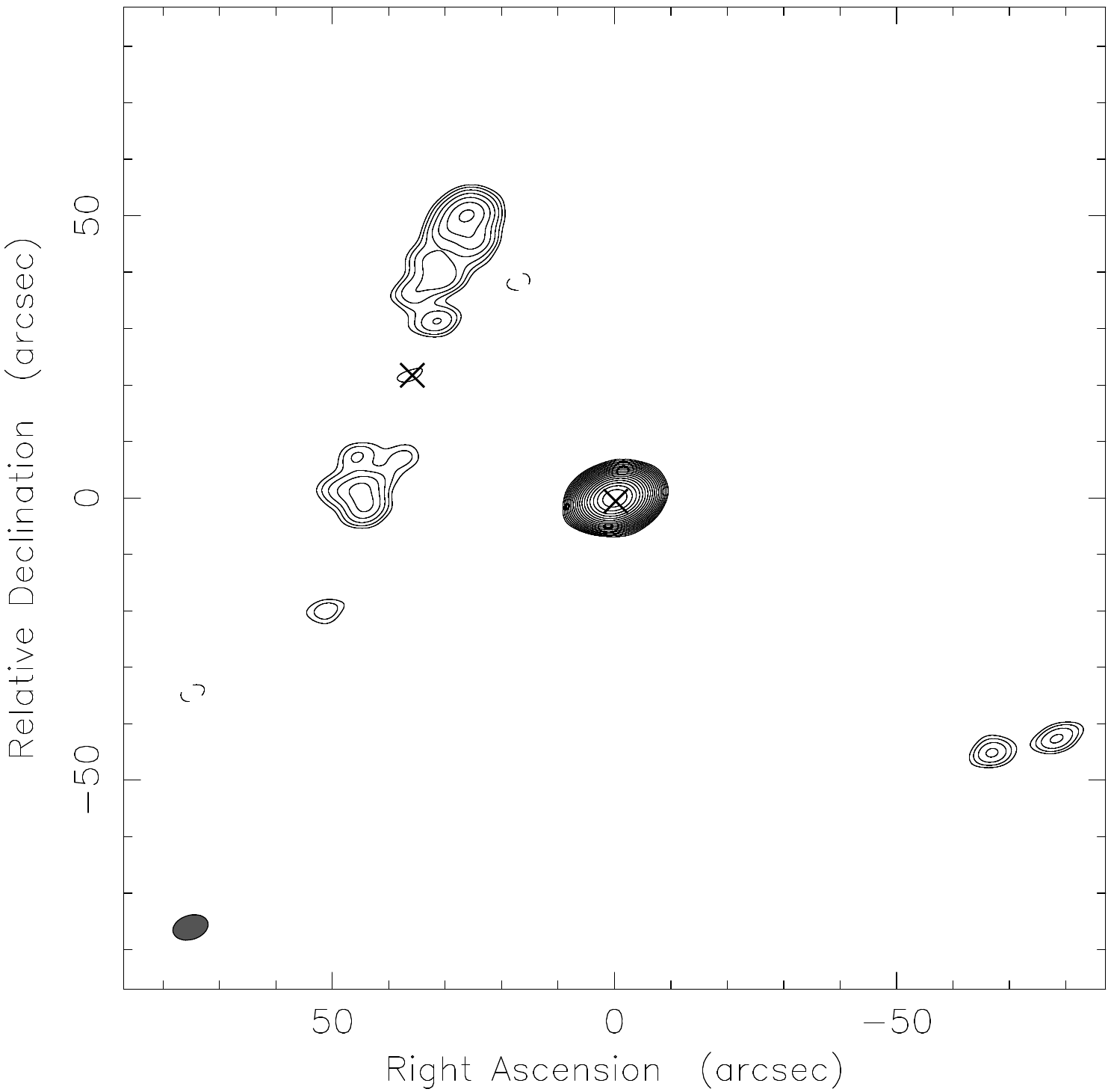} \label{fig:Images13-4}} 
\subfloat[Part 8][J1540+4738 at 1.490 GHz]{\includegraphics[width=2.2in]{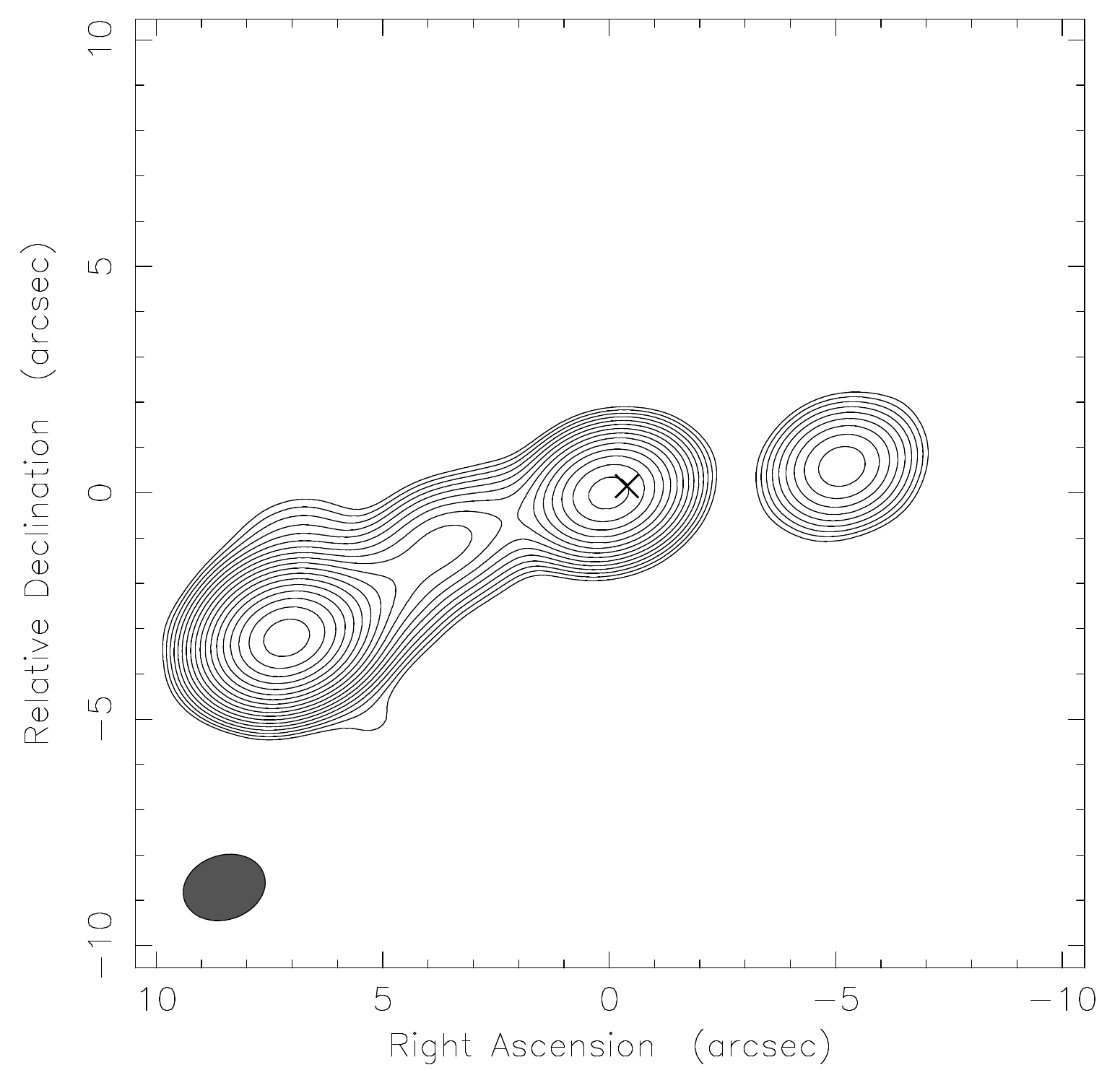} \label{fig:Images13-5}} 
\subfloat[Part 9][J1559+0304 at 1.425 GHz]{\includegraphics[width=2.2in]{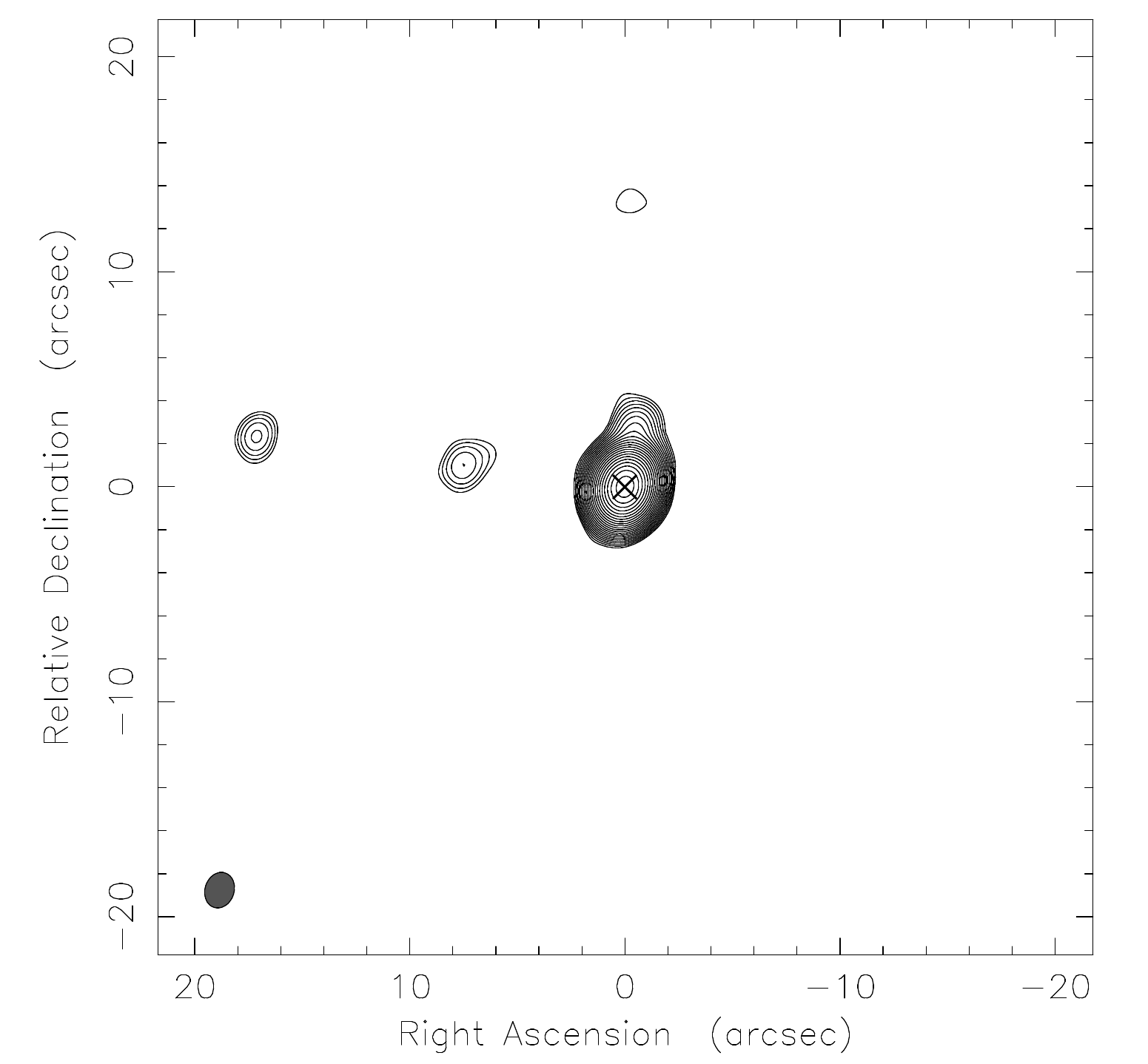} \label{fig:Images14-1}} 
\caption{}
\label{fig:Images7}
\end{figure}

\begin{figure}
\figurenum{8}
\centering
\subfloat[Part 1][J1602+2410 at 4.860 GHz]{\includegraphics[width=2.2in]{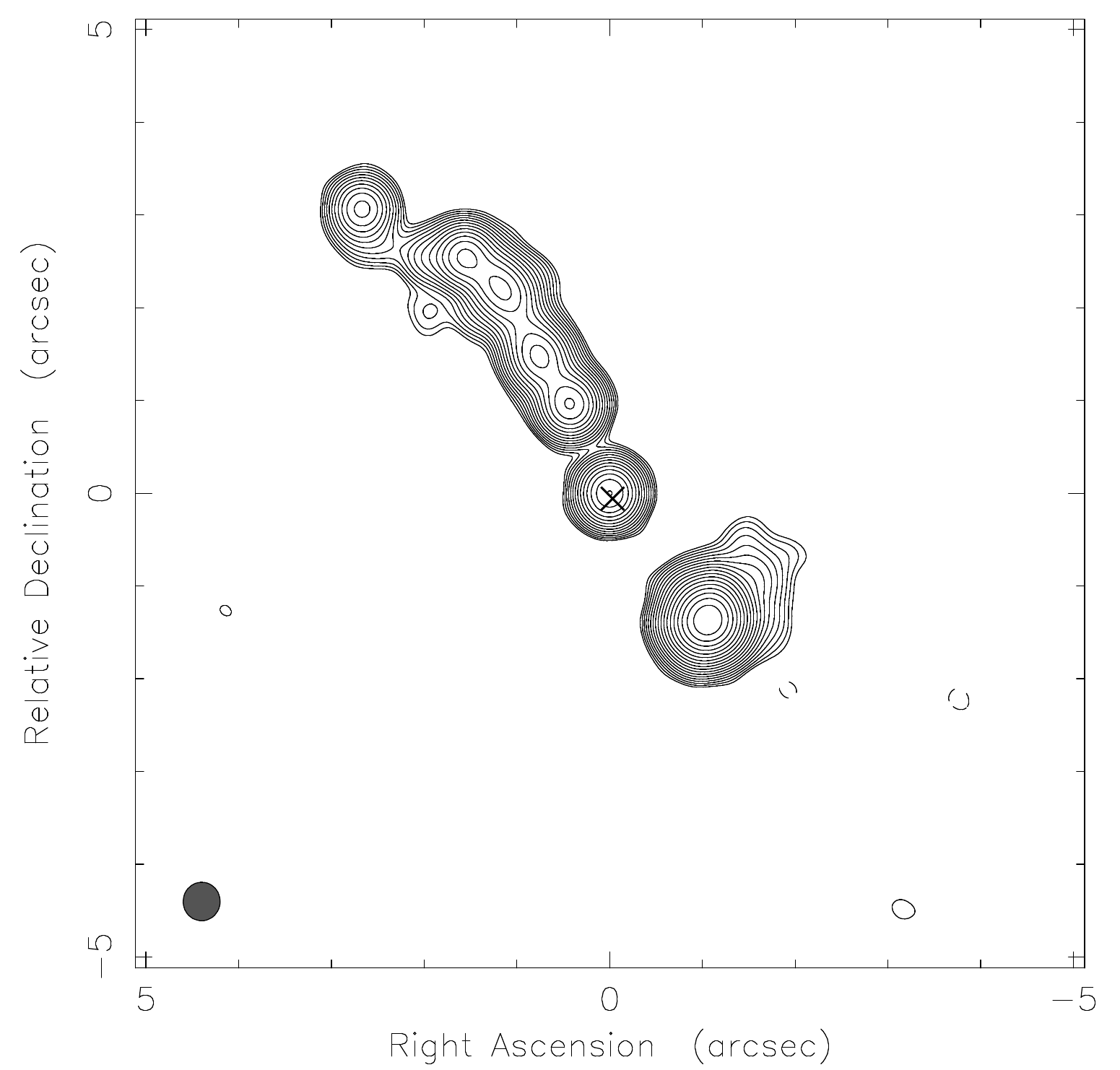} \label{fig:Images14-2}} 
\subfloat[Part 2][J1610+1811 at 4.860 GHz]{\includegraphics[width=2.2in]{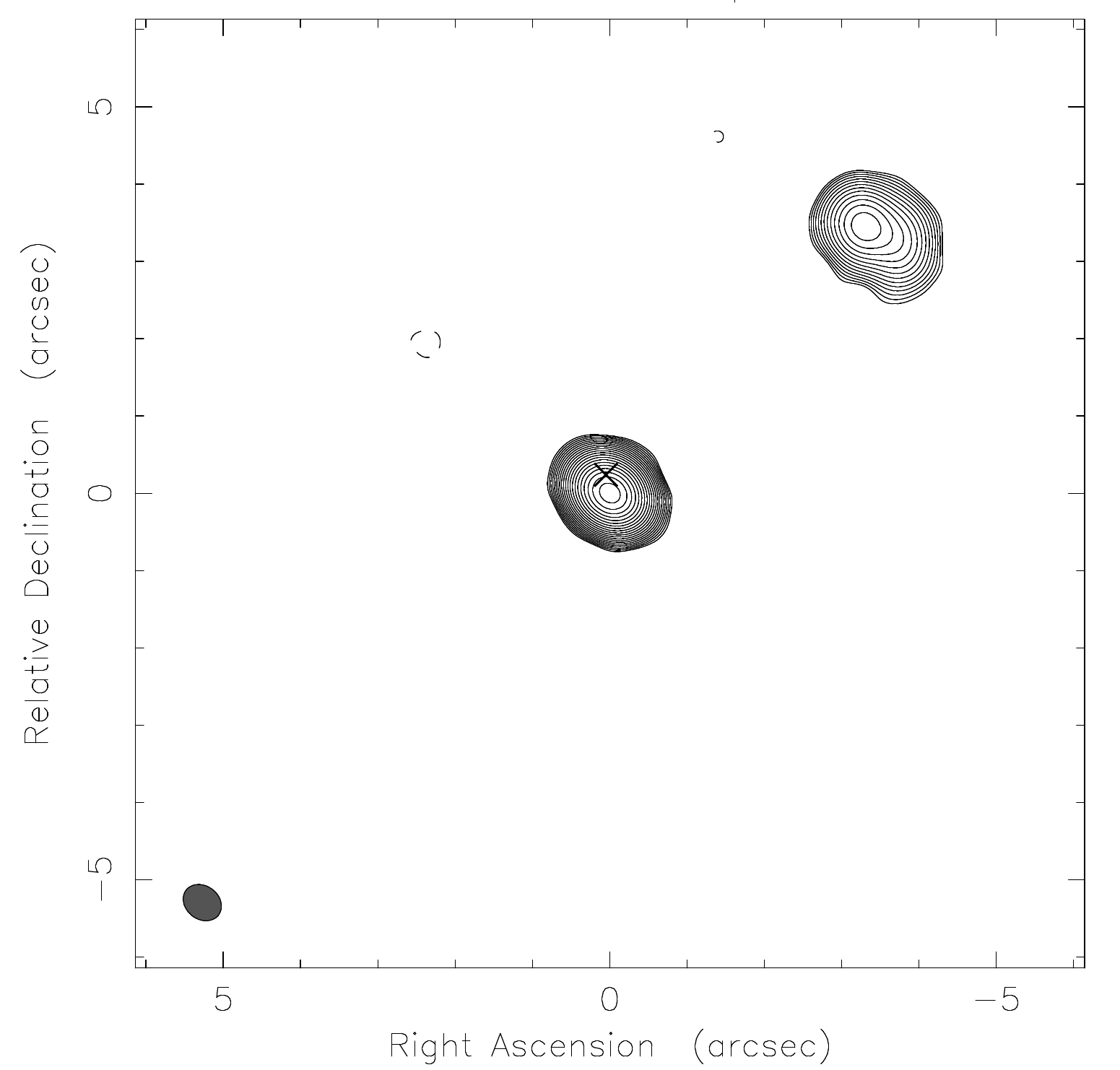} \label{fig:Images14-4}} 
\subfloat[Part 3][J1612+2758 at 4.860 GHz]{\includegraphics[width=2.2in]{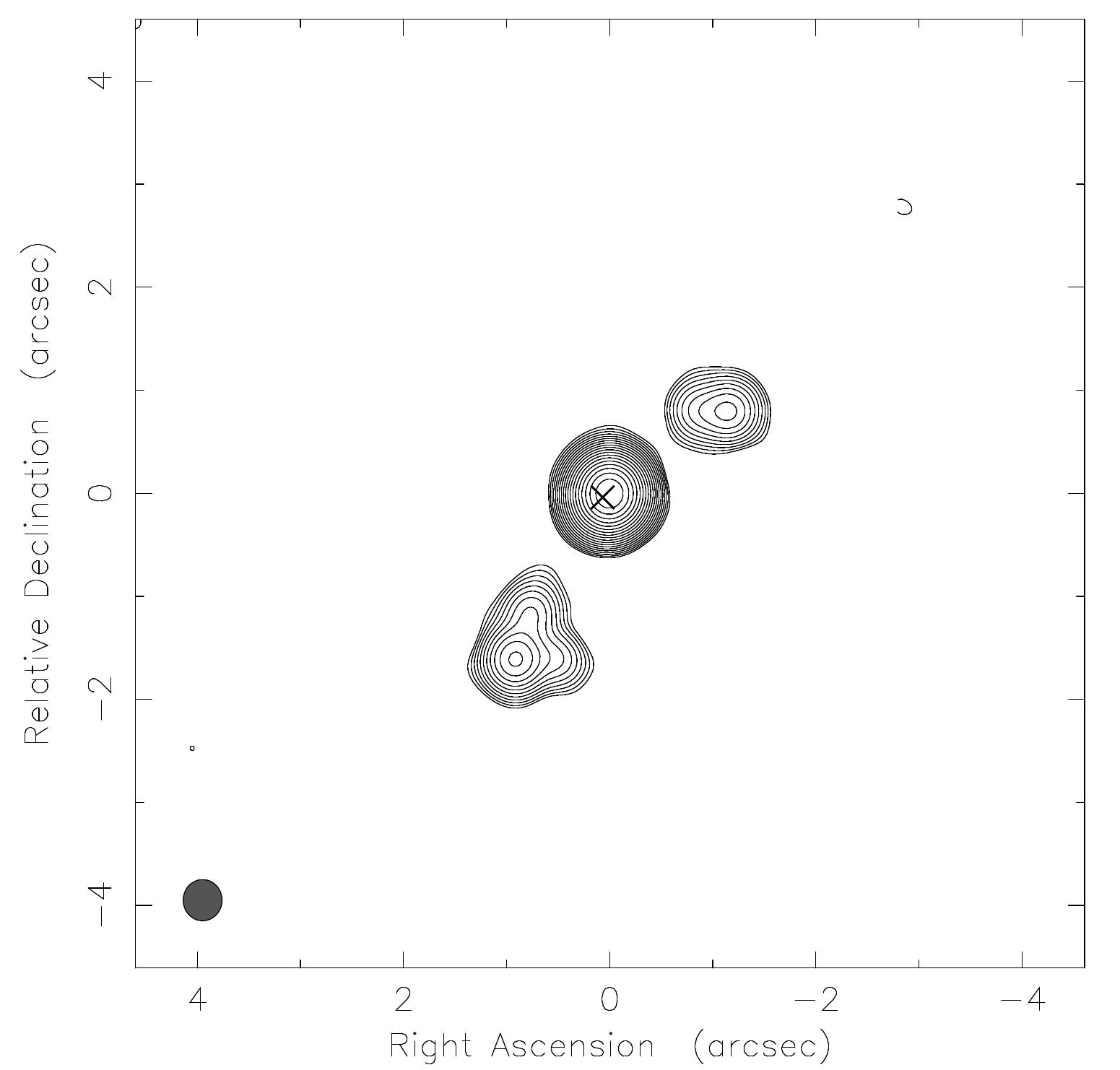} \label{fig:Images15-1}} \\
\subfloat[Part 4][J1616+0459 at 8.440 GHz]{\includegraphics[width=2.2in]{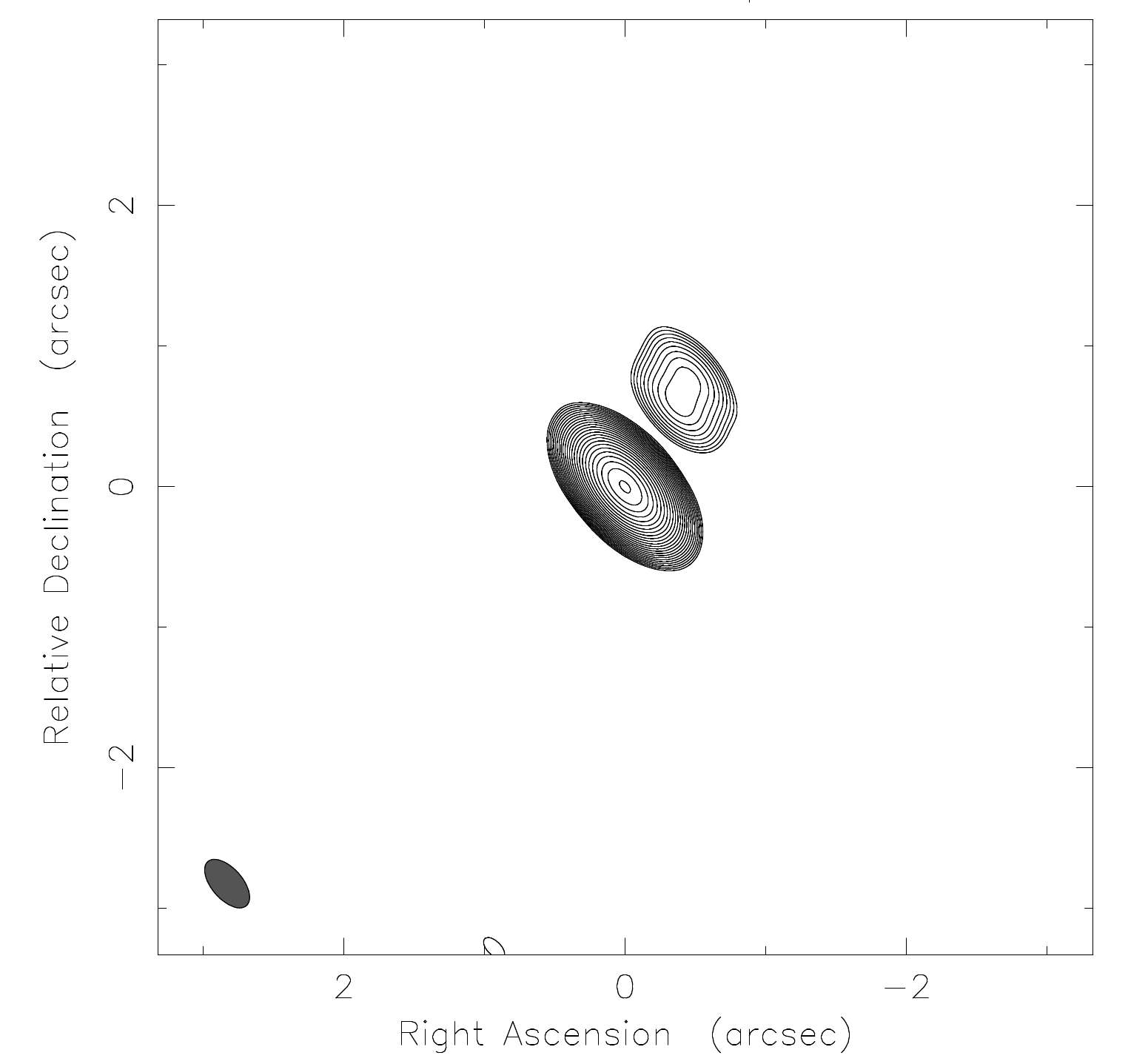} \label{fig:Images15-2}} 
\subfloat[Part 5][J1625+4134 at 4.985 GHz]{\includegraphics[width=2.2in]{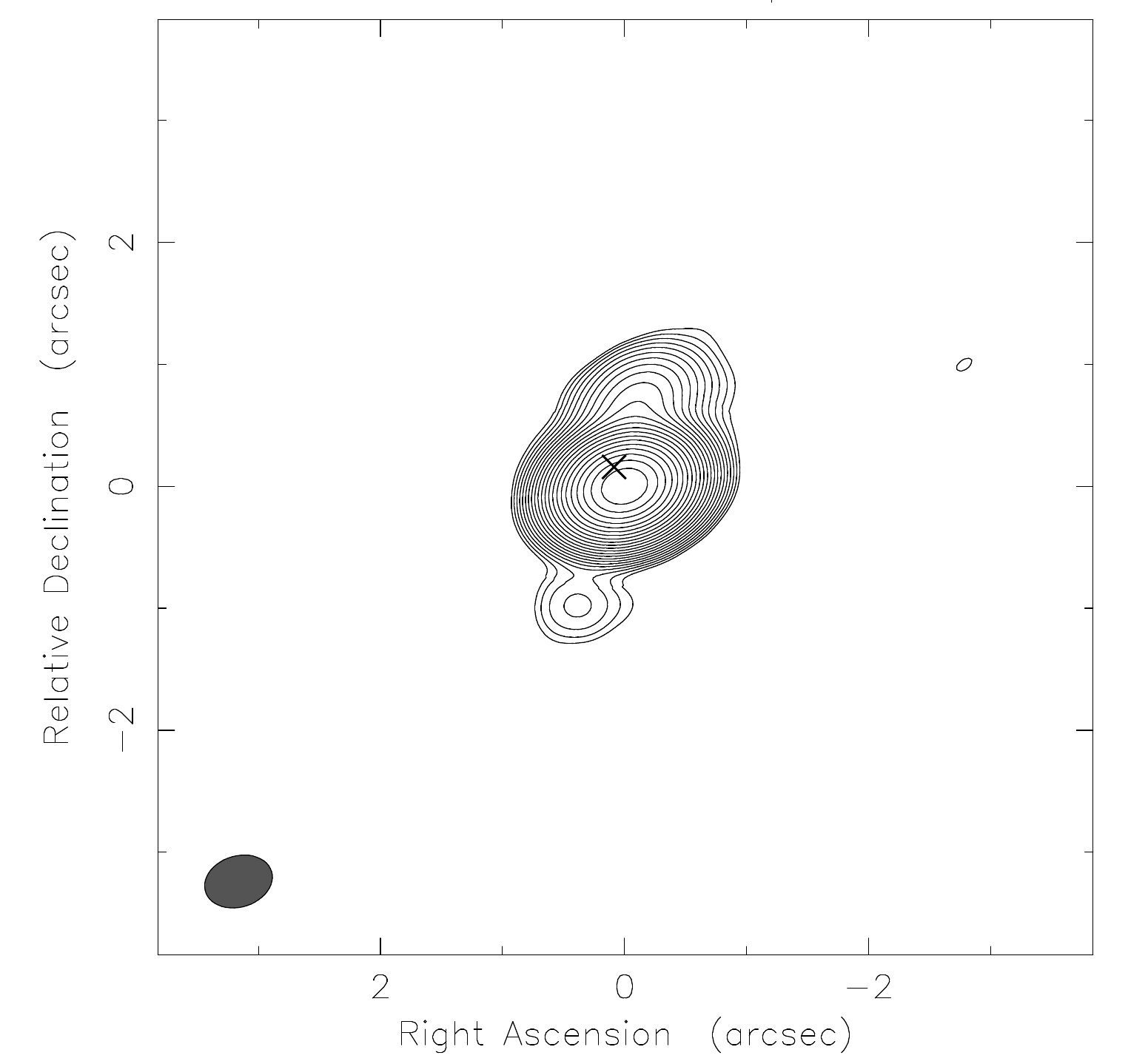} \label{fig:Images15-3}} 
\subfloat[Part 6][J1655+3242 at 4.860 GHz]{\includegraphics[width=2.2in]{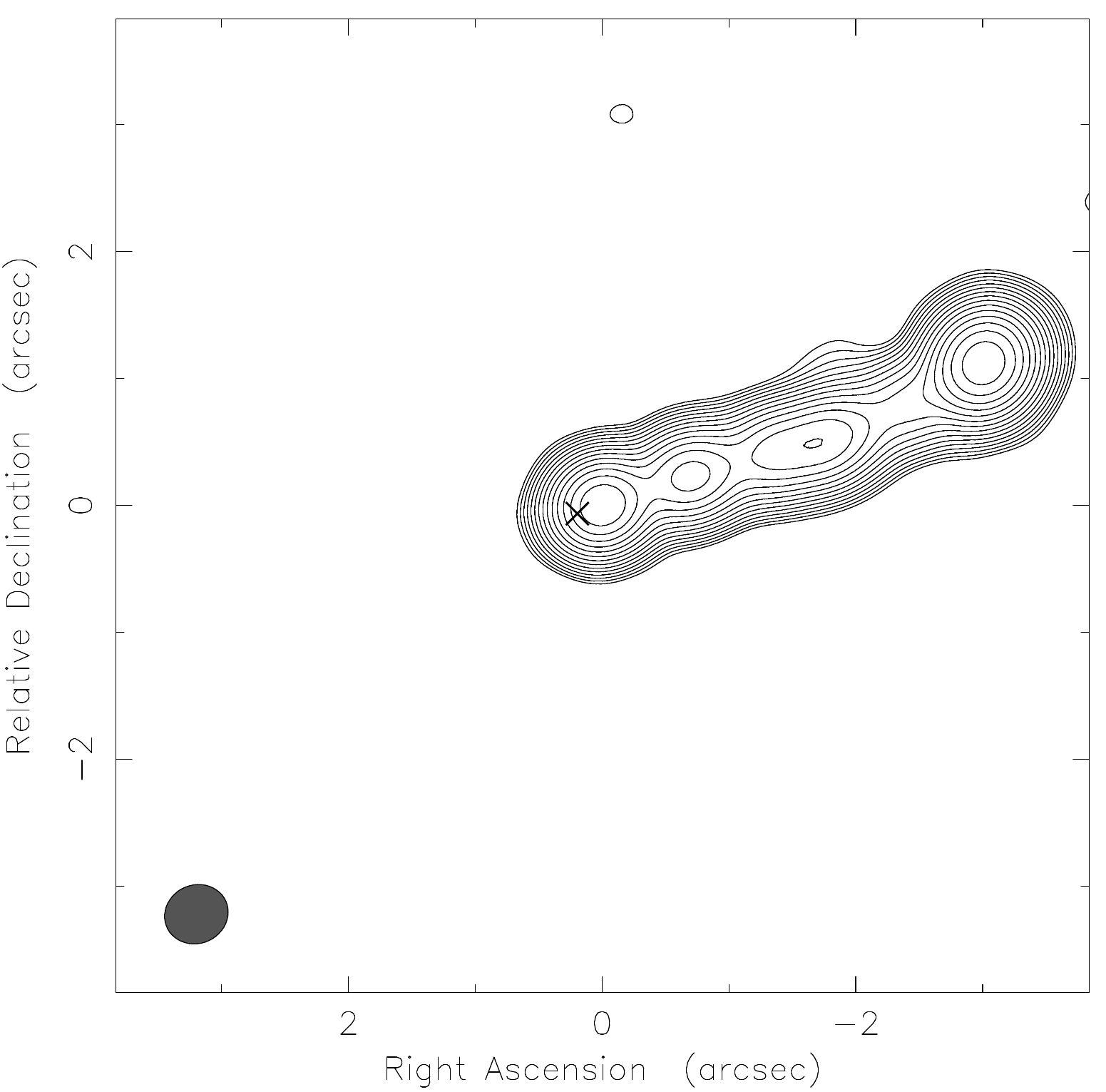} \label{fig:Images15-4}} \\
\subfloat[Part 7][J1704+0134 at 1.425 GHz]{\includegraphics[width=2.2in]{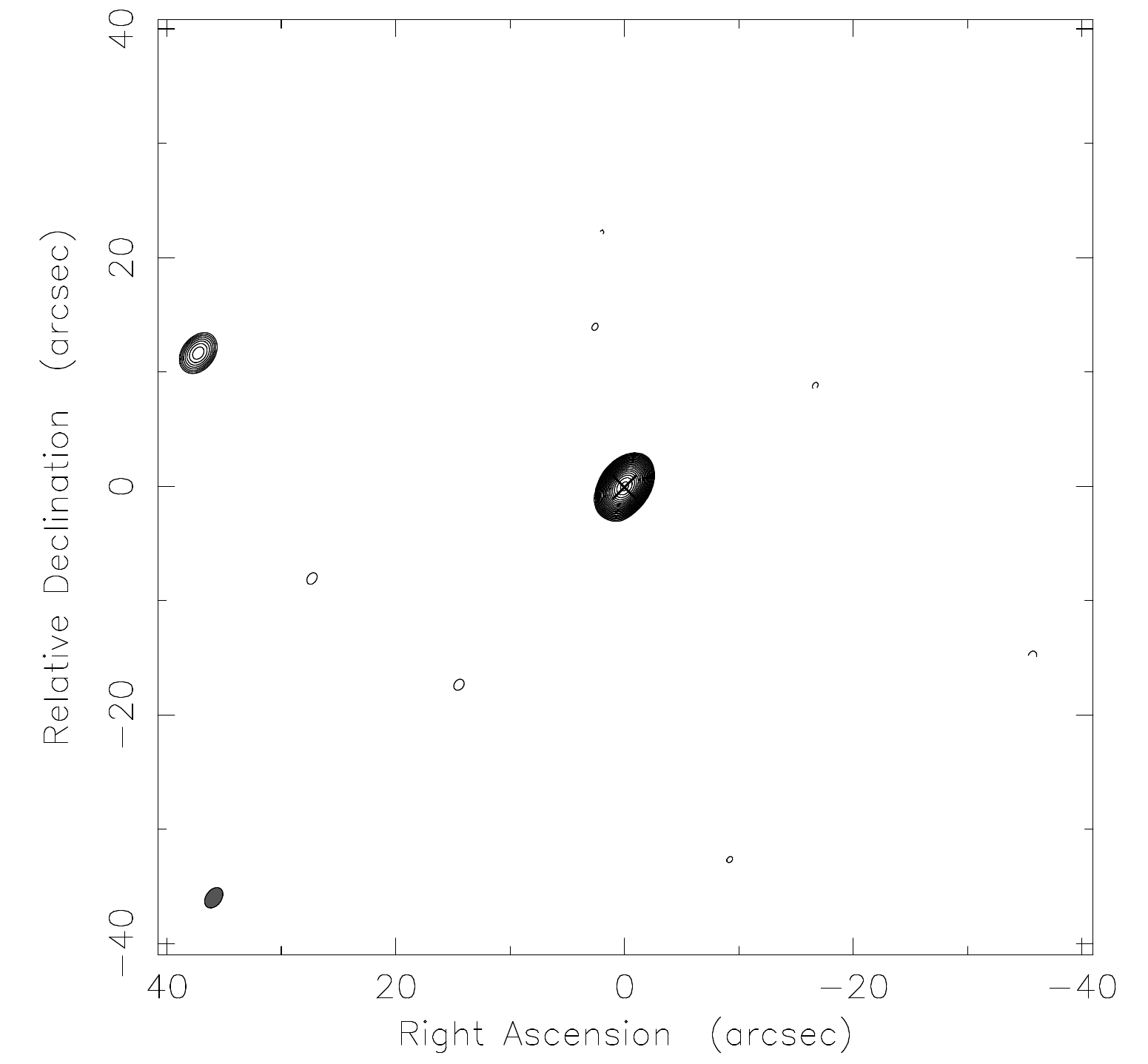} \label{fig:Images16-1}} 
\subfloat[Part 8][J1715+2145 at 8.440 GHz]{\includegraphics[width=2.2in]{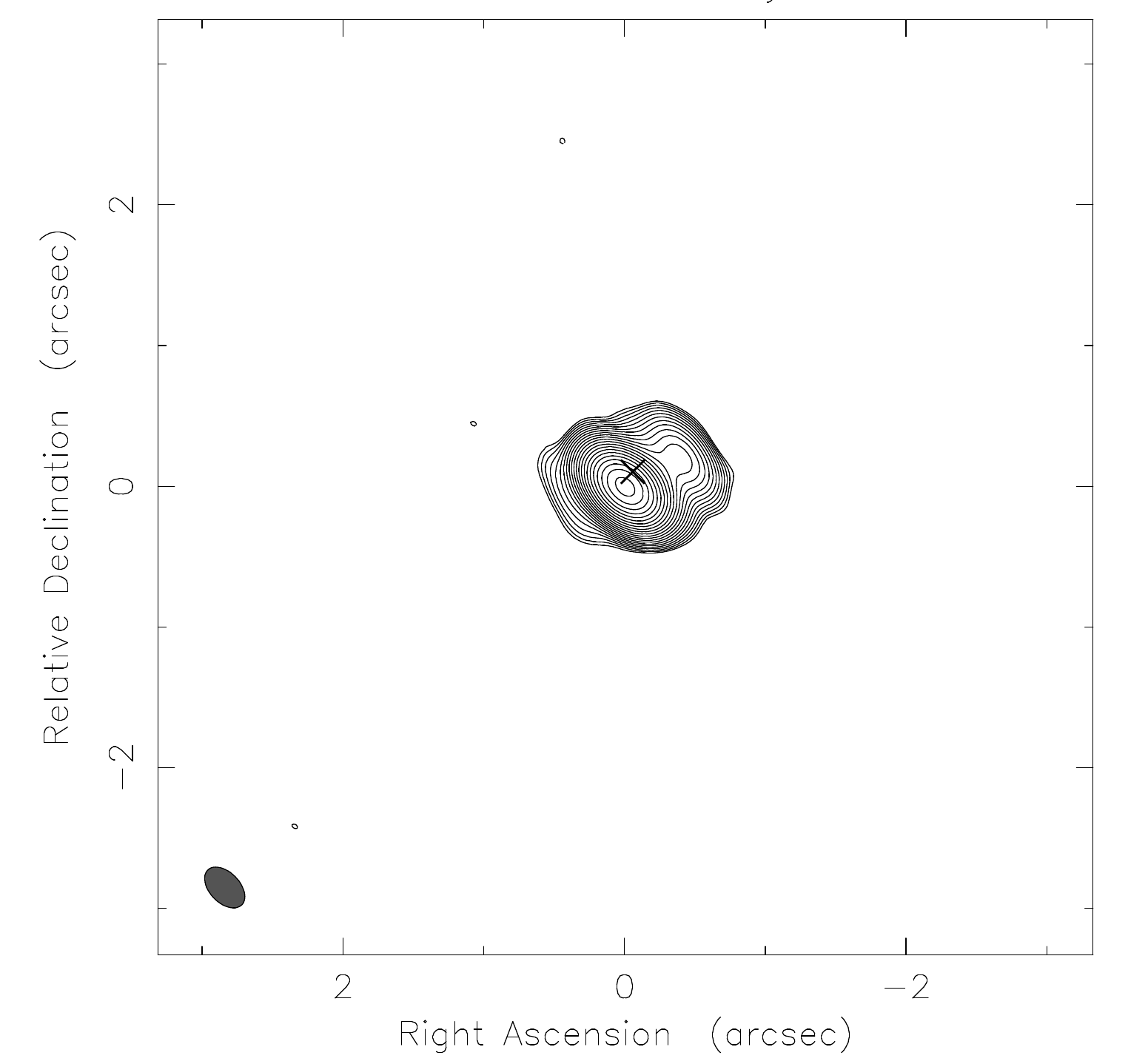} \label{fig:Images16-2}} 
\caption{}
\label{fig:Images8}
\end{figure}

\end{document}